\documentclass[12pt,a4paper]{article}

\usepackage{amsmath}
\usepackage{mathtools}
\usepackage{amsfonts}
\usepackage{amssymb}
\usepackage{graphicx}
\usepackage{color}
\usepackage{booktabs}
\usepackage{inputenc}
\usepackage[T1]{fontenc}
\usepackage{mathrsfs}
\usepackage{enumerate}
\usepackage{xcolor,url}
\usepackage{cancel,dsfont} 
\usepackage[normalem]{ulem} 

\newcommand{\eea}{\end{eqnarray}}
\newcommand{\bea}{\begin{eqnarray}}
\newcommand{\be}{\begin{equation}}
\newcommand{\ee}{\end{equation}}

\newcommand{\nuo}{\nu_1}
\newcommand{\nut}{\nu_2}

\newcommand{\kvec}{\vec{k}}
\newcommand{\qvec}{\vec{q}}
 \newcommand{\momspmeas}[1]{\frac{d^3 #1}{(2 \pi)^3}}

\def\be{\begin{equation}}
\def\ee{\end{equation}}

\def\knl{k_{\rm NL}}
\renewcommand{\(}{\left(}
\renewcommand{\)}{\right)}

\newcommand{\vk}{\vec k}

\newcommand{\vs}{\mathbf{s}} 
\newcommand{\km}{k_{\rm M}}

\def\hinvMpc{h\,{\rm Mpc}^{-1}}
\def\Mpcinvh{{\rm Mpc}/h}

\newcommand{\eV}{{\rm eV}}

\newcommand{\kmax}{k_{\rm max }}
\newcommand{\smin}{s_{\rm min }}

\newcommand{\code}[1]{\texttt{#1}}

\usepackage[colorlinks,bookmarks]{hyperref}
\definecolor{linkblue}{rgb}{0,0,0.8}
\definecolor{linkgreen}{rgb}{0,0.5,0}

\hypersetup{pdfpagemode=UseNone, pdfstartview=FitH, linkcolor=linkblue,
            citecolor=linkgreen, urlcolor=linkblue}

\font\BF=cmmib10
\def\k{{\hbox{\BF k}}}
\def\q{{\hbox{\BF q}}}
\def\z{{\hbox{\BF z}}}

\begin{document}

\begin{center}

{\Large \bf BOSS Correlation Function Analysis\\[0.3cm] from the Effective Field Theory of Large-Scale Structure}  \\[0.7cm]

{\large  Pierre Zhang${}^{1,2,3}$, Guido D'Amico${}^{4,5}$,  Leonardo Senatore${}^{6}$,  \\[0.3cm]
Cheng Zhao${}^{7}$, Yifu Cai${}^{1,2,3}$ \\[0.7cm]}

\end{center}

\begin{center}

\vspace{.0cm}

{\normalsize { \sl $^{1}$ Department of Astronomy, School of Physical Sciences, \\
University of Science and Technology of China, Hefei, Anhui 230026, China}}\\
\vspace{.3cm}

{\normalsize { \sl $^{2}$ CAS Key Laboratory for Research in Galaxies and Cosmology, \\
University of Science and Technology of China, Hefei, Anhui 230026, China}}\\
\vspace{.3cm}

{\normalsize { \sl $^{3}$ School of Astronomy and Space Science, \\
University of Science and Technology of China, Hefei, Anhui 230026, China}}\\
\vspace{.3cm}

{\normalsize { \sl $^{4}$ Department of Mathematical, Physical and Computer Sciences,\\ University of Parma, 43124 Parma, Italy}}\\
\vspace{.3cm}

{\normalsize { \sl $^{5}$ INFN Gruppo Collegato di Parma, 43124 Parma, Italy}}\\
\vspace{.3cm}

{\normalsize { \sl $^{6}$ Institut fur Theoretische Physik, ETH Zurich,
8093 Zurich, Switzerland}}\\
\vspace{.3cm}

{\normalsize { \sl $^{7}$ Institute of Physics, Laboratory of Astrophysics,\\
 \'Ecole Polytechnique F\'ed\'erale de Lausanne (EPFL), Observatoire de Sauverny, CH-1290 Versoix, Switzerland}}\\
\vspace{.3cm}

\vspace{.3cm}

\end{center}

\hrule \vspace{0.3cm}
{\small  \noindent \textbf{Abstract} After calibrating the predictions of the Effective Field Theory of Large-Scale Structure against several sets of simulations, as well as implementing a new method to assert the scale cut of the theory without the use of any simulation, we analyze the Full Shape of the BOSS Correlation Function.
Imposing a prior from Big Bang Nucleosynthesis on the baryon density, we are able to measure all the parameters in $\Lambda$CDM + massive neutrinos in normal hierarchy, except for the total neutrino mass, which is just bounded.
When combining the BOSS Full Shape with the Baryon Acoustic Oscillation measurements from BOSS, 6DF/MGS and eBOSS, we determine the present day Hubble constant, $H_0$, the present day matter fraction, $\Omega_m$, the amplitude of the {primordial} power spectrum, $A_s$, and the tilt of the primordial power spectrum, $n_s$,
to $1.4 \%, 4.5 \%, 23.5\%$ and $7.6\%$ precision, respectively, at $68 \%$-confidence level, finding $H_0=68.19 \pm 0.99$ (km/s)/Mpc, $\Omega_m=0.309\pm 0.014$, $\ln (10^{10}A_{s })=3.12^{+0.21}_{-0.26}$ and $n_s=0.963^{+0.062}_{-0.085}$, and we bound the total neutrino mass to $0.87 \, \eV$ at $95 \%$-confidence level. 
These constraints are fully consistent with Planck results and the ones obtained from BOSS power spectrum analysis. 
In particular, we find no tension in $H_0$ or $\sigma_8$ with Planck measurements, finding consistency at $1.2\sigma$ and $0.6\sigma$, respectively. 
\noindent

\vspace{0.3cm}}
\hrule

\vspace{0.3cm}
\newpage

\tableofcontents

\section{Introduction and Summary\label{sec:intro}}

\paragraph{Introduction:} In the last couple of years, the Effective Field Theory of Large-Scale Structure~(EFTofLSS) has been applied to the analysis of the Full Shape (FS) of the Power Spectrum (PS) of the BOSS galaxy-clustering data by using the {EFTofLSS prediction at one-loop order}~\cite{DAmico:2019fhj,Ivanov:2019pdj,Colas:2019ret}.
Ref.~\cite{DAmico:2019fhj} has also analyzed the BOSS galaxy-clustering bispectrum monopole using the tree-level prediction.
These analyses have produced measurements of all the $\Lambda$CDM cosmological parameters using just  a prior from Big Bang Nucleosynthesis (BBN), achieving extremely good measurements for some parameters such as the present amount of matter, $\Omega_m$, or the Hubble constant (see also~\cite{Philcox:2020vvt,DAmico:2020kxu} for subsequent refinements), whose error bars are not far from the ones obtained from the Cosmic Microwave Background (CMB)~\cite{Planck:2018vyg}.
Quintessence models have also been investigated, finding $\lesssim 5\%$ limits on the dark energy equation of state $w$ parameter using only late-time measurements~\cite{DAmico:2020kxu,DAmico:2020tty}, which is again not far from the ones obtained with the CMB~\cite{Planck:2018vyg}. 

In particular, the measurements of the Hubble constant represent a novel, CMB-independent, way of determining this parameter~\cite{DAmico:2019fhj}, which is already comparable in precision with the measurements obtained from  the cosmic ladder~\cite{Riess:2019cxk}.
In fact, very recently, this capability has been employed to show how some models that were proposed to alleviate the discrepancy between the CMB and cosmic-ladder measurements of the Hubble constant (the so called Hubble tension~\cite{Verde:2019ivm}) do not actually significantly improve the concordance once the BOSS data are analyzed with a controlled model such as the EFTofLSS~\cite{DAmico:2020ods,Ivanov:2020ril} (see also~\cite{Niedermann:2020qbw,Smith:2020rxx}). 

Of course, these results did not come effortlessly.
An intense and years-long line of study was needed to develop the EFTofLSS from the initial formulation to the level {that allows it} to be applied to data.
We therefore find it fair to add the following footnote in every paper where the EFTofLSS is used to analyze observational data.
Even though some of the mentioned papers are not strictly required to analyze the data, we believe that we, and probably {anybody} else, would {not} have applied the EFTofLSS to data without all these intermediate results~\footnote{The initial formulation of the EFTofLSS was performed in Eulerian space in~\cite{Baumann:2010tm,Carrasco:2012cv}, and subsequently extended to Lagrangian space in~\cite{Porto:2013qua}.
The dark matter power spectrum has been computed at one-, two- and three-loop orders in~\cite{Carrasco:2012cv, Carrasco:2013sva, Carrasco:2013mua, Carroll:2013oxa, Senatore:2014via, Baldauf:2015zga, Foreman:2015lca, Baldauf:2015aha, Cataneo:2016suz, Lewandowski:2017kes,Konstandin:2019bay}.
These calculations were accompanied by some  theoretical developments of the EFTofLSS, such as a careful understanding of renormalization~\cite{Carrasco:2012cv,Pajer:2013jj,Abolhasani:2015mra} (including rather-subtle aspects such as lattice-running~\cite{Carrasco:2012cv} and a better understanding of the velocity field~\cite{Carrasco:2013sva,Mercolli:2013bsa}), of several ways for extracting the value of the counterterms from simulations~\cite{Carrasco:2012cv,McQuinn:2015tva}, and of the non-locality in time of the EFTofLSS~\cite{Carrasco:2013sva, Carroll:2013oxa,Senatore:2014eva}.
These theoretical explorations also include an enlightening study in 1+1 dimensions~\cite{McQuinn:2015tva}.
An IR-resummation of the long displacement fields had to be performed in order to reproduce the Baryon Acoustic Oscillation (BAO) peak, giving rise to the so-called IR-Resummed EFTofLSS~\cite{Senatore:2014vja,Baldauf:2015xfa,Senatore:2017pbn,Lewandowski:2018ywf,Blas:2016sfa}. 
Accounts of baryonic effects were presented in~\cite{Lewandowski:2014rca,Braganca:2020nhv}. The dark-matter bispectrum has been computed at one-loop in~\cite{Angulo:2014tfa, Baldauf:2014qfa}, the one-loop trispectrum in~\cite{Bertolini:2016bmt}, and the displacement field in~\cite{Baldauf:2015tla}.
The lensing power spectrum has been computed at two loops in~\cite{Foreman:2015uva}.
Biased tracers, such as halos and galaxies, have been studied in the context of the EFTofLSS in~\cite{ Senatore:2014eva, Mirbabayi:2014zca, Angulo:2015eqa, Fujita:2016dne, Perko:2016puo, Nadler:2017qto,Donath:2020abv} (see also~\cite{McDonald:2009dh}), the halo and matter power spectra and bispectra (including all cross correlations) in~\cite{Senatore:2014eva, Angulo:2015eqa}. Redshift space distortions have been developed in~\cite{Senatore:2014vja, Lewandowski:2015ziq,Perko:2016puo}. 
Neutrinos have been included in the EFTofLSS in~\cite{Senatore:2017hyk,deBelsunce:2018xtd}, clustering dark energy in~\cite{Lewandowski:2016yce,Lewandowski:2017kes,Cusin:2017wjg,Bose:2018orj}, and primordial non-Gaussianities in~\cite{Angulo:2015eqa, Assassi:2015jqa, Assassi:2015fma, Bertolini:2015fya, Lewandowski:2015ziq, Bertolini:2016hxg}.
Faster evaluation schemes for the calculation of some of the loop integrals have been developed in~\cite{Simonovic:2017mhp}.
Comparison with high-quality $N$-body simulations to show that the EFTofLSS can accurately recover the cosmological parameters have been performed in~\cite{DAmico:2019fhj,Colas:2019ret,Nishimichi:2020tvu,Chen:2020zjt}}.

In this paper, after calibrating the scale cut of the model against several sets of simulations as well as implementing a new method to assert the scale cut of the theory without the use of any simulation in Section~\ref{sec:scalecut}, in Section~\ref{sec:results} we analyze the FS of the BOSS Correlation Function~(CF).
The reason why it is worthwhile to investigate the CF is twofold. First, all the information available in the Baryon Acoustic Oscillation (BAO) peak from the two point function is easily recovered in full in the CF analysis, while, in the PS, part of the signal resides at wavenumbers too high for a straightforward analysis to apply.
Second, observational systematics can play a different role in this observable. 
We therefore analyze the FS of BOSS CF pre-reconstructed multipoles using the EFTofLSS.
The results are summarized below and discussed in Section~\ref{sec:results}. 
A careful comparison with results obtained fitting the FS of BOSS PS measurements using the EFTofLSS is also presented there. 
We provide formulas and details on the evaluation of the redshift-space galaxy CF at one loop in the EFTofLSS and on the posterior sampling in App.~\ref{app:redcf}.
In App.~\ref{app:nnlo_ps}, we determine the scale cut for BOSS PS FS analysis without relying on simulations as put forward for the CF in Section~\ref{sec:scalecut}.
The CF best fits are given in App.~\ref{app:bestfit}.
Finally, we check in App.~\ref{app:selection} how our results are affected by line-of-sight selection effects. 

\paragraph{Data sets:}
We separate the BOSS DR12 data into two redshift bins, $0.2<z<0.43$ and $0.43<z<0.7$, respectively named LOWZ and CMASS. 
The CF FS data are measured using the Landy-Szalay estimator~\cite{Landy:1993yu} in fine bins ($1 \, \Mpcinvh$) in separation $s$ and $\mu = \mathbf{\hat{s}} \cdot \mathbf{\hat{z}}$, where $\mathbf{\hat{z}}$ is the line-of-sight direction. 
Systematic effects have been corrected by appropriate weights, as described in~\cite{Reid:2015gra}.
We bin in separation $s$ by $\Delta s = 5 \, \Mpcinvh$ and into two multipoles, the monopole and the quadrupole. 
The covariances are built from measurements on 2048 patchy mocks~\cite{Kitaura:2015uqa}. 
We have checked that fitting the data with a covariance built with about half of the mocks (1000) instead of the 2048 mocks leads to the same cosmological constraints, with at most $0.2\sigma$ shift in $\Omega_m$, and $\lesssim 0.1 \sigma$ in the other parameters. 
As those shifts are negligibly small, this validates our estimation of the covariance. 

In order to perform a careful comparison between the results of the BOSS CF and PS analyses, we measure the BOSS PS multipoles on the same catalog and with same redshift selection, using the estimator described in~\cite{Yamamoto:2005dz}.  
We use Piecewise Cubic Spline (PCS) particle assignment scheme, with grid interlacing as described in~\cite{Sefusatti:2015aex}, and a grid size consisting of $512^3$ cells.

We also include the baryon acoustic oscillations (BAO) of BOSS DR12 post-reconstructed power spectrum measurements~\cite{Gil-Marin:2015nqa} obtained in~\cite{DAmico:2020kxu} using standard BAO extraction analysis. When quoting results as just `BOSS', we refer to BOSS pre-reconstructed CF FS combined with BOSS post-reconstructed BAO. Moreover, we will consider various combinations with other experiments: measurements at small redshift from 6DF~\cite{Beutler:2011hx} and SDSS DR7 MGS~\cite{Ross:2014qpa}, as well as high redshift Lyman-$\alpha$ forest auto-correlation and cross-correlation with quasars from eBOSS DR14 BAO measurements~\cite{Agathe:2019vsu, Blomqvist:2019rah}, that we will collectively refer as `ext. BAO'; Supernovae (SN) measurements from the Pantheon sample~\cite{Scolnic:2017caz}; and finally, Planck2018 TT,TE,EE+lowE + lensing~\cite{Planck:2018vyg}. 
The inclusion of post-reconstructed BOSS BAO measurements gives a non-negligible improvement because the reconstruction amounts to using higher $n$-point functions.
Importantly, the pre- and post-reconstruction BOSS BAO measurements are correlated.
This is taken into account as in~\cite{DAmico:2020kxu} (see also~\cite{Philcox:2020vvt}).
When combining with other experiments, we simply add the log-likelihoods, since all the measurements refer to separate redshift bins. 
The small cross-correlation of the galaxy clustering data with the Planck weak lensing and integrated Sachs-Wolfe effect is neglected. 

\begin{figure}[h]
  \centering
  \includegraphics[width=0.32\textwidth]{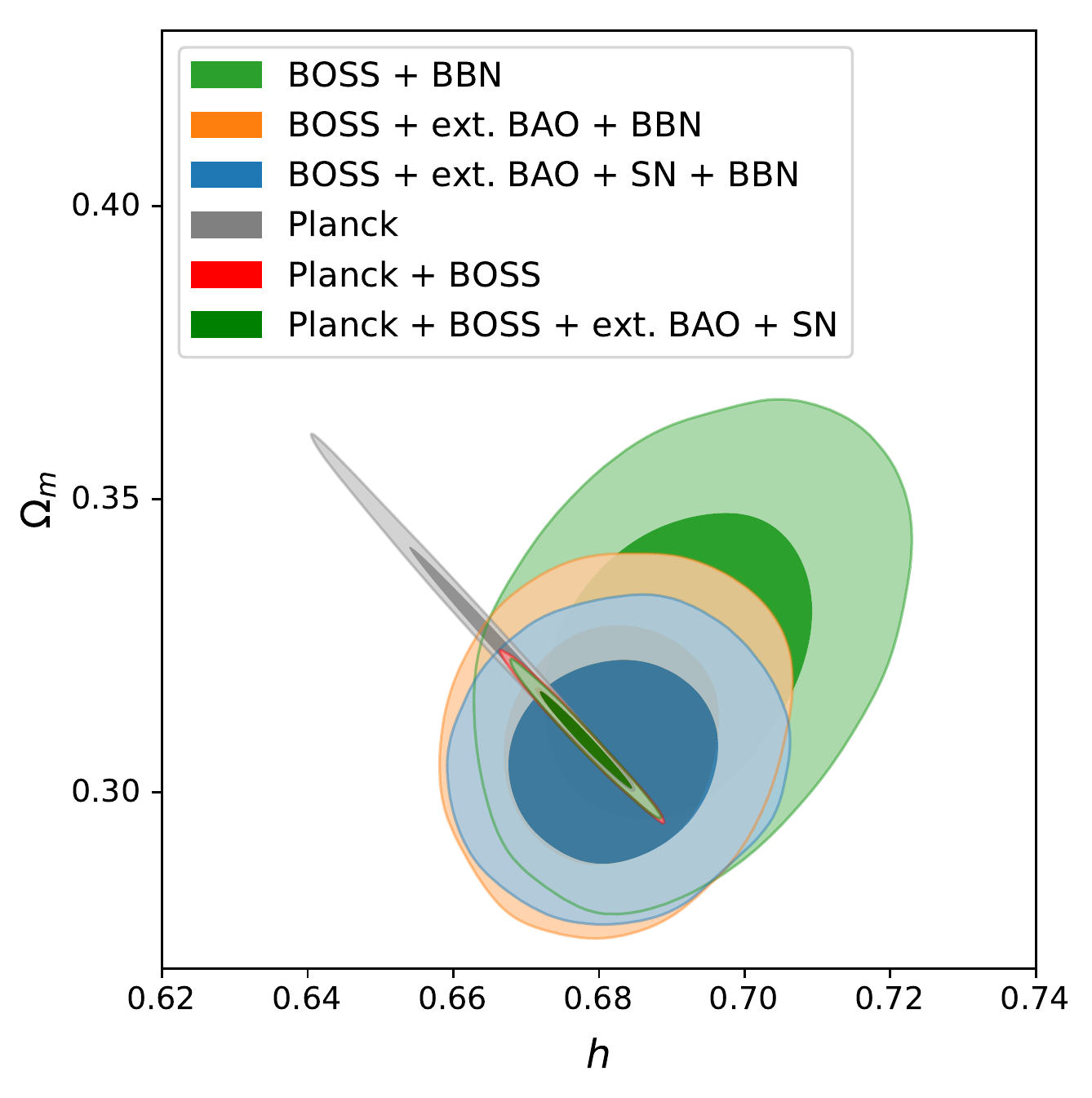}
  \includegraphics[width=0.32\textwidth]{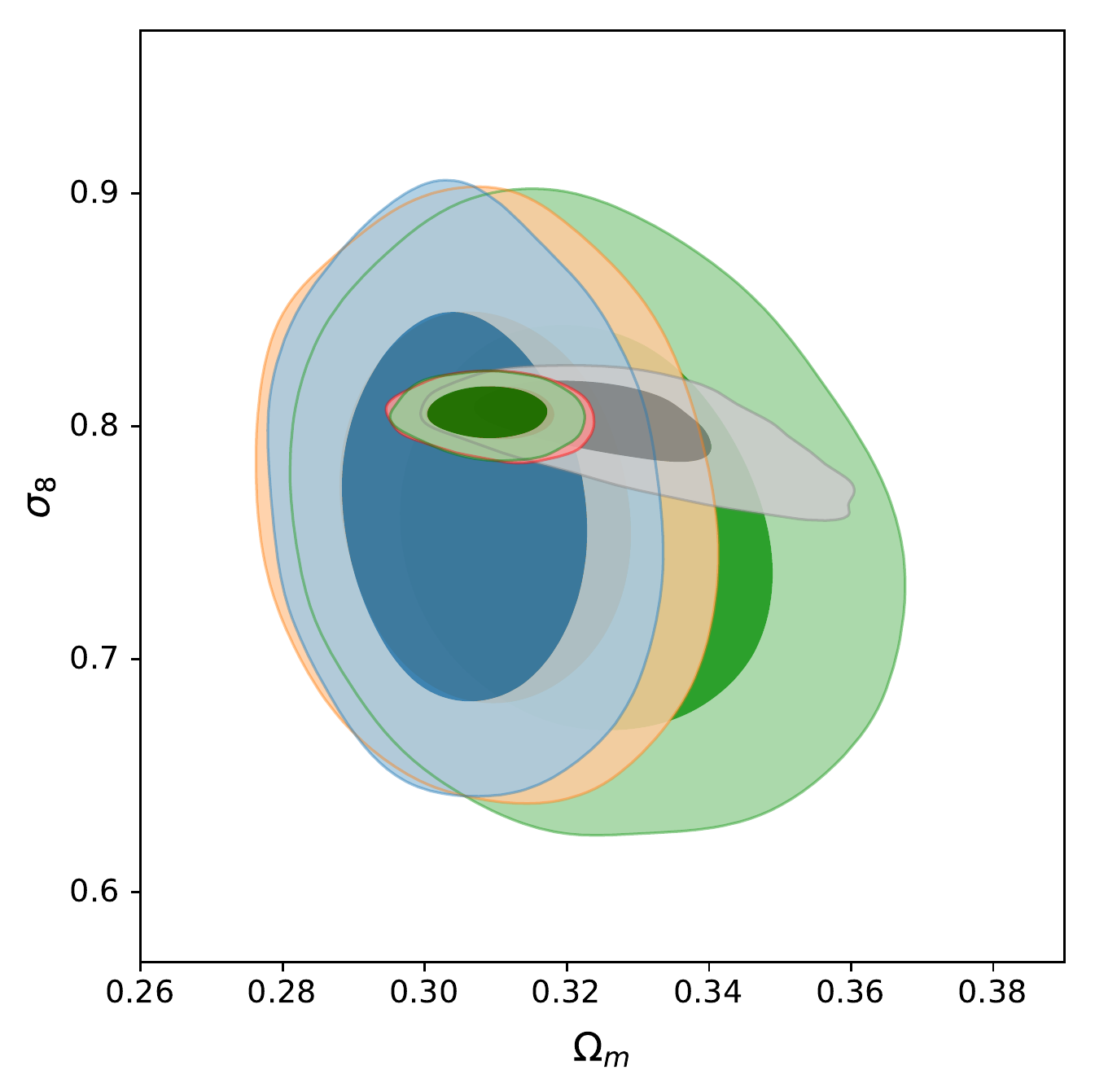}
    \includegraphics[width=0.32\textwidth]{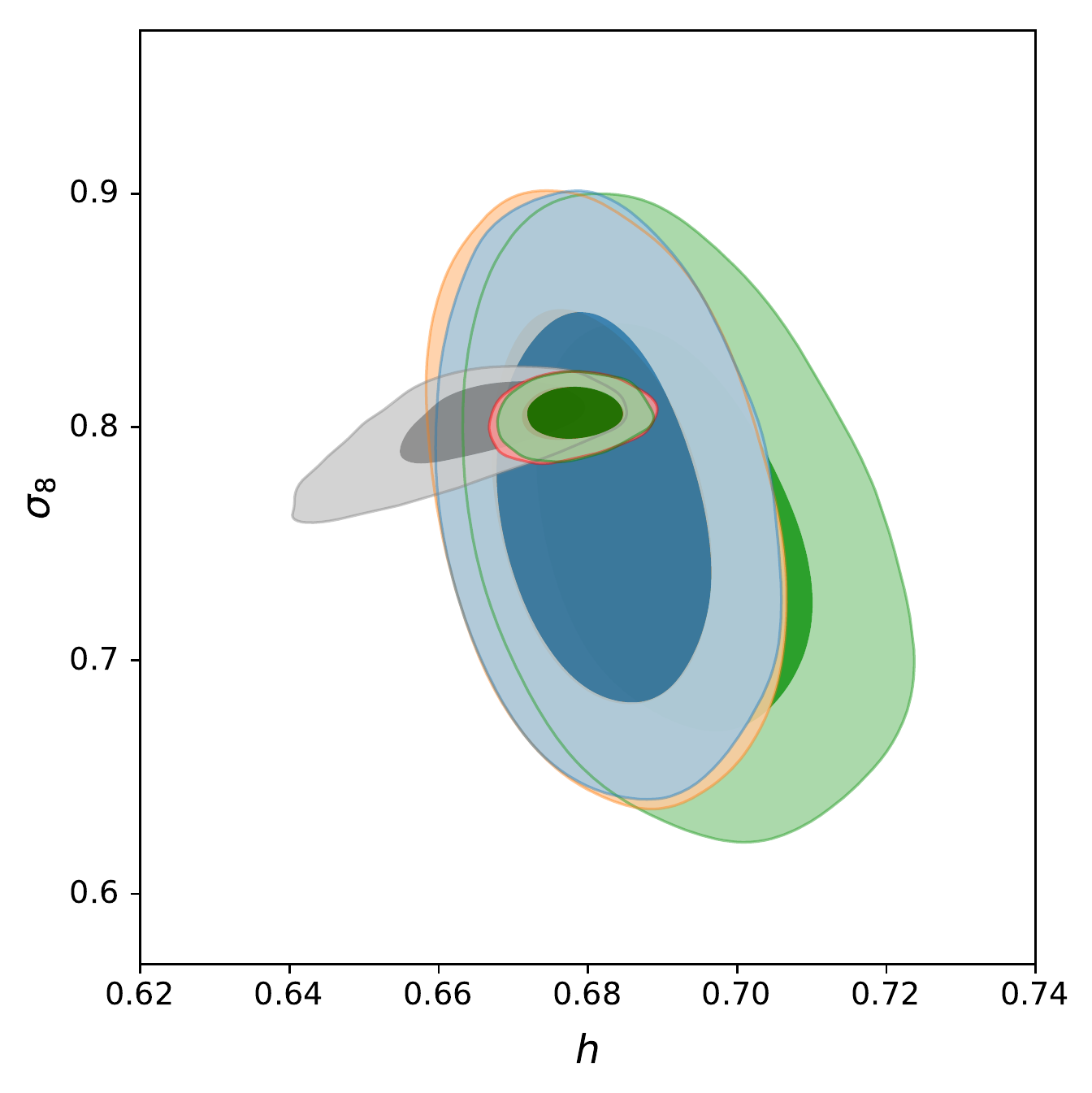}\\
    \vspace{0.3em}
    \scriptsize
  \begin{tabular}{c|c|c|c|c|c|c} 
  \hline
  Limits		& BOSS+BBN	 		& \begin{tabular}{@{}c@{}}BOSS+BBN\\+ext. BAO \end{tabular} & \begin{tabular}{@{}c@{}}BOSS+BBN\\+ext. BAO+SN \end{tabular} 		& Planck	 		& Planck+BOSS & \begin{tabular}{@{}c@{}}Planck+BOSS\\+ext. BAO+SN \end{tabular} \\ \hline
  $h$ 			& $0.691 \pm 0.012$ 		& $0.6819 \pm 0.0099$ 					& $0.6822 \pm 0.0096$ 		& $0.6655_{-0.0067}^{+0.0110}$ 	& $0.6776 \pm 0.0046$ 		& $0.6780 \pm 0.0043$ \\ 
  $\Omega_m$ 		& $0.323 \pm 0.018$			& $0.309 \pm 0.014$						& $0.306 \pm 0.012$			& $0.3262_{-0.0150}^{+0.0092}$	& $0.3097 \pm 0.0060$		& $0.3090 \pm 0.0056$ \\
  $\sigma_8$ 		& $0.756 \pm 0.058$			& $0.766 \pm 0.057$						& $0.766 \pm 0.056$ 		& $0.8004 \pm 0.012$			& $0.8052 \pm 0.0075$		& $0.8054 \pm 0.0071$ \\
  $\sum m_\nu$ 	& $< 1.15 {\rm eV} (2\sigma)$	& $< 0.87 {\rm eV} (2\sigma)$				& $< 0.74 {\rm eV} (2\sigma)$ 	& $< 0.26 {\rm eV} (2\sigma)$		& $< 0.14 {\rm eV} (2\sigma)$	& $< 0.14 {\rm eV} (2\sigma)$\\
  \hline
   \end{tabular}\\    \vspace{0.3em}
  
  \caption{\small $h-\Omega_m-\sigma_8$ contours and their 68\%-confidence intervals from the various analyses performed in this work. Here BOSS refers to BOSS pre-reconstructed CF FS combined with BOSS post-reconstructed BAO. For the total neutrino mass, we instead quote the 95\%-confidence bound. }
  \label{fig:ncdm_h_Omega_m}
  \end{figure}

\paragraph{Main Results:}
Using these data sets in various combinations, we measure all parameters in $\Lambda$CDM + massive neutrinos in normal hierarchy ($\nu\Lambda$CDM model).
When not analyzed with Planck, we use a BBN prior centered on $\omega_{b, {\rm BBN}} = 0.02233$ of width $\sigma_{\rm BBN} = 0.00036$~\cite{Mossa:2020gjc}. 
We also impose a flat prior of $[0.06, 1.5]$ eV on the sum of neutrino masses, that plays a negligible role. 
The main results of our analyses are maybe best represented by Fig.~\ref{fig:ncdm_h_Omega_m}. 
Fitting BOSS CF FS, we determine at $68\%$-confidence level (CL) $h$ to $1.8 \%$ precision, $\Omega_m$ to $5.7 \%$ precision, $A_s$ to $25\%$ precision and $n_s$ to $8.8\%$ precision, and also get a bound on the total neutrino mass of about $1.1 \, \eV$ at $95 \%$ CL.
Combining with BAO measurements from BOSS in cross-correlation and from 6DF/MGS and eBOSS, $h$ is determined to $1.4 \%$ precision and $\Omega_m$ to $4.5 \%$ precision. 
Notice that the precision of the measurements on $h$ and $\Omega_m$ is very close to the one of Planck for the $\nu\Lambda$CDM model, see Table~\ref{tab:planck} (see also~\cite{Planck:2018vyg}). 
Finally, adding SN data from Pantheon, the constraint on $\Omega_m$ improves to $3.8 \%$ precision, and the total neutrino mass is bounded to about $0.74 \, \eV$ at 95\% CL.
App.~\ref{app:selection} suggests that our results are robust to line-of-sight selection effects, once physical priors on the size of these terms are imposed. 

We find that these constraints from late-time probes, that are independent from Planck or the cosmic distance ladder, are completely consistent with Planck results: all parameters are consistent within $\lesssim 1.2\sigma$. 
In particular, we find no tension in $h$ or $\sigma_8$. 
Combining Planck and BOSS FS+BAO, we find that the constraints on $\Omega_m$ and $h$ are improved by $\sim 50\%$ with respect to the results of Planck alone, and we bound the neutrino total mass to $< 0.14$eV at $95\%$ CL.

We end this summary of the main results with a note of warning. It should be emphasized that in performing this analysis, as well as the preceding ones using the EFTofLSS by our group~\cite{DAmico:2019fhj,Colas:2019ret,DAmico:2020kxu,DAmico:2020ods,DAmico:2020tty}, we have assumed that the observational data are not affected by any unknown systematic error, such as, for example, selection effects beyond the ones we discuss in app.~\ref{app:selection} or undetected foregrounds. In other words, we have simply analyzed the publicly available data for what they were declared to be: the two-point function of the galaxy density in redshift space. Given the additional cosmological information that the theoretical modeling of the EFTofLSS allows us to exploit in BOSS data, it might be worthwhile to investigate if potential undetected systematic errors might affect our results. We leave an investigation of these  issues to future work.

\paragraph{Public Codes:}  
The predictions for the FS of the galaxy CF and PS in the EFTofLSS are obtained using \code{PyBird}: Python code for Biased tracers in Redshift space~\cite{DAmico:2020kxu}~\footnote{\href{https://github.com/pierrexyz/pybird}{https://github.com/pierrexyz/pybird}}.
The linear power spectra were computed with the \code{CLASS} Boltzmann code~\cite{Blas_2011}~\footnote{ \href{http://class-code.net}{http://class-code.net}}.
The posteriors were sampled using the \code{MontePython} cosmological parameter inference code~\cite{Brinckmann:2018cvx, Audren:2012wb}~\footnote{ \href{https://github.com/brinckmann/montepython\_public}{https://github.com/brinckmann/montepython\_public}}.
The plots have been obtained using the \code{GetDist} package~\cite{Lewis:2019xzd}. 
The FS of BOSS CF and PS, as well as the ones from the patchy mocks for the covariance, are measured using \code{FCFC} and \code{powspec}, respectively~\cite{Zhao:2020bib}~\footnote{\href{https://github.com/cheng-zhao/FCFC}{https://github.com/cheng-zhao/FCFC} ; \href{https://github.com/cheng-zhao/powspec}{https://github.com/cheng-zhao/powspec}}. 
The PS window functions have been measured as described in~\cite{Beutler:2018vpe} using \code{nbodykit}~\cite{Hand:2017pqn}~\footnote{\href{https://github.com/bccp/nbodykit}{https://github.com/bccp/nbodykit}}.

\section{Scale cuts}
\label{sec:scalecut}

Although the EFTofLSS has already been extensively tested against simulations for the power spectrum (see e.g. \cite{DAmico:2019fhj,Colas:2019ret,Nishimichi:2020tvu}), due to the different correlations in the data there is no direct translation between the Fourier space scale cuts and the configuration space ones.
We therefore repeat here, for the correlation function, the series of tests performed in \cite{DAmico:2019fhj,Colas:2019ret}, by fitting various sets of simulations.

We also present yet another way to assert the scale cut of the theory without relying on simulations but directly fitting the data, by measuring the shift in the posteriors upon adding an estimate for some of the next-to-next-to-leading-order (NNLO) terms, {{\it i.e.} the terms that we do not include in our predictions}.
Both calibration methods give the same (or a very close) answer and for BOSS CMASS data we find that we can fit the multipoles down to $s_{\rm min} = 20 \, \Mpcinvh$ without a significant theoretical systematic error ({\it i.e.} less than about $1/3$ of the error bars of the cosmological parameters measured on BOSS).  

\subsection{Tests against simulations}
We analyze several sets of simulations as described in \cite{DAmico:2019fhj}: the `lettered' challenges and the patchy mocks. 
In practice, we fit the CF FS monopole and quadrupole measured from those mocks, that are in redshift space. 
The covariance matrices are computed from the measurements of patchy mocks: when analyzing the FS of the  `lettered' challenges or the mean of the periodic patchy mocks, we will use the patchy periodic mocks of side length $2.5 \, {\rm Gpc}/h$, while when analyzing the mean of the patchy lightcone, we will use the patchy lightcone mocks, as described in Section~(1). 
In the following, we will be interested to find the scale cut, i.e. the minimal scale at which we fit those simulations, with a controlled theory-systematic error. We will find that we can fit them down to $s_{\rm min} = 20 \Mpcinvh$ such that the theory error is under control for BOSS data. 
For the longest scale, we will always use the maximal scale made available to us in the measurements, $s_{\rm max} = 180 \Mpcinvh$. 
We have checked that using instead $s_{\rm max} = 150 \Mpcinvh$ does not affect significantly our results, as discussed in App.~C. 
Let us now describe in details how we determine the scale cut using these simulations. 

First, we consider two independent realizations of side length $2.5 \, {\rm Gpc}/h$, one being populated by 4 different {halo occupation distribution (HOD)} models, labelled A, B, F, G, and the other one labelled D, populated by a different HOD model. 
These `lettered' challenge boxes are high-fidelity simulations that we use to calibrate the scale cut of the EFTofLSS. 
Details on the HOD models and other specifics of those simulations can be found in~\cite{BOSS:2016wmc}. 
{Since A, B, F, and G are correlated, we fit them separately and average the posteriors of the cosmological parameters over the 4 boxes, instead of taking the product of the posteriors}. 
We fit D {separately}.
Using one or either realization, we measure the theory-systematic error for each cosmological parameter as the distance of the 68\%-confidence interval of the 1D posterior to the truth.
In particular, if the truth lies within the $1\sigma$-region, we have no statistical evidence for the detection of a theory error, and we thus do not report one in this case. 
This already allows us to measure the theory-systematic error quite well, given the size of the simulations with respect to the data. 
However, we can do even better. 
As ABFG and D are independent realizations, we can combine them, allowing us to measure the theory-systematic error with a better precision by another factor $\sim \sqrt{2}$. 
In practice, we combine the individual 1D posteriors of the shifts from the truth of the mean of the 1D posteriors from boxes A, B, F and G with the independent realization D (as the product of two Gaussians). 
The theory-systematic error for each $\Lambda$CDM parameter is the distance from zero of the 68\%-confidence interval of the resulting 1D posteriors of the shifts. 
Given the number of cosmological parameters we actually measure, this represents a conservative requirement: 
after all, it is not extraordinary to find the mean of one or few parameters farther than $1\sigma$ to the truth in our multi-dimensional analysis. 
The combination of ABFG+D allows us to measure the theory systematics using a volume about $14$ times larger than the BOSS effective volume.
In practice, the minimally-measurable systematic errors that this procedure allows us to detect are the following fraction of the errors that we obtain on BOSS data: $0.33$, $0.42$, $0.26$, $0.33$ for $\Omega_{m}$, $h$, $\ln (10^{10} A_s)$ and $n_s$ respectively~\footnote{A fraction of systematic error equal to $0.42\sigma_{\rm data}$ on $h$ might not appear negligible. We however stress that we are not detecting such a large systematic error, we are simply unable to detect an error smaller than this. In fact, from the analysis of the subsequent section, we find indications that the systematic error is indeed smaller. Furthermore, if we were to correct our findings on the data by the  offset measured in simulations, we would need to add in quadrature the statistical error of the simulations to the one of the data.  Then, as we can see from Table~\ref{tab:challenge}, we would need to add $0.001\pm 0.006$ to our observed value of $h$. This would increase the error on $h$ just by a relative fraction of about 8\%, which is certainly negligible, and produce a negligible shift of the central value.  Similar considerations apply to the other cosmological parameters. 

{We do not perform these shifts in the posteriors on the data also for another reason. While it is believed that we can trust simulations to measure the overall size of `average$-$truth', it is unclear if we can trust them for the actual shift. This would motivate an alternative procedure to account for the theoretical systematic error measured from simulation: to simply add in quadrature  `average$-$truth' to our statistical errors on the data. As we said, this is imprecise because it would consider as systematic error deviations from truth that are within the $68\%$ C.L.; still, even doing this, for the worst case, which is given by $n_s$, would just degrade our error bars by $\sim15\%$, which is small.} We thank Chia-Hsun Chuang for stimulating discussions on this point.}.

\begin{figure}
\centering
\includegraphics[width=0.49\textwidth]{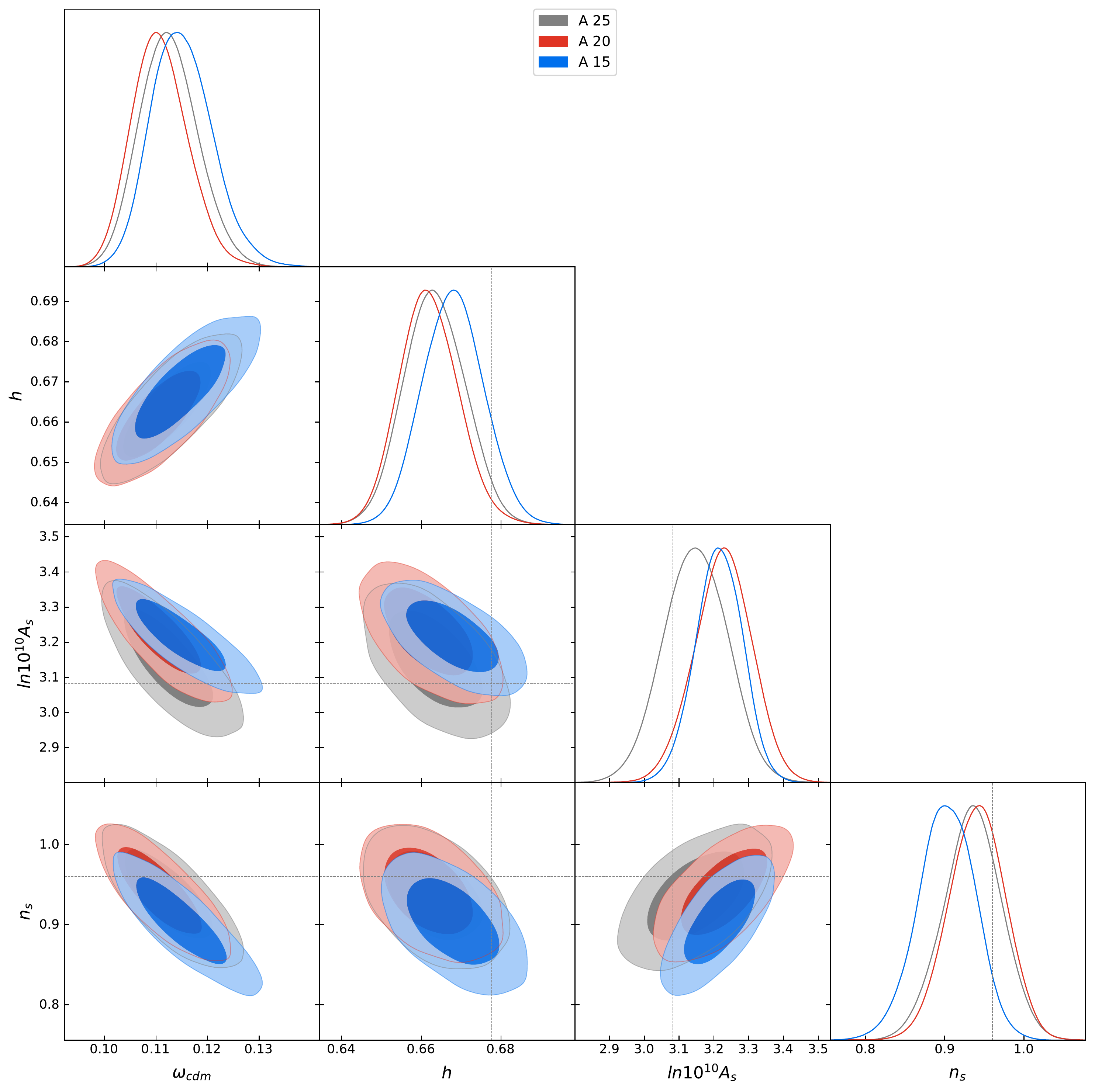}
\includegraphics[width=0.49\textwidth]{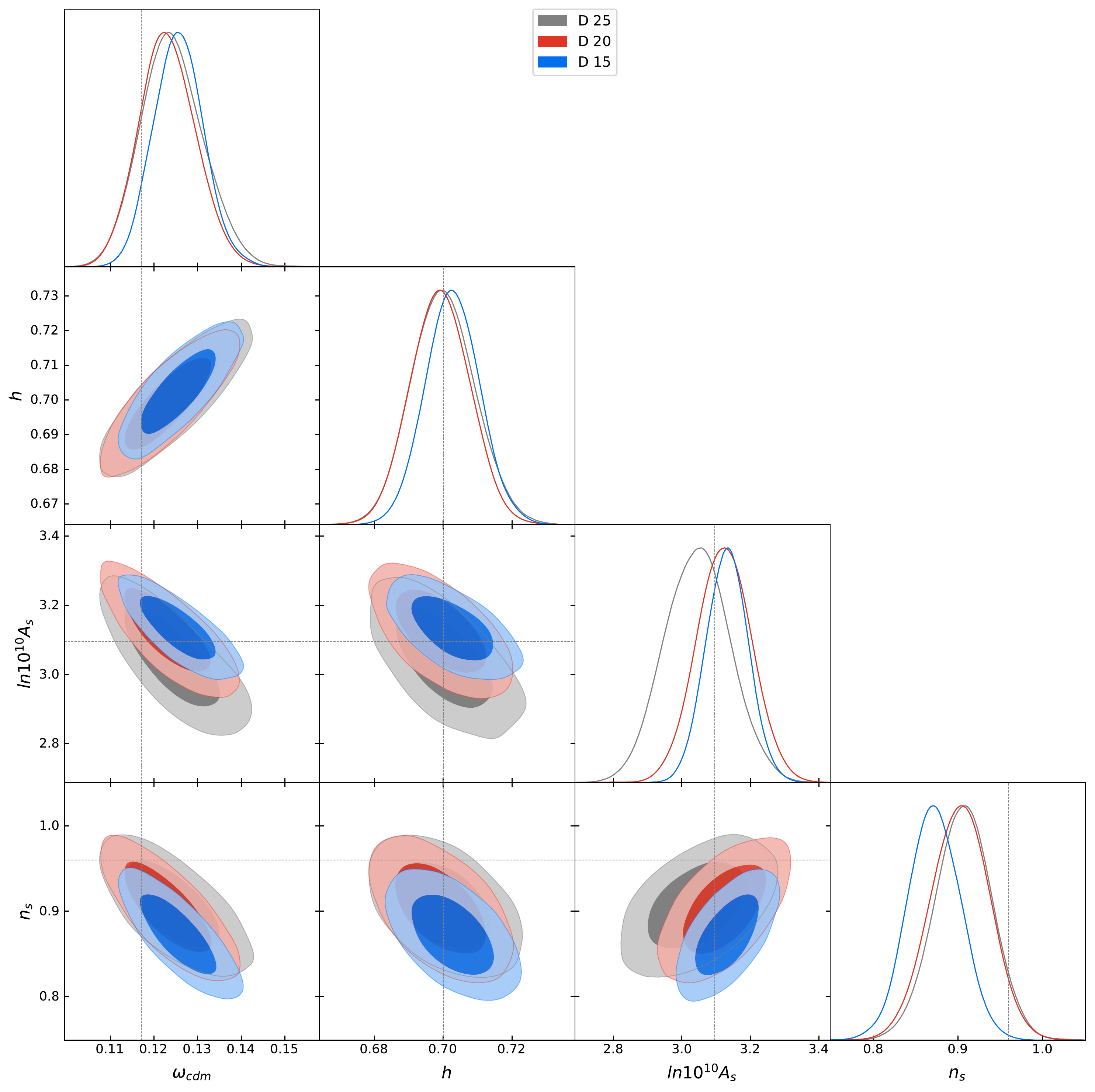}
\includegraphics[width=0.49\textwidth]{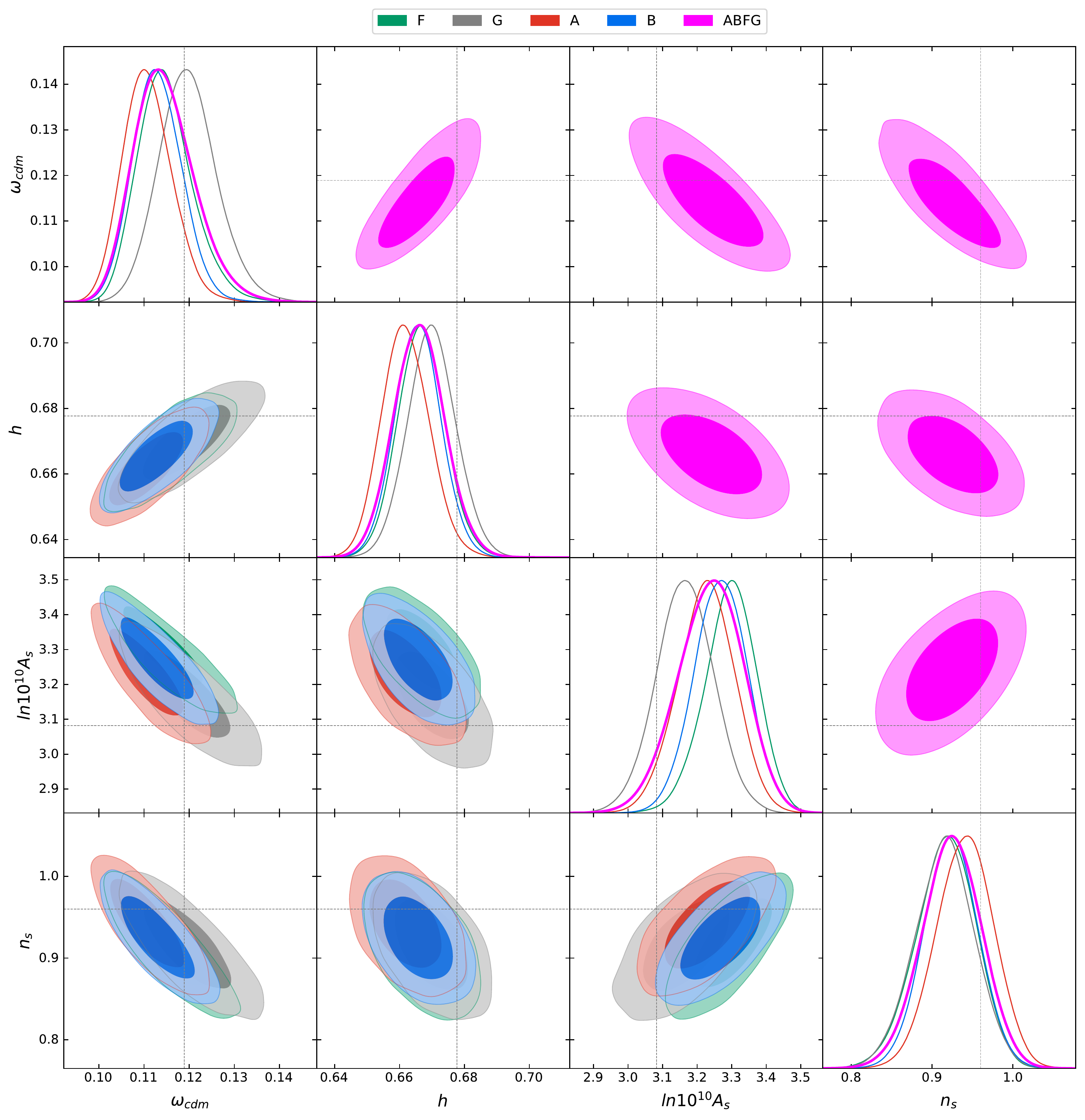}
\includegraphics[width=0.49\textwidth]{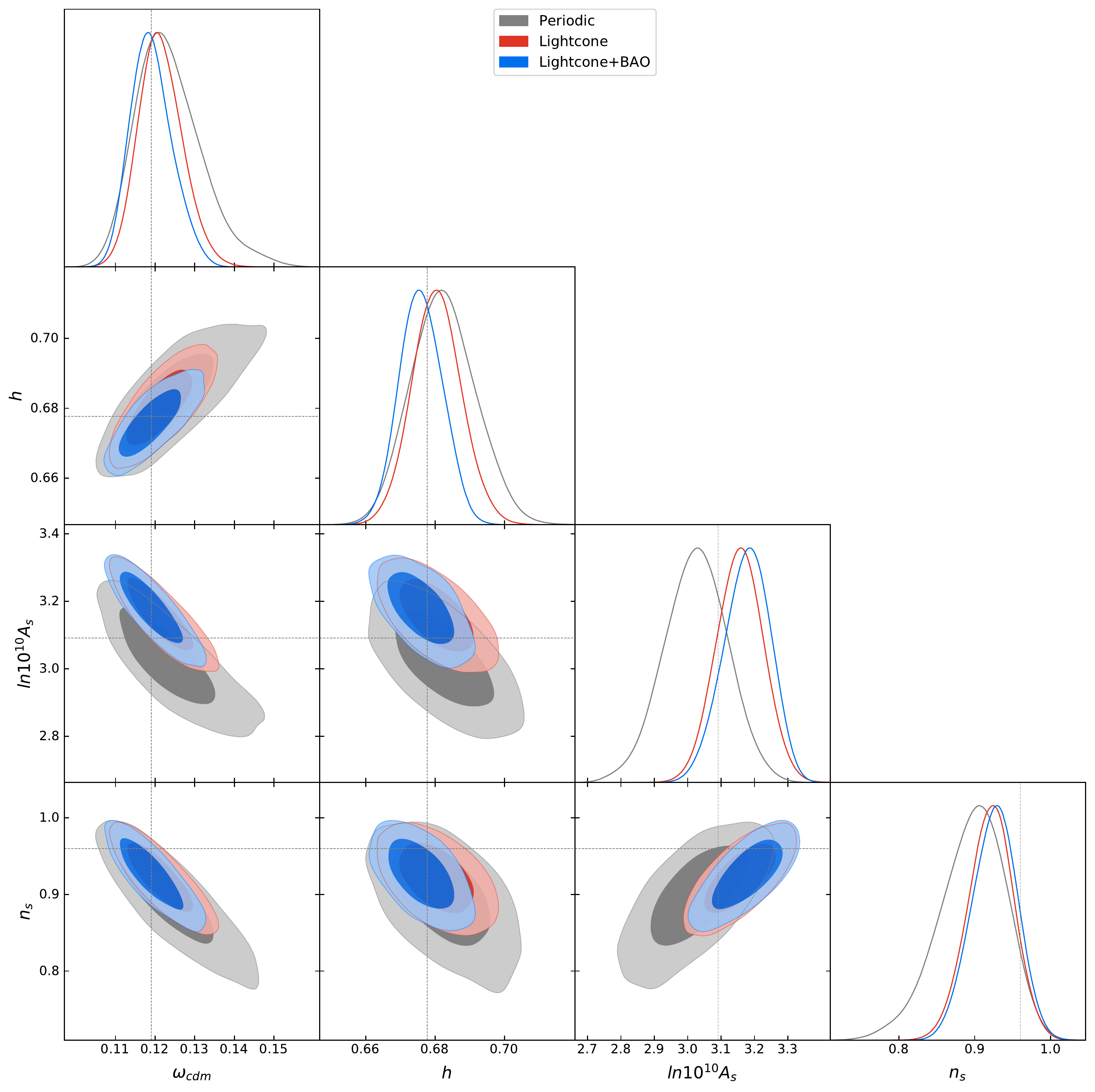}
\caption{\small Triangle plots obtained fitting the lettered challenge or patchy multipoles on $\Lambda$CDM with a BBN prior. 
{\it Upper left}: box A for various scale cuts~$\smin$. 
{\it Upper right}: box D for various scale cuts~$\smin$.
{\it Lower left}: boxes ABFG at scale cut $\smin = 20 \Mpcinvh$.
{\it Lower right}: Patchy periodic, patchy CMASS NGC lightcone, and patchy CMASS NGC lightcone combined with reconstructed BAO, at scale cut $\smin = 20 \Mpcinvh$. The patchy lightcone covariance, as well as the BAO covariance and the cross-covariance, are rescaled by 16, see main text for details.
}
\label{fig:challenge_l2}
\end{figure}

\begin{table}
\scriptsize
\centering
\begin{tabular}{l|c|c|c|c|c|c} 
 \hline 
 $\left({\rm mean-truth}\right)\pm \sigma_{\rm stat}$ 		& $\Delta \omega_{cdm}$ & $\Delta h$ & $\Delta \ln (10^{10} A_s)$ & $\Delta n_s$ 	& $\Delta \Omega_m$		& $\Delta \sigma_8$		\\ \hline
ABFG 		& $-0.004\pm0.007 $ & $-0.011\pm 0.008$ 	& $0.16\pm 0.09$ 	& $-0.036\pm 0.037$ 	&	$0.001\pm 0.010$ &	$0.021\pm 0.028$\\ \hline 
D  			& $0.006\pm 0.006 $ & $-0.001\pm 0.009$ 	& $0.03\pm 0.08$ 	& $-0.057\pm 0.035$ 	&	$0.013\pm 0.009$ &	$-0.011\pm 0.025$\\ \hline 
ABFG+D  		& $0.001\pm 0.005 $ & $-0.007\pm 0.006$ 	& $0.08\pm0.06$ 	& $-0.047\pm 0.025$ 	&	$0.008\pm 0.007$ &	$0.003\pm 0.019$ \\ \hline 
P  			& $0.005\pm 0.008 $ & $0.005\pm 0.009$ 	& $-0.07\pm 0.09$ 	& $-0.063\pm 0.044$ 	&	$0.005\pm 0.012$ &	$-0.033\pm 0.027$ \\ \hline 
L 			& $0.002\pm 0.006 $ & $0.003\pm 0.007$ 	& $0.07\pm 0.07$ 	& $-0.039\pm 0.030$ 	&	 $0.003\pm 0.008$ &	$0.021\pm 0.019$ \\ \hline 
L +BAO 		& $0.000\pm 0.005 $ & $-0.002\pm 0.006$ 	& $0.08\pm 0.07$ 	& $-0.034\pm 0.030$ 	&	$0.002\pm 0.008$ & $0.017\pm 0.018$ \\ \hline 
 \end{tabular}
 \caption{\small 68\%-confidence intervals of  `${\rm mean-truth}$' found fitting simulation FS with a BBN prior. The simulations are: ABFG: mean of lettered challenge boxes A, B, F, G; lettered challenge box D;  ABFG+D: ABFG combined with D; P: patchy Periodic; L: patchy lightcone; L+BAO: patchy lightcone FS combined with reconstructed BAO. We define the minimally-measurable (given the simulations available to us)  theoretical systematic error, $\sigma_{\rm sys}$, as the $\max{[\left|{\rm mean-truth}\right|-\sigma_{\rm stat},0]}$, so that if the truth lies within the 68\% confidence interval around the mean, we do not report a systematic error.}
 \label{tab:challenge}
\end{table}

The triangle plots for the posteriors are shown in Fig.~\ref{fig:challenge_l2}.  
The 68\%-confidence intervals as well as the theory systematic errors are summarized in Table~\ref{tab:challenge}. 
{Using the combination of ABFG+D, we find, relative to BOSS volume} and down to $s_{\rm min} = 20 \, \Mpcinvh$, zero or negligible theory-systematic errors on all cosmological parameters but a marginal one on $n_s$ of less than $\sigma_{\rm data} / 3$, which we consider still negligible for the purpose of data analysis.
This test is particularly reassuring given that our criterion to measure the systematic error is very stringent once we take into account that we measure four cosmological parameters.
These tests on the lettered challenge boxes tell us that we can confidently fit the BOSS data down to those scales. 

We perform further tests using the patchy mocks~\cite{Kitaura:2015uqa}, as they allow us to test for some observational effects.
Triangle plots for the posteriors of the cosmological parameters are shown in Fig.~\ref{fig:challenge_l2} and results are in Table~\ref{tab:challenge}.
First, we check that we find no or negligible theory-systematic errors for all cosmological parameters on the periodic box of side length $2.5 {\rm Gpc}/h$ at redshift $z = 0.5763$ by fitting the mean over all realizations, {but keeping the covariance for the volume of one box.} Similar results hold when fitting the mean over all realizations of CMASS NGC lightcones patchy mocks, with the covariance corresponding to the volume of CMASS NGC rescaled by 16. Both fits to the patchy periodic box or the patchy lightcone with rescaled covariance amount to fitting a volume approximately {equal} to the one of a lettered challenge box~\footnote{
We remind that the patchy lightcone are constructed from $5$ snapshots of the patchy periodic at different redshifts, which, given their volume ratio of about $16$, induce a very small correlation between patchy periodic and patchy lightcone simulations: thus in practice, the patchy lightcone and the patchy periodic can be considered to be uncorrelated. 
Then, there is no reason to worry about the eventual shifts in the parameters measured from those two `independent' realizations: we compare their results only with the truth to assess the theory-systematic error. }.
Upon addition of the reconstructed BAO measurements with error bars rescaled by 4, there is no significant shift of the systematic error. We also checked that fitting with a covariance built from half of the mocks gives similar results, validating our covariance measurements. 
To sum up, these tests using the patchy mocks show that we can confidently fit the BOSS data given the lightcone geometry and upon addition of the reconstructed BAO measurements. 

As the sets of simulations we used have effective redshift $z\sim z_{\rm CMASS}$, we rescale the scale cut for LOWZ as done in \cite{DAmico:2019fhj}. We use for LOWZ $s_{\rm min} = 23 \Mpcinvh$ (instead of $s_{\rm min} = 20 \Mpcinvh$ for CMASS).

\subsection{Adding NNLO}\label{sec:nnlo}
\begin{figure}[]
\centering
\includegraphics[width=.49\textwidth]{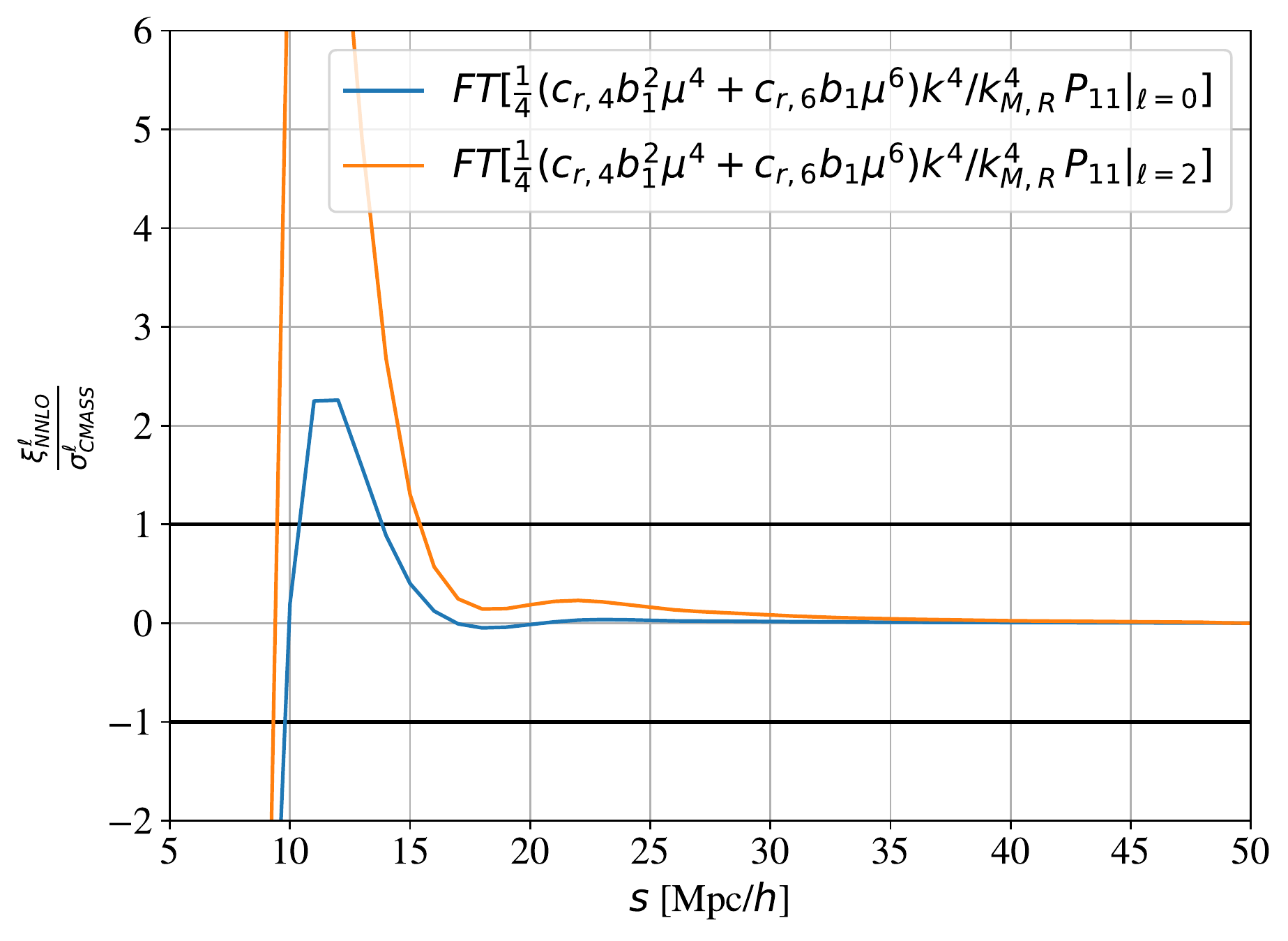}
\includegraphics[width=.49\textwidth]{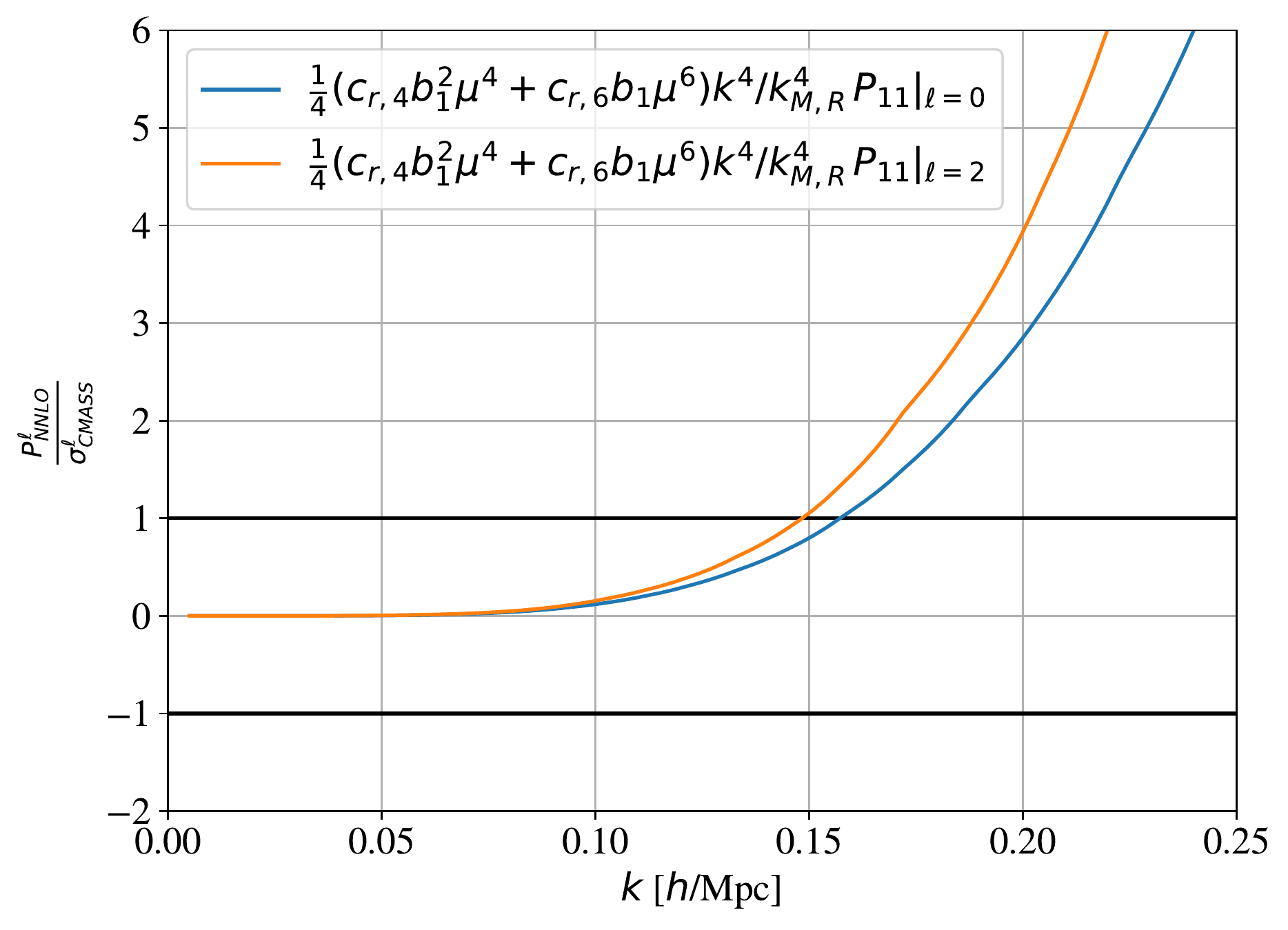}
\caption{\small Size of the NNLO estimates for the multipoles of the CF (\emph{left}) and the PS (\emph{right}) with respect to BOSS CMASS error bars. 
Here we set $c_{r,4}$ and $c_{r,6}$ to $1$. 
$FT$ denotes the Fourier transform. 
}
\label{fig:nnlo_spectra}
\end{figure}

\begin{figure}[h]
\centering
\includegraphics[width=.9\textwidth]{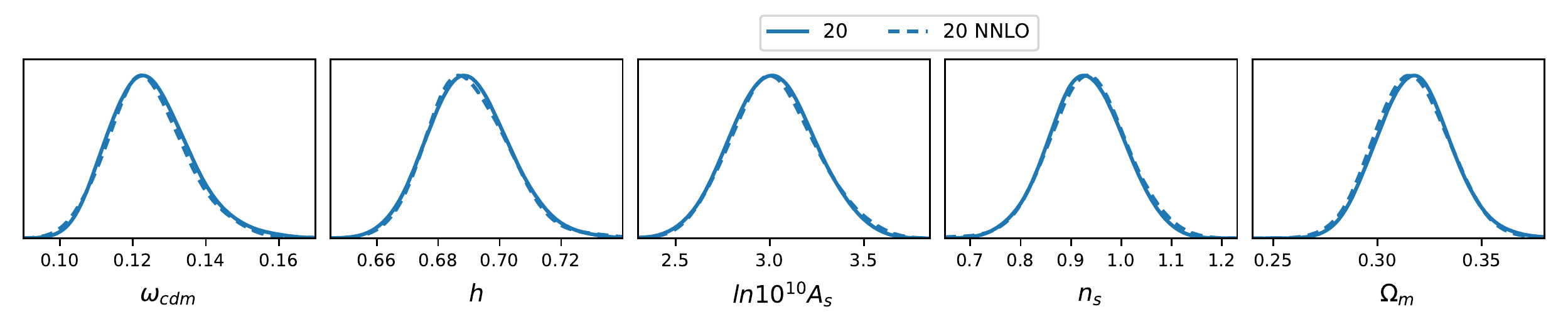}
\includegraphics[width=.9\textwidth]{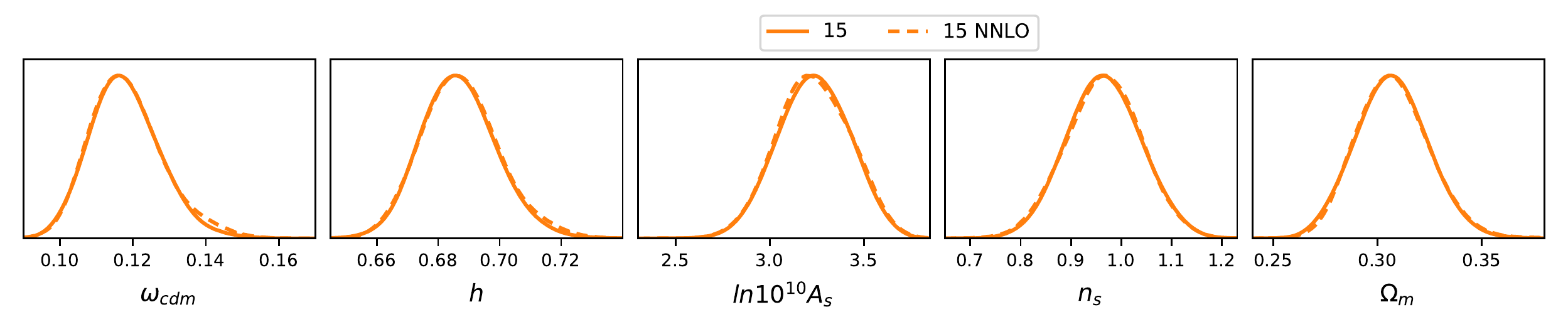}
\includegraphics[width=.9\textwidth]{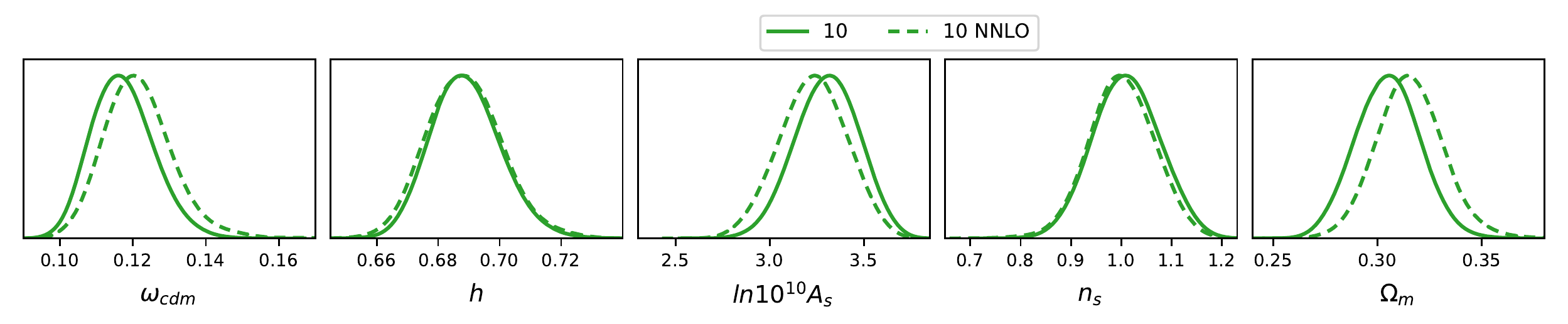}
\caption{\small Posteriors obtained fitting BOSS CF on $\nu\Lambda$CDM with a BBN prior, with or without NNLO term, at $\smin = 20, 15, 10 \Mpcinvh$. 
}
\label{fig:nnlo}
\end{figure}

If using state-of-the-art simulations is a standard, and quite tested, way for model calibration, it is desirable to have other means to corroborate the answer that we can get from simulations, in order to be robust against potential undetected systematics in the simulations (for example, these can range from lack of modeling baryonic effects, satellites, etc., to issues in the estimators for the observables).
In particular, increasing number of tests should be performed given the possible varieties of HOD populations, tracer masses, etc. (see discussions in~\cite{Fujita:2016dne,Eggemeier:2020umu}). 
Furthermore, it is notoriously hard to simulate some extensions to $\Lambda$CDM. 
Given these considerations, we present here another way to determine the scale cut of the theory relying solely on the data and without comparing with simulations. 
This is possible as we use a controlled perturbative approach to LSS: at a given order, one should stop fitting at the scales where the size of the contribution that comes at the next order in perturbation theory and that was not included in the model prediction becomes relevant (with respect to the error bars of the data).
In the data analysis we do in this paper, we include the linear power spectrum, which is the Leading Order (LO) prediction, and the one-loop correction, which is the Next-to-Leading Order (NLO) term. Therefore, the first order in perturbation theory that we do not include is the Next-to-Next-to-Leading-Order (NNLO) correction.
To quantitatively identify the scale at which the NNLO contribution becomes relevant, we fit the BOSS data adding an estimate of the NNLO term {to our EFTofLSS prediction}. 
The scale cut is then chosen as the smallest analyzed scale where the shift in the cosmological parameters induced by the mistake of not including the NNLO is safely small. 
A plot of the NNLO estimates used in this analysis is shown in Fig.~\ref{fig:nnlo_spectra}. 
The difference in the posteriors obtained fitting with and without NNLO terms are shown in Fig.~\ref{fig:nnlo}. 
In App.~\ref{app:nnlo_ps}, the same is shown for BOSS PS analysis.  

For each skycut and each multipole $\ell$, our NNLO estimate consists in:
\begin{align}\label{eq:nnlo}
\xi_{\rm NNLO}^{\ell}(s) & = i^\ell \int \, \frac{dk}{2\pi^2} k^2 P_{\rm NNLO}^{\ell}(k) j_\ell(ks) \, , \\
P_{\rm NNLO}^{\ell}(k) & = \frac{1}{4}c_{r,4} b_1^2 \mu^4  \frac{k^4}{k_{\rm M,R}^4} P_{11}(k) \Big|_{\ell} + \frac{1}{4}c_{r,6} b_1 \mu^6 \frac{k^4}{k_{\rm M,R}^4} P_{11}(k) \Big|_{\ell}  \, ,
\end{align}
where $k_{\rm M,R}^2 = k_{\rm M}^2/8$, $P_{11}$ is the matter linear power spectrum, and $\star |_{\ell}$ denotes the multipole of order $\ell$ of $\star$. 
Here $c_{r,4}$ and $c_{r,6}$ are free `NNLO' parameters. 
They can be thought as some of the higher-order counterterms appearing at NNLO, respectively. 
This estimate is obviously not the full NNLO expression, that contains many more terms. 
However, it can serve as a good proxy of the potential impact of the NNLO on our analysis, as it is let free to vary within physical range. 
We put Gaussian priors on $c_{r,4}$ and $c_{r,6}$ centered on $0$ of width $1$, as discussed in~\cite{DAmico:2021ymi}. 
Note that here we are keeping track of the factorial and symmetrization factors in our NNLO estimate, making for an overall factor of $1/4$. 

Let us give some practical details regarding the evaluation of the NNLO terms in Eq.~\eqref{eq:nnlo}. 
First, as our baseline LO+NLO is consistently IR-resummed up to NLO, we do not want to add extra (spurious) contributions to the BAO signal. 
We therefore use the smooth Eisenstein-Hu linear power spectrum~\cite{Eisenstein:1997ik} as input to evaluate the NNLO terms. 
Second, we damp the high-$k$ tails by changing the powers of $k$ entering in Eq.~\eqref{eq:nnlo} by Pad\'e approximants: 
$k^4 \rightarrow k^4 / (1+k^4/({1 \hinvMpc})^4)$, in order to perform (analytically) the FFTLog from the power spectrum to the correlation function, as described in App.~\ref{app:redcf}. 
This amounts to add terms that are of even higher order than NNLO, and are thus under parametric control. 
Therefore, adding similar damping terms at high wavenumbers, such as Gaussian with variance larger than $\knl$, will lead to equivalent results.

Let us now comment {on} the posteriors. 
We see that, upon addition of the NNLO terms, the shift in the cosmological parameters is practically zero ($\lesssim 0.1 \sigma$) at $s_{\rm min} = 20 \,  \Mpcinvh$ and $s_{\rm min} = 15 \, \Mpcinvh$. 
At $s_{\rm min} = 10  \, \Mpcinvh$, in contrast, the shifts {become} significant: in particular, $\Omega_m$ is shifted by about $2\sigma/3$. 
These observations tell us that we should stop fitting the BOSS data multipoles at about $s_{\rm min} = 15 \, \Mpcinvh$. 
Simulations instead show that we should stop at $s_{\rm min} = 20 \, \Mpcinvh$, which is a close answer (for our binning, it corresponds to fitting one less bin). 
Keeping the most conservative scale cut, we fit the data multipoles down to $s_{\rm min} = 20 \, \Mpcinvh$.

It would be interesting to see if similar conclusions hold for other data sets or observables.
We leave this for future work (see also~\cite{Chudaykin:2019ock,Philcox:2020vvt, Foreman:2015lca} for different methods based on similar considerations).

\section{Cosmological results}
\label{sec:results}

\subsection{BOSS CF and combined probes}

\begin{figure}[t]
\centering
\includegraphics[width=0.99\textwidth]{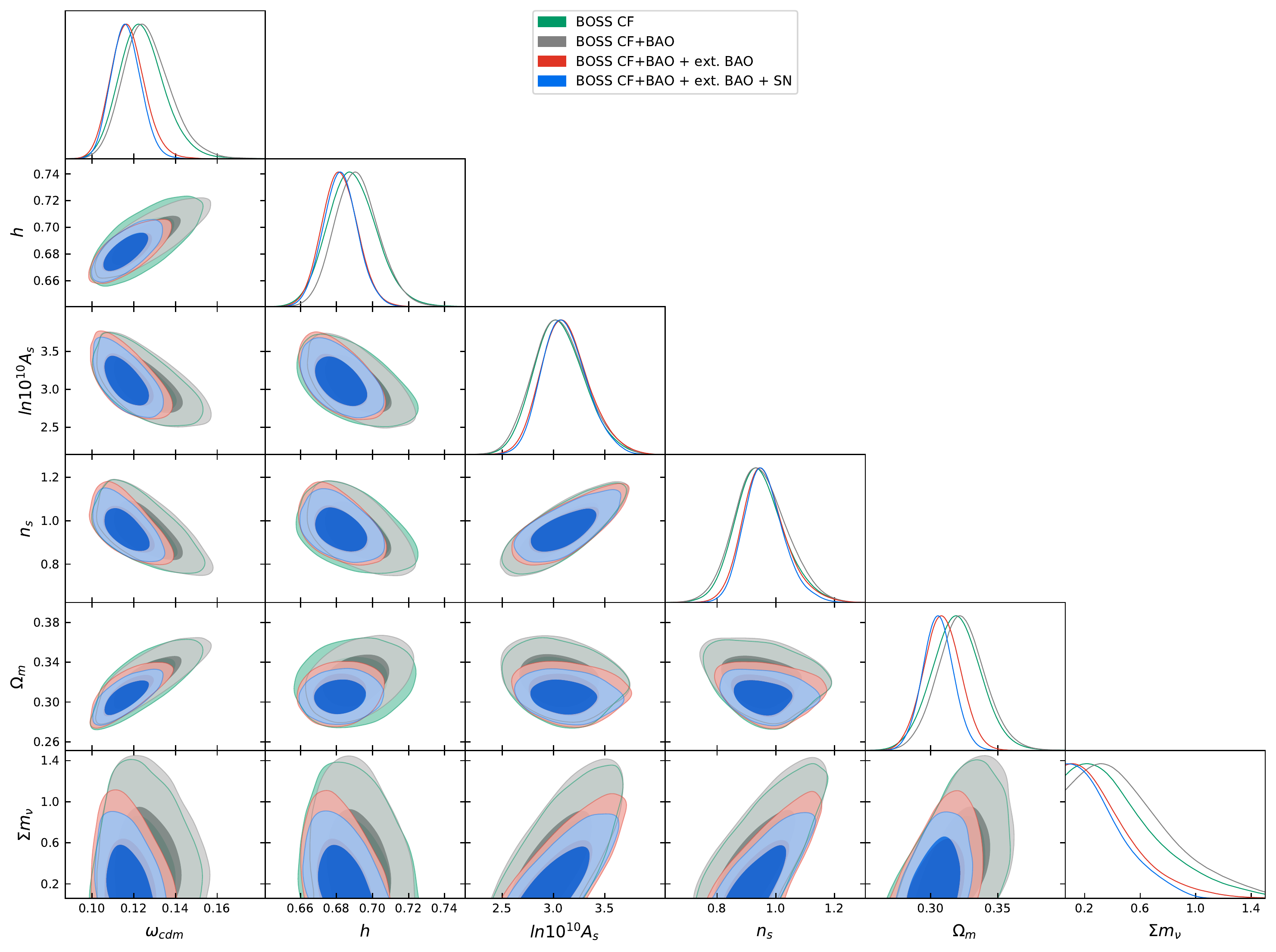}
\caption{\small Triangle plots obtained fitting BOSS CF, BOSS CF + BAO, and in combination with other late-time experiments, on $\nu \Lambda$CDM with a BBN prior. }
\label{fig:results}
\end{figure}

In Fig.~\ref{fig:results} and Table~\ref{tab:results} we show our results for the $\nu \Lambda$CDM fit of BOSS CF, BOSS CF + BAO, and in combination with other late-time experiments, namely small-z BAO and Ly-$\alpha$ BAO (collectively designated as ext. BAO) and Pantheon supernovae (SN). 
Results from BOSS PS and BOSS PS + BAO are inserted for comparison. 
The best fits are provided in App.~\ref{app:bestfit}, as well as a discussion of the contribution of some potential systematic errors, finding that these do not affect significantly our cosmological parameter determination.
In App.~\ref{app:selection}, we argue that line-of-sight selection effects, given physical priors on their size, do not change significantly our results. 

To combine the CF FS and post-reconstructed BAO measurements, we follow our methodology described in~\cite{DAmico:2020kxu}. 
The post-reconstructed BAO parameters are the two usual best-fit BAO scaling parameters $\alpha_{\parallel}$ and $\alpha_{\perp}$, parallel and perpendicular to the line of sight, obtained by fitting the reconstructed PS with a fixed template~\cite{BOSS:2016hvq}. 
Their covariance is built from the best-fit BAO parameters obtained fitting the 2048 post-reconstructed patchy mocks. 
To account for the cross-covariance with CF FS, we first combine in one vector, for each patchy mock, the CF FS pre-reconstructed data, with the best-fit BAO parameters from the post-reconstructed data from the same mock, then compute the full covariance using those vectors.

\begin{table}[h]
\scriptsize
\centering

\begin{tabular}{|l|c|c|} 
 \hline 
CF & best-fit & mean$\pm\sigma$  \\ \hline 
$\omega_{cdm }$ &$0.1118$ & $0.1243_{-0.012}^{+0.0095} $  \\ 
$h$ &$0.6781$ & $0.689_{-0.014}^{+0.012}$   \\ 
$\ln (10^{10}A_{s })$ &$3.287$ & $3.06_{-0.27}^{+0.23}$  \\ 
$n_{s }$ &$0.9542$ & $0.948_{-0.094}^{+0.074}$ \\ 
$\sum m_{\nu}$ [eV]  &$0.35$ & $< 1.09 (2\sigma)$  \\ 
\hline 
$\Omega_m$ & $0.3004$ & $0.319_{-0.019}^{+0.018}$   \\
$\sigma_8$ &$0.8055$ & $0.754_{-0.06}^{+0.055}$ \\
$S_8$ & $0.806$ & $0.777^{+0.055}_{-0.062}   $\\
\hline
 \end{tabular} 
 \begin{tabular}{|l|c|c|} 
 \hline 
CF+BAO & best-fit & mean$\pm\sigma$  \\ \hline 
$\omega_{cdm }$ &$0.1167$ & $0.1266_{-0.013}^{+0.0098}$  \\ 
$h$  & $0.6817$ & $0.692_{-0.013}^{+0.011}$ \\ 
$\ln (10^{10}A_{s })$ & $3.235$ & $3.06_{-0.28}^{+0.24}$ \\ 
$n_{s }$ &$0.9743$ & $0.950_{-0.098}^{+0.082}$  \\ 
$\sum m_{\nu}$ [eV]  &$0.52$  &  $< 1.15 (2\sigma)$ \\ 
\hline 
$\Omega_m$ &$0.3113$ & $0.323_{-0.019}^{+0.017}$\\
$\sigma_8$ &$0.7796$ & $0.756_{-0.062}^{+0.054}$ \\
$S_8$ & $0.794$ & $0.784^{+0.056}_{-0.063}   $\\
\hline 
 \end{tabular} \\ 
  
   \begin{tabular}{|l|c|c|} 
 \hline 
 \begin{tabular}{@{}c@{}}  CF+BAO\\+ext.BAO \end{tabular} & best-fit & mean$\pm\sigma$  \\ \hline 
$\omega_{cdm }$ & $0.1124$ & $0.1172_{-0.0088}^{+0.0075}$  \\ 
$h$  & $0.6806$ & $0.6819_{-0.01}^{+0.0097}$ \\ 
$\ln (10^{10}A_{s })$ & $3.255$ & $3.12_{-0.26}^{+0.21}$ \\ 
$n_{s }$ & $0.9421$ & $0.963_{-0.085}^{+0.062}$  \\ 
$\sum m_{\nu}$ [eV]  &$0.31$  &  $< 0.87 (2\sigma)$ \\ 
\hline 
$\Omega_m$ &$0.298$ & $0.3085_{-0.014}^{+0.014}$\\
$\sigma_8$ & $0.8033$ & $0.766_{-0.059}^{+0.054}$\\
$S_8$ & $0.801$ &  $0.776\pm 0.056            $\\
\hline 
 \end{tabular}
   \begin{tabular}{|l|c|c|} 
 \hline 
 \begin{tabular}{@{}c@{}}  CF+BAO\\+ext.BAO+SN \end{tabular} & best-fit & mean$\pm\sigma$  \\ \hline 
$\omega_{cdm }$ & $0.1111$ & $0.1163_{-0.0078}^{+0.0066}$ \\ 
$h$  & $0.6755$ & $0.6822_{-0.0097}^{+0.0095}$\\ 
$\ln (10^{10}A_{s })$ & $3.217$ & $3.11_{-0.24}^{+0.19}$ \\ 
$n_{s }$ & $0.939$ & $0.961_{-0.076}^{+0.058}$ \\ 
$\sum m_{\nu}$ [eV]  & $0.32$ &  $< 0.74 (2\sigma)$ \\ 
\hline 
$\Omega_m$ &$0.2997$ & $0.306_{-0.012}^{+0.011}$\\
$\sigma_8$ & $0.7777$ & $0.766_{-0.058}^{+0.053}$\\
$S_8$ & $0.777$ &  $0.773^{+0.052}_{-0.059}   $\\
\hline 
 \end{tabular}\\
 
  \begin{tabular}{|l|c|c|} 
 \hline 
PS & best-fit & mean$\pm\sigma$  \\ \hline 
$\omega_{cdm }$ & $0.1174$ & $0.1258_{-0.012}^{+0.009}$  \\ 
$h$  & $0.677$ & $0.683_{-0.014}^{+0.012}$\\ 
$\ln (10^{10}A_{s })$ &$3.081$ & $3.08_{-0.26}^{+0.19}$ \\ 
$n_{s }$ &$0.9469$ & $0.970_{-0.098}^{+0.07}$  \\ 
$\sum m_{\nu}$ [eV]  &$0.08$  &  $< 1.25 (2\sigma)$ \\ 
\hline 
$\Omega_m$ &$0.3067$ & $0.329_{-0.023}^{+0.019}$\\
$\sigma_8$ &$0.8064$ & $0.770_{-0.053}^{+0.046}$ \\
$S_8$ & $0.815$ &  $0.806^{+0.048}_{-0.060}   $\\
\hline 
 \end{tabular} 
\begin{tabular}{|l|c|c|} 
 \hline 
PS+BAO & best-fit & mean$\pm\sigma$  \\ \hline 
$\omega_{cdm }$ &$0.1181$ & $0.1279^{+0.0085}_{-0.013} $  \\ 
$h$ &$0.6851$ & $0.690^{+0.011}_{-0.013}   $  \\ 
$\ln (10^{10}A_{s })$ &$2.998$ & $3.07^{+0.21}_{-0.26}      $  \\ 
$n_{s }$ &$0.9362$ & $0.972^{+0.074}_{-0.10}    $  \\ 
$\sum m_{\nu}$ [eV] &$0.07$ & $< 1.30(2\sigma)$   \\ 
\hline 
$\Omega_{m }$ &$0.3006$ & $0.327^{+0.017}_{-0.021}   $  \\ 
$\sigma_8$ &$0.7789$ & $0.767^{+0.045}_{-0.051}   $  \\ 
$S_8$ & $0.780$ &  $0.801^{+0.049}_{-0.060}   $\\
\hline 
 \end{tabular} \\ 
 
\caption{\small Results obtained fitting BOSS CF, BOSS CF+BAO, and in combination with other late-time experiments, on $\nu \Lambda$CDM with a BBN prior. 
For the total neutrino mass we quote the 95\%-confidence bound instead of the 68\%-confidence interval.
For comparison are also shown the results obtained fitting BOSS PS and BOSS PS+BAO. }
\label{tab:results}
\end{table}

Looking at Table~\ref{tab:results} and focusing first on BOSS, we can clearly see that we can measure all cosmological parameters. This is expected given the fact that the FS has enough information to determine all of them (see the discussion of sec.~4.3 of~\cite{DAmico:2019fhj}). Furthermore, we see that adding BOSS BAO slightly decreases the error on $h$ by about $10\%$. 
Notice that the gain from adding post-reconstructed BAO to the (pre-reconstructed) CF is less than when it is added to the (pre-reconstructed) PS, analyzed with a sharp scale cut: the gain in $\Omega_m$ is $\sim 5\%$ for the CF, while $\sim 10\%$ for the PS, and the gain in $h$ is about the same for CF and PS.  
In the CF analysis all the BAO information available in the two-point function is automatically included. 
Thus, only the BAO information from the higher-order $n$-point function adds up to the CF, whereas the PS receives also information from the two-point function.

When combining BOSS data with other datasets, the main improvement comes from adding ext. BAO, with some gain coming from the SN data.
In particular, there is an improvement of about $\sim 15\%$ on the error bars of $h$ (mainly from ext. BAO), and a $\sim 35\%$ improvement on $\Omega_m$ (mainly from SN but also ext. BAO).
The combination of all datasets also provides a $\sim 20\%$ and $25\%$ better constraint on $\ln(10^{10}A_s)$ and $n_s$, respectively, and a $\sim 35\%$ tighter bound on neutrino masses. 

\begin{figure}[h]
\centering
\includegraphics[width=0.99\textwidth]{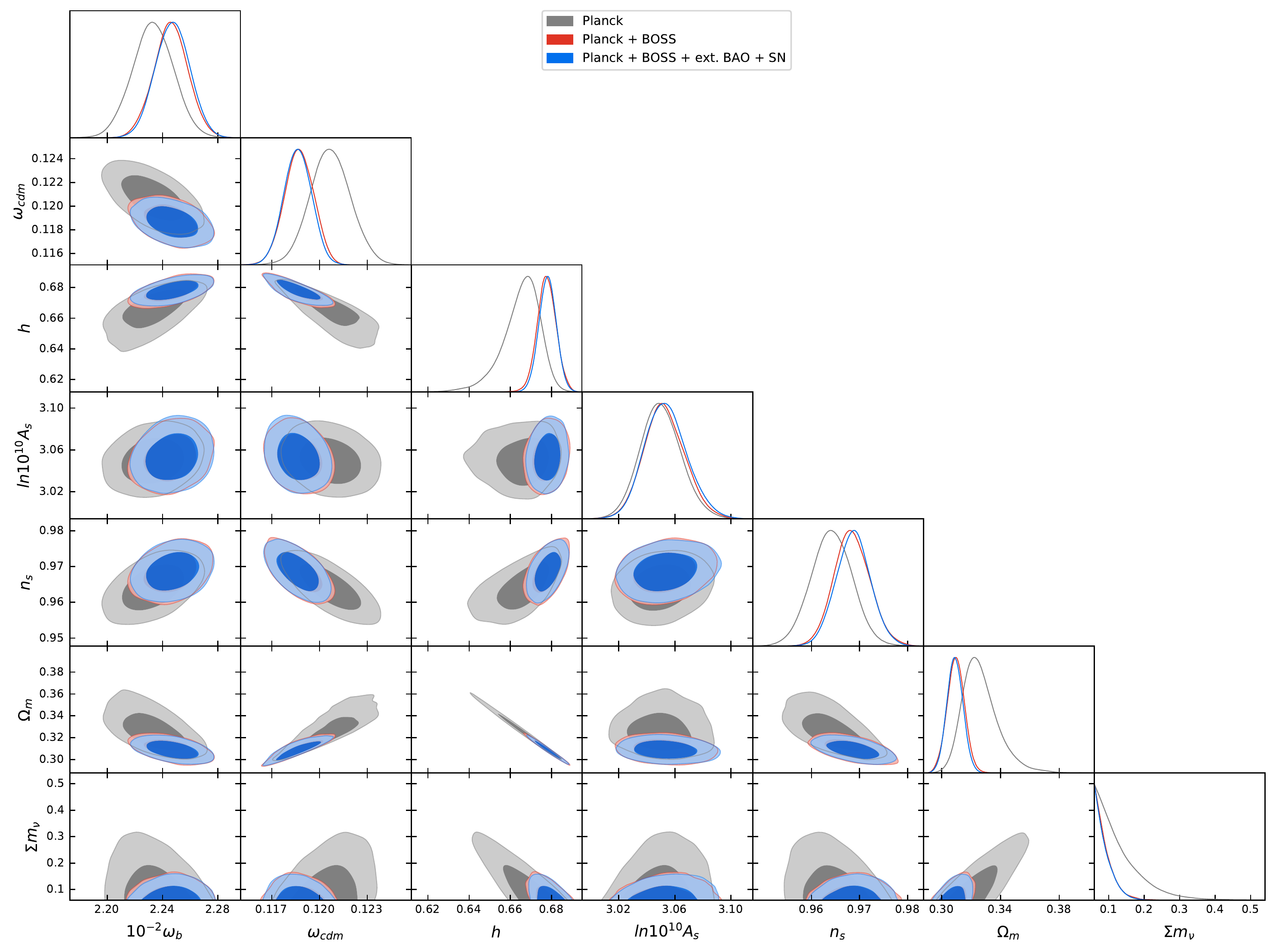}
\caption{\small Triangle plots obtained fitting Planck, Planck + BOSS CF+BAO, and in combination with other late-time experiments, on $\nu \Lambda$CDM.  } 
\label{fig:planck}
\end{figure}

\begin{table}[h]
\scriptsize
\centering
  
 \begin{tabular}{|l|c|c|} 
 \hline 
Planck & best-fit & mean$\pm\sigma$  \\ \hline 
 $100~\omega_{b }$ &$2.236$ & $2.233_{-0.015}^{+0.015}$ \\
$\omega_{cdm }$ & $0.1202$ & $0.1206_{-0.0013}^{+0.0013}$  \\ 
$100*\theta_{s }$ &$1.042$ & $1.042_{-0.0003}^{+0.00029}$\\
$\ln (10^{10}A_{s })$ & $3.041$ & $3.05_{-0.015}^{+0.015}$ \\ 
$n_{s }$ &$0.9654$ & $0.9643_{-0.0043}^{+0.0042}$   \\ 
$\tau_{reio }$ &$0.05238$ & $0.05597_{-0.0081}^{+0.0073}$ \\
$\sum m_{\nu}$ [eV]  &$0.06$ &  $< 0.26 (2\sigma)$ \\ 
\hline 
$h$  & $0.6731$ & $0.6655_{-0.0067}^{+0.011}$ \\ 
$\Omega_m$ &$0.3162$ & $0.3262_{-0.015}^{+0.0092}$\\
$\sigma_8$ & $0.8101$ & $0.8004_{-0.008}^{+0.016}$\\
$S_8$ & $0.8317$ & $0.835\pm 0.013            $\\
\hline 
 \end{tabular}
 \begin{tabular}{|l|c|c|} 
 \hline 
Planck+BOSS & best-fit & mean$\pm\sigma$  \\ \hline 
 $100~\omega_{b }$ &$2.244$ & $2.246_{-0.013}^{+0.013}$ \\
$\omega_{cdm }$ & $0.1192$ & $0.1187_{-0.00089}^{+0.00097}$  \\ 
$100*\theta_{s }$ &$1.042$ & $1.042_{-0.00029}^{+0.00029}$\\
$\ln (10^{10}A_{s })$ & $3.048$ & $3.053_{-0.016}^{+0.014}$ \\ 
$n_{s }$ &$0.9689$ & $0.9684_{-0.0039}^{+0.0037}$   \\ 
$\tau_{reio }$ &$0.05623$ & $0.05908_{-0.0079}^{+0.0071}$ \\
$\sum m_{\nu}$ [eV]  &$0.06$ &  $< 0.14 (2\sigma)$ \\ 
\hline 
$h$  & $0.6779$ & $0.6776_{-0.0045}^{+0.0047}$  \\ 
$\Omega_m$ &$0.3097$ & $0.3097_{-0.0061}^{+0.0059}$\\
$\sigma_8$ &$0.8103$ & $0.8052_{-0.0069}^{+0.0086}$\\
$S_8$ & $0.8233$ & $0.818\pm 0.010            $\\
\hline 
 \end{tabular} \\

 \begin{tabular}{|l|c|c|} 
 \hline 
 \begin{tabular}{@{}c@{}}  Planck+BOSS\\+ext.BAO \end{tabular} & best-fit & mean$\pm\sigma$  \\ \hline 
 $100~\omega_{b }$ &$2.242$ & $2.247_{-0.013}^{+0.013}$ \\
$\omega_{cdm }$ & $0.1186$ & $0.1187_{-0.00087}^{+0.00091}$  \\ 
$100*\theta_{s }$ &$1.042$ & $1.042_{-0.00028}^{+0.00029}$\\
$\ln (10^{10}A_{s })$ & $3.05$ & $3.053_{-0.016}^{+0.014}$\\ 
$n_{s }$ &$0.9667$ & $0.9685_{-0.0036}^{+0.0037}$  \\ 
$\tau_{reio }$ &$0.05638$ & $0.0592_{-0.0078}^{+0.0073}$ \\
$\sum m_{\nu}$ [eV]  &$0.06$&  $< 0.14 (2\sigma)$ \\ 
\hline 
$h$  & $0.6791$ & $0.6776_{-0.0045}^{+0.0042}$  \\ 
$\Omega_m$ &$0.3073$ & $0.3097_{-0.0056}^{+0.0057}$\\
$\sigma_8$ &$0.8092$ & $0.8051_{-0.007}^{+0.0088}$\\
$S_8$ & $0.8189$ & $0.818\pm 0.011            $\\
\hline 
 \end{tabular} 
 \begin{tabular}{|l|c|c|} 
 \hline 
 \begin{tabular}{@{}c@{}}  Planck+BOSS\\+ext.BAO+SN \end{tabular} & best-fit & mean$\pm\sigma$  \\ \hline 
 $100~\omega_{b }$ &$2.247$ & $2.247_{-0.013}^{+0.013}$ \\
$\omega_{cdm }$ & $0.1189$ & $0.1186_{-0.00089}^{+0.0009}$  \\ 
$100*\theta_{s }$ &$1.042$ & $1.042_{-0.00028}^{+0.00029}$\\
$\ln (10^{10}A_{s })$ & $3.05$ & $3.054_{-0.017}^{+0.014}$ \\ 
$n_{s }$ &$0.9672$ & $0.9687_{-0.0037}^{+0.0036}$   \\ 
$\tau_{reio }$ &$0.05869$ & $0.05976_{-0.0084}^{+0.007}$  \\
$\sum m_{\nu}$ [eV]  &$0.06$ &  $< 0.14 (2\sigma)$ \\ 
\hline 
$h$  & $0.6787$ & $0.678_{-0.0043}^{+0.0042}$ \\ 
$\Omega_m$ &$0.3085$ & $0.309_{-0.0056}^{+0.0055}$\\
$\sigma_8$ & $0.8102$ & $0.8054_{-0.0069}^{+0.0082}$ \\
$S_8$ & $0.8215$ & $0.817\pm 0.010            $\\
\hline 
 \end{tabular}\\

\caption{\small Results obtained fitting Planck, Planck + BOSS CF+BAO, and in combination with other late-time experiments, on $\nu \Lambda$CDM. 
For the total neutrino mass we quote the 95\%-confidence bound instead of the 68\%-confidence interval.} 
\label{tab:planck}
\end{table}

In Fig.~\ref{fig:planck} and Table~\ref{tab:planck}, we show the results obtained fitting Planck, and in combination with BOSS CF+BAO and other late-time probes, on $\nu \Lambda$CDM. 
First, we notice that the results from BOSS+BBN and Planck are consistent: all posteriors of the cosmological parameters are consistent at $< 1\sigma$, with the exception {of} $h$, {whose posteriors are consistent} at $\sim 1.7\sigma$. 
If we instead compare the results from BOSS+BBN+ext.BAO to Planck, the measurements on $h$ are then consistent at $\sim 1.2\sigma$, as well as those for the other parameters. 
We find no tension on $h$ or $\sigma_8$: 
Planck measures $h=0.6655_{-0.0067}^{+0.011}$ and $\sigma_8 = 0.8004_{-0.008}^{+0.016}$ at $68\%$CL, while we obtain $h=0.6819_{-0.01}^{+0.0097}$ and $\sigma_8=0.766_{-0.059}^{+0.054}$. 
This is also true on $S_8 \equiv \sigma_8 \sqrt{\Omega_m / 0.3}$, for which Planck gets $S_8 = 0.835 \pm 0.013$ at $68\%$CL, while we get $S_8 = 0.776 \pm 0.056$~(\footnote{The results from Planck and LSS are still in close agreements if we put a prior bound on the neutrino total mass of $<0.25 \eV$ instead of $<1.5 \eV$. 
In this case, we obtain at $68\%$CL: 
\begin{itemize}
\item BOSS+BBN: $\Omega_m=0.314\pm 0.017$, $h=0.694\pm 0.012$, $\ln (10^{10}A_{s }) = 2.87\pm 0.19$, $n_s=0.881\pm 0.063$, $\sigma_8=0.733\pm 0.053$, and $S_8=0.750\pm 0.052$; 
\item BOSS+ext.BAO+BBN: $\Omega_m=0.304\pm 0.013$, $h=0.6844\pm 0.0095$, $\ln (10^{10}A_{s }) = 2.97\pm 0.17$, $n_s=0.917\pm 0.052$, $\sigma_8=0.748\pm 0.051$, and $S_8=0.752\pm 0.050$;
\item Planck: $\Omega_m=0.3267^{+0.0083}_{-0.014}$, $h=0.665^{+0.010}_{-0.0063}$, $\ln (10^{10}A_{s }) = 3.050\pm 0.015$, $n_s=0.9642\pm 0.0043$, $\sigma_8=0.800^{+0.016}_{-0.0072}$, and $S_8=0.835\pm 0.013$. 
\end{itemize}
Thus, BOSS+BBN constraints are consistent with the ones of Planck at $0.6\sigma, 1.8\sigma, 1.0\sigma, 1.3\sigma, 1.2\sigma, 1.6\sigma$, respectively, while BOSS+ext.BAO+BBN constraints are consistent with the ones of Planck at $1.2\sigma, 1.4\sigma, 0.5\sigma, 0.9\sigma, 1.0\sigma, 1.6\sigma$, respectively. }).

Second, we find that the combination with BOSS improves the error bars over Planck alone mainly on $\Omega_m$ and $h$, by about $50\%$. 
This is because LSS data can help break the degeneracy in the $\Omega_m - h$ plane present in the CMB analysis. 
The neutrino $2\sigma$-bound is also decreased from $0.26$eV to $0.14$eV. 
This bound is comparable to one obtained combining Planck with BOSS BAO + RSD~\cite{Planck:2018vyg} (see also~\cite{Ivanov:2019hqk,DAmico:2020kxu,DAmico:2020tty,DAmico:2020ods} for a combination of Planck with BOSS PS FS analyzed with the EFTofLSS, on $\nu \Lambda$CDM and various other cosmological models).

\subsection{Comparison to BOSS PS}\label{sec:ps_cf}

\begin{figure}[h!]
\centering
\includegraphics[width=0.99\textwidth]{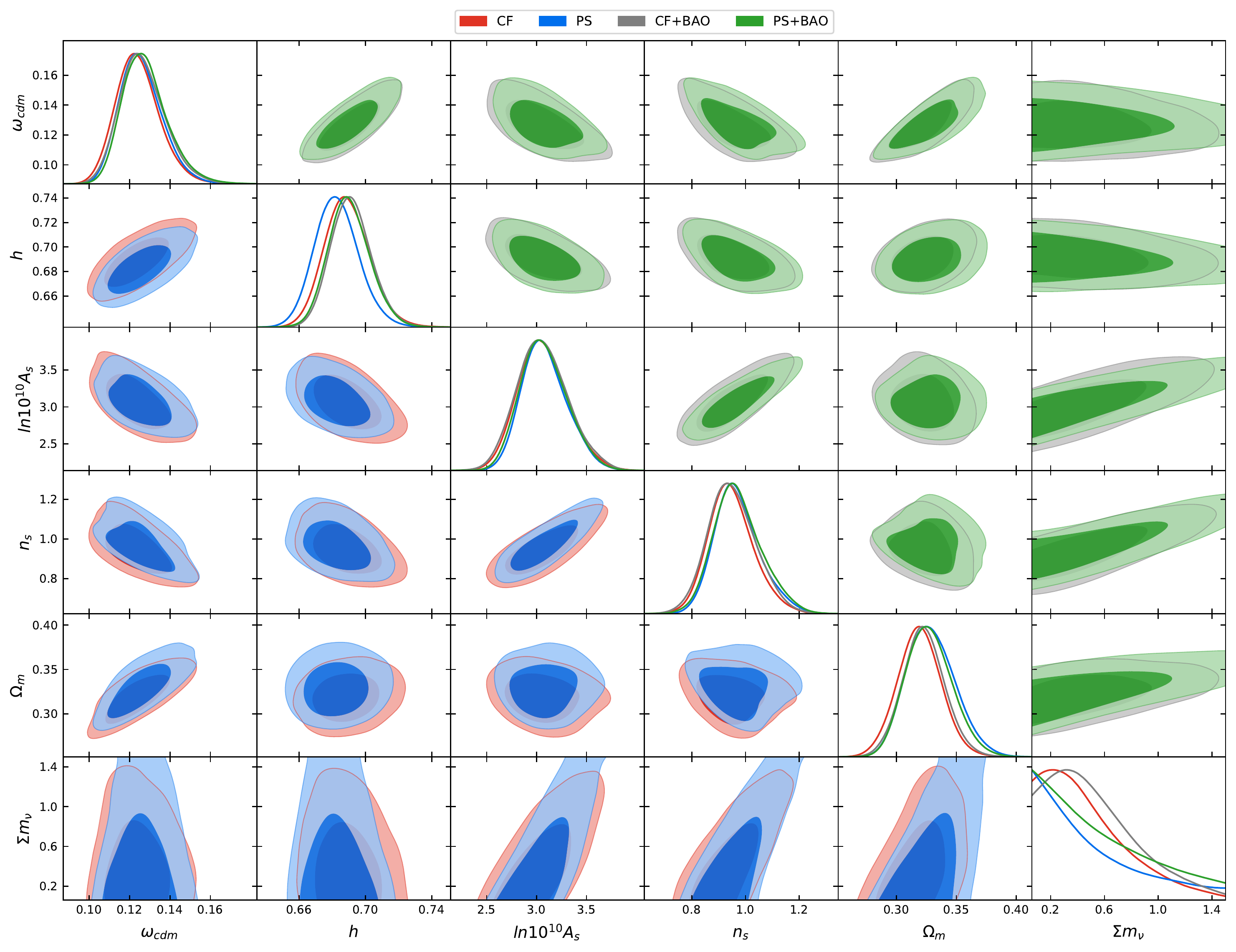}
\includegraphics[width=0.99\textwidth]{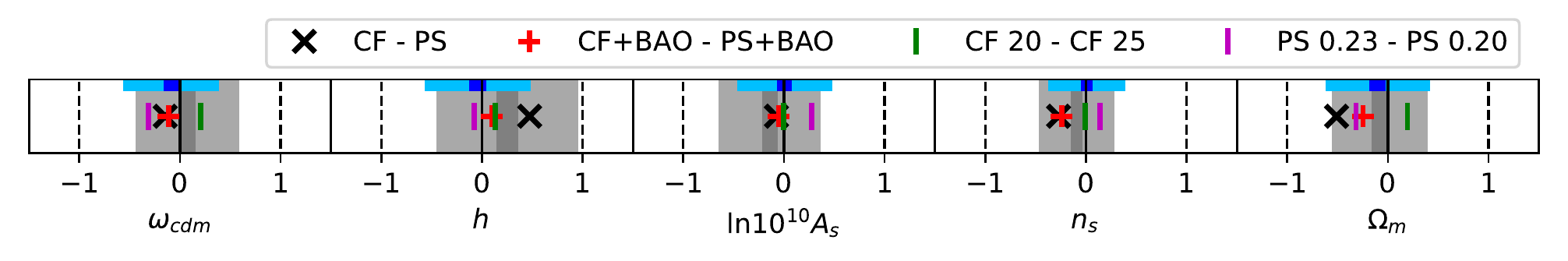}
\caption{\small \emph{Top:} Triangle plots obtained fitting BOSS CF and BOSS PS, at $\smin = 20 \Mpcinvh$ and $\kmax = 0.23 \hinvMpc$, respectively, and with their cross-correlation with BOSS reconstructed BAO. 
\emph{Bottom: } Relative shifts of the mean of the cosmological parameters between the fits of CF and PS (black cross), of CF and PS both combined with BAO (red cross), of CF at $s_{\min} = 20\Mpcinvh$ and $25\Mpcinvh$ (green line), and of PS at $\kmax = 0.23\hinvMpc$ and $0.20\hinvMpc$ (purple line). 
The grey band represents the standard deviation of the shifts measured among 40 patchy boxes, where the darker grey region represents the uncertainty on the mean of the shifts, that is equal to the standard deviation divided by $\sqrt{40} \sim 6$. 
Similarly, the blue band represents the standard deviation of the shifts measured among 40 patchy boxes when combined with BAO. } 
\label{fig:cfps}
\end{figure}

We compare our results obtained fitting the FS of BOSS CF and BOSS PS using the EFTofLSS on $\nu \Lambda$CDM with a BBN prior. 
BOSS PS were analyzed using the EFTofLSS in~\cite{DAmico:2019fhj,Colas:2019ret}, and in cross-correlation with reconstructed BAO in~\cite{DAmico:2020kxu}. 
These previous analyses were based on the PS FS measurements of~\cite{Gil-Marin:2015sqa}. 
For the present work, we re-measure the BOSS PS multipoles, in order to perform a careful comparison between the results of the BOSS CF and PS analyses. 
The reason is twofold: first, the catalog versions used in~\cite{Gil-Marin:2015sqa} are older than the one we use here. 
Especially, \cite{Gil-Marin:2015sqa} uses the split LOWZ and CMASS samples, covering redshift ranges $0.15 < z < 0.43$ and $0.43 < z < 0.7$, respectively, whereas we use the final BOSS DR12 sample combining both the LOWZ and CMASS galaxies, that we then split into two samples by selecting the redshift ranges $0.2 < z < 0.43$ and $0.43 < z < 0.7$ (that we `abusively' name LOWZ and CMASS). 
We therefore use the exact same catalog and redshift selection function for both the PS and CF measurements. 
Second, the PS measurements can be subject to various hyper-parameter choices. 
In particular, we here use PCS particle assignment scheme with grid interlacing, and use $k$-bin size of $\Delta k = 0.01$. 
The window functions are measured as described in~\cite{Beutler:2018vpe} on the randoms catalogs with selections corresponding to the ones used on the data catalogs. 
We have also checked that the measurements obtained with \code{powspec} and \code{nbodykit} give excellent agreements (see Section~\ref{sec:intro} for references). 

For completeness, let us mention that when comparing the present and previous analyses on BOSS PS~\cite{Colas:2019ret,DAmico:2020kxu}, there are other minor differences in the analyses: we here analyze all four BOSS skycuts, whereas before LOWZ SGC {was} left over; the BBN constraint on $\omega_b$ has been updated since ($\sigma_{\rm BBN} = 0.0005 \rightarrow 0.00036$); we enlarge slightly the prior on the sum of the neutrino masses, from $0.06 < \sum m_\nu / e{\rm V} < 1 \rightarrow 1.5$ such that the prior plays practically no role anymore.  
We have checked that these last minor changes do not affect the results significantly ($\lesssim 0.2\sigma$ in shifts and $20\%$ in error bars). 
In particular, to test the accuracy of the EFTofLSS predictions that we use here in the high neutrino masses regime, we find that the results, both for the CF and the PS, are barely changed when instead bounding the sum of the neutrino masses to $0.06 < \sum m_\nu / e{\rm V} < 1$.
However, we find that the differences at the level of the catalog and, possibly, the measurements, make a significant difference in the determination of the cosmological parameters, in particular in $\ln (10^{10}A_s)$, which shifts by about $2 \sigma /3$~(\footnote{See discussions about potential issues in previous BOSS measurements in~\cite{deMattia:2020fkb,Wadekar:2020hax}. }). 

In Fig.~\ref{fig:cfps}, we show the triangle plots obtained fitting BOSS CF and PS, at $(s_{\rm min}, s_{\rm max}) = (20, 200) \Mpcinvh$ and $(k_{\rm min}, k_{\rm max}) = (0.01, 0.23) \hinvMpc$, respectively, and the relative differences between the two analyses in the cosmological parameters. 
We can observe two differences: one, the mean of the cosmological parameters are different, and second, the error bars are slightly different. 
Let us discuss them. 

 \paragraph{Differences in mean} From Table~\ref{tab:results}, we see that {the means of} $\Omega_m$, $h$, $\ln (10^{10} A_s)$, and $n_s$ are different between the CF and PS fits  by about $0.5, 0.5, 0.1$, and $0.3$ in unit of their error bars. 
Although not large, these differences are, still, not completely negligible, keeping in mind that the actual data are mostly the same. 
To understand what causes such differences, let us scrutiny the difference in the CF and PS data. 
There are two main differences: the BAO information and the analyzed scales.
Schematically, the BAO information in the two-point function is encoded in the BAO peak that is fully analyzed in the CF, while it is encoded in the BAO wiggles that are \emph{partially} analyzed in the PS given that we stop at $k_{\rm max} = 0.23 \hinvMpc$. 
Furthermore, the analyzed scales are not exactly similar given sharp scale cuts in either configuration space or Fourier space, as the transformation between the CF and the PS involves a spherical-Bessel function that has support with non-vanishing finite width in $k$ or $s$. 
In order to assess the impact on the cosmological parameters coming from those differences, we perform the following tests.

First, we compare the CF and PS results both cross-correlated with reconstructed BAO. 
The reconstructed BAO are analyzed up to $k_{\rm max} = 0.3 \hinvMpc$. 
Thus, besides the information coming from the higher-$n$ point functions in the reconstructed BAO, that is added equally to the CF and the PS analyses, the BAO information from the two-point function in the reconstructed BAO, that is already saturated in the CF analysis, adds up to the PS analysis. 
As such, if the BAO plays a role in the difference between the CF and PS results, we expect that adding reconstructed BAO to both analyses should make the results closer. 
Besides, we expect the BAO to play a role only on $\Omega_m$ and $h$, see e.g.~\cite{DAmico:2019fhj}. 
As we can see from Fig.~\ref{fig:cfps} and Table~\ref{tab:results}, the differences in $\Omega_m$, $h$, $\ln (10^{10} A_s)$, and $n_s$ between the CF and PS fit  are reduced to about $0.25, 0.10, 0.05$, and $0.24$ in unit of their error bars, when reconstructed BAO are added. 
Interestingly, we now see that the difference in $\Omega_m$ and $h$ between the CF and PS fit {is} now relatively small. 
More precisely, adding the reconstructed BAO does not shift $\ln (10^{10} A_s)$ nor $n_s$ significantly, but shifts $h$ and $\Omega_m$, as expected: 
$h$ is shifted by about $0.2$ and $0.6$ in the same direction for the CF and the PS, respectively, while $\Omega_m$ is shifted by about in opposite direction for the CF and for the PS, respectively. 
The fact that the shift in $\Omega_m$ is in opposite direction, or that the sizes of the shifts in $h$ are significantly different, clearly indicates that the BAO information from the two-point function that is localized between $k \sim 0.23 - 0.3 \hinvMpc$ plays a role. 
Therefore, it seems that the difference in the BAO information between the CF and the PS analysis can explain a large part of the differences that we see in $\Omega_m$ and $h$, that are reduced when adding the reconstructed BAO. 
Eventually, the residual small difference in $\Omega_m$ and $h$ could be explained by the residual difference in the BAO between the two fits after the addition of the reconstructed BAO, which mainly consists in the BAO information from the two-point function above $k \sim 0.3 \hinvMpc$ not included in the PS analysis, and in the scale cuts, as we discuss next. 

We now compare the results obtained at different scale cuts. 
We see  from Fig.~\ref{fig:cfps} that changing the scale cut in either CF or PS analysis does not change their results significantly, as the cosmological parameters are shifted by at most $\sim 0.2$ or $\sim 0.3$, respectively, in unit of their error bars. 
Nevertheless, those shifts, when compared to the standard deviation of the differences found in patchy between CF and PS (see below), are not so small: difference in the effective scale cut can thus explain, in part, the difference we see between the CF and the PS results.~\footnote{We warn however that it is difficult to interpret the size of the shifts as we change the scale cut, as the choice of the change in scale cut is somewhat arbitrary. 
Moreover, the fact that we see the shift in the PS fit to be slightly bigger than in the CF fit might also be attributed to the BAO information that is cut out in the PS fit when reducing the scale cut. }

Finally, we perform the following rather powerful test. 
By fitting the CF and PS of 40 patchy mocks, we find that, on average among those many realizations, the difference in the cosmological parameters between the CF and PS results is consistent with $0$, or negligibly small. 
This tells us that there is no particular systematics related to our theoretical modeling, nor in the way the data are measured. 
Importantly, furthermore, the differences observed in the BOSS data lie within the standard deviation of the differences observed among the 40 patchy boxes. 
This tells us that the differences observed in BOSS are typical. 
We can also observe that the typical differences in patchy are less than the size of the error bars, $\lesssim 50\%$ with the exception of $h$ where the typical difference is {$\sim 70\%$}. 
When adding reconstructed BAO, the typical differences are reduced by about $5\%$ for $\omega_{cdm}$ and $\ln (10^{10}A_s)$ and $25\%$ for $h$, while there is no reduction on $n_s$ and an increase of about $5\%$ in $\Omega_m$. Moreover, on average among the 40 patchy mocks, the differences in the cosmological parameters between the CF and PS results is now fully consistent with $0$. 

In conclusion, the marginally-significant differences between the CF and the PS results in the cosmological parameters can be attributed to the difference in the BAO information analyzed and in the scales analyzed, and appear to be consistent with what we measure in simulations.
The size of the residual differences gives a measure of the independent information contained in CF and PS.
It would be interesting therefore to perform a combined analysis of CF and PS. 
This however requires a careful measurement {or modeling} of the covariance. 
Work is in progress to perform such an analysis.

\paragraph{Error bars} Looking at Table~\ref{tab:results}, we see that the error bars between PS and CF on $n_s$ and $h$ are similar. 
$\ln (10^{10} A_s)$ however is slightly larger by $\sim 10\%$, in the CF fit, with respect to the PS fit. 
This difference might be due to the fact that we effectively analyze {a bit less modes} in configuration space than in Fourier space. 
In contrast, the error bars on $\Omega_m$ is slightly tighter by about $10\%$, in the CF fit, with respect to the PS fit. 
This reflects that the CF contains more BAO information than the PS. \\

\noindent {\bf Note Added:} While we were carefully finalizing this paper, Ref.~\cite{Chen:2021wdi}, which overlaps in parts with this work, appeared.

\section*{Acknowledgements}
We wish to thank Chia-Hsun Chuang for discussions on the systematic errors in the data and for comments on the draft, Ashley Ross for discussions on tidal alignments, and Jeremy Tinker for support with the BOSS `lettered' challenge simulations. 

\noindent 
PZ is grateful for support from the ANSO/CAS-TWAS Scholarship. 
For a fraction of the time when this project was carried over, LS was partially supported by the Simons Foundation Origins of the Universe program (Modern Inflationary Cosmology collaboration) and by NSF award 1720397. 
YFC is supported in part by the NSFC (Nos. 11722327, 11653002, 11961131007, 11421303), by the CAST-YESS (2016QNRC001), by CAS project for young scientists in basic research (YSBR-006), by the National Youth Talents Program of China, and by the Fundamental Research Funds for Central Universities. 
PZ thanks Ellio Schneider for hospitality during completion of this work. 

\noindent The analysis was performed part on the Sherlock cluster at the Stanford University, for which we thank the support team, part on the HPC (High Performance Computing) facility of the University of Parma, whose support team we thank, and part on the computer clusters LINDA \& JUDY in the particle cosmology group at USTC. 

\appendix

\section{Redshift-space one-loop galaxy correlation function}  \label{app:redcf}

\paragraph{Main formulas:} The redshift-space galaxy correlation function at one loop in the EFTofLSS is the (inverse) Fourier transform of the power spectrum:
\begin{equation}
\xi_g(s, \mu_s) =  \int \, \frac{d^3 \k}{(2\pi)^3} \, e^{i \k \cdot \q} P_g(k, \mu_k) 
\end{equation}
where $\mu_s$, $\mu_k$ are, respectively, the cosines of the angles between the line-of-sight and the separation $\vs$ and of the wavenumber $\k$, and the power spectrum reads \cite{Perko:2016puo,DAmico:2019fhj}:
\begin{align}\label{eq:powerspectrum}
& P_{g}(k,\mu) =  Z_1(\mu)^2 P_{11}(k)  + 2 Z_1(\mu) P_{11}(k)\left( c_\text{ct}\frac{k^2}{{ k^2_\textsc{m}}} + c_{r,1}\mu^2 \frac{k^2}{k^2_\textsc{m}} + c_{r,2}\mu^4 \frac{k^2}{k^2_\textsc{m}} \right) \\  
& + 2 \int \frac{d^3q}{(2\pi)^3}\; Z_2(\q,\k-\q,\mu)^2 P_{11}(|\k-\q|)P_{11}(q) + 6 Z_1(\mu) P_{11}(k) \int\, \frac{d^3 q}{(2\pi)^3}\; Z_3(\q,-\q,\k,\mu) P_{11}(q), \nonumber 
\end{align}
where $k^{-1}_\textsc{m} (\simeq k^{-1}_\textsc{nl})$ controls the bias (dark matter) derivative expansion. 
In the first line, the first term is the linear contribution, and the next ones are the counterterms: the term in $c_\text{ct}$ is a linear combination of the dark matter speed of sound~\cite{Baumann:2010tm,Carrasco:2012cv} and a higher derivative bias~\cite{Senatore:2014eva}, and the terms in $c_{r,1}$ and $c_{r,2}$ represent the redshift-space counterterms~\cite{Senatore:2014vja}. 
The second line is the 1-loop contribution.
Here we dropped the stochastic terms, as in configuration space there are none: in Fourier space, they are integer power laws, and therefore yield to Laplacians of the delta function in configuration space, that can be dropped for all practical purposes 
\footnote{Instead of Taylor expanding in powers of $k/k_{\rm M}$, one can instead write the stochastic contributions in terms of their Pad\'e approximant, e.g. $\bar n^{-1}(k/\km)^0 \rightarrow \tfrac{\bar n^{-1}}{1+k^2/\km^2}$, which upon Fourier transform gives $-\bar n^{-1} (4\pi s)^{-1} \km^2 e^{-\km s}$. These `Yukawa' potentials decrease exponentially fast as $s > 1/\km$ and are therefore very short range.
We checked that adding them does not change our results.}. 

The redshift-space galaxy density kernels $Z_1,Z_2$ and $Z_3$ are given by (see e.g.~\cite{Perko:2016puo}):
\begin{align}\label{eq:redshift_kernels}\nonumber
    Z_1(\q_1) & = K_1(\q_1) +f\mu_1^2 G_1(\q_1) = b_1 + f\mu_1^2,\\ \nonumber
    Z_2(\q_1,\q_2,\mu) & = K_2(\q_1,\q_2) +f\mu_{12}^2 G_2(\q_1,\q_2)+ \, \frac{1}{2}f \mu q \left( \frac{\mu_2}{q_2}G_1(\q_2) Z_1(\q_1) + \text{perm.} \right),\\ \nonumber
    Z_3(\q_1,\q_2,\q_3,\mu) & = K_3(\q_1,\q_2,\q_3) + f\mu_{123}^2 G_3(\q_1,\q_2,\q_3) \nonumber \\ 
    &+ \frac{1}{3}f\mu q \left(\frac{\mu_3}{q_3} G_1(\q_3) Z_2(\q_1,\q_2,\mu_{123}) +\frac{\mu_{23}}{q_{23}}G_2(\q_2,\q_3)Z_1(\q_1)+ \text{cyc.}\right),
\end{align}
where $\mu= \q \cdot \hat{\z}/q$, $\q = \q_1 + \dots +\q_n$, and $\mu_{i_1\ldots  i_n} = \q_{i_1\ldots  i_n} \cdot \hat{\z}/q_{i_1\ldots  i_n}$, $\q_{i_1 \dots i_m}=\q_{i_1} + \dots +\q_{i_m}$, with $\hat{\z}$ being the unit vector in the direction of the line of sight, $f$ is the growth rate, and $n$ is the order of the kernel $Z_n$. 
$G_i$ are the standard perturbation theory velocity kernels, while $K_i$ are the galaxy density kernels given by~\footnote{Notice that here we write explicitly the symmetrized version of the kernels, correcting a typo (missing factor $1/2$ in $K_2$) appearing previously in our series of works~\cite{DAmico:2019fhj,DAmico:2020kxu,Nishimichi:2020tvu}}: 
 \begin{align}
     K_1 & = b_1, \\
     K_2(\q_1,\q_2) & = b_1 \frac{\q_1\cdot \q_2 (q_1^2 + q_2^2)}{2 q_1^2 q_2^2}+ b_2\left( F_2(\q_1,\q_2) -  \frac{\q_1\cdot \q_2 (q_1^2 + q_2^2)}{2 q_1^2 q_2^2} \right) + b_4 \, , \\
     K_3(\q,-\q,\k) & = \frac{b_1}{504 k^3 q^3}\left( -38 k^5q + 48 k^3 q^3 - 18 kq^5 + 9 (k^2-q^2)^3\log \left[\frac{k-q}{k+q}\right] \right) \nonumber \\
    &+ \frac{b_3}{756 k^3 q^5} \left( 2kq(k^2+q^2)(3k^4-14k^2q^2+3q^4)+3(k^2-q^2)^4 \log \left[\frac{k-q}{k+q}\right]  \right) \nonumber \\
    & +\frac{b_1}{36 k^3 q^3} \left( 6k^5 q + 16 k^3 q^3 - 6 k q^5 + 3 (k^2 - q^2)^3 \log \left[\frac{k-q}{k+q}\right] \right) \, ,
 \end{align}
 where $F_2$ is the symmetrized standard perturbation theory second-order density kernel (for explicit expressions see~e.g.~\cite{Bernardeau:2001qr}), and the third-order kernel is written in its UV-subtracted version and is integrated over $k \cdot \hat q$. We work in the basis of descendants \cite{Senatore:2014eva,Angulo:2015eqa,Fujita:2016dne}: at the one-loop order, all kernels can be described with 4 galaxy bias parameters~$b_i$.

Writing $\xi_g(s, \mu_s) = \sum_\ell \xi_g^\ell(s) \mathcal{L}_\ell(\mu_s)$, where $\mathcal{L}_\ell(\mu_s)$ is the Legendre polynomial of order $\ell$, one can relate the correlation function multipoles $\xi_g^\ell(s)$ to the power spectrum multipoles $P_g^\ell(k)$ through a spherical-Bessel transform:
\begin{equation}\label{eq:xi_to_ps_multipole}
\xi_g^\ell(s) = i^\ell \int \, \frac{dk}{2\pi^2} k^2 P_g^\ell(k) j_\ell(ks),
\end{equation}
where $j_\ell$ is the spherical Bessel function of order $\ell$.

\paragraph{Evaluation strategy:} The correlation function, including the loop and the counterterms, can be analytically evaluated using the FFTLog decomposition of the matter power spectrum in a sum of complex power laws, following \cite{Simonovic:2017mhp,Lewandowski:2018ywf}. In the following we detail this procedure, and later we discuss the IR-resummation in configuration space.

First, the linear matter power spectrum can be decomposed using the FFTLog~\cite{Hamilton:1999uv}:
\begin{equation} \label{eq:plindef}
P_{11} ( k_n ) = \sum_{m=-N_{\rm max}/2}^{N_{\rm max}/2}  c_m k_n^{-2 \nu_m}
\end{equation}
where $-2 \nu_m \equiv \nu + i \eta_m$, with $\nu$ a real number, and
\begin{equation} \label{eq:cmetam}
c_m = \frac{1}{N_{\rm max}} \sum_{l = 0 }^{N_{\rm max} -1} P_{11} ( k_l ) k_l^{-\nu} k_{\rm min}^{- i \eta_m } e^{-2 i m l /N} \ , \quad \quad \eta_m = \frac{2 \pi m}{\log ( k_{\rm max} / k_{\rm min})} \ . 
\end{equation}
{In the equations above, we denote by $k_n$ the $N_{\rm max}$ sampling points, which are chosen logarithmically spaced from $k_{\rm min}$ to $k_{\rm max}$.
Once the $c_m$ have been computed with this sampling choice, we can interpolate the equations below at every value of $k$, which are therefore written without a subscript.}
For the power spectrum, each one-loop contribution $P_{13}$ or $P_{22}$ (below $\sigma \in \{13,22\}$), of diagram type `13' or `22' respectively, can be written as simple matrix multiplications~\cite{Simonovic:2017mhp}:  
\begin{equation} \label{eq:p1loopdecomp}
P_{\sigma} ( k ) = k^3 \sum_{m_1, m_2} c_{m_1} k^{-2 \nu_1} M_{\sigma} ( \nu_1,\nu_2 )  k^{-2 \nu_2} c_{m_2}
\end{equation}
where $\nu_i \equiv \nu_{m_i}$. 
This expression comes from the following integral evaluated using dimensional regularization~\cite{Scoccimarro:1996se,Pajer:2013jj}:
\begin{equation} \label{eq:dimreg}
\int \momspmeas{q} \frac{1}{q^{ 2 \nuo} |\kvec - \qvec|^{2 \nut}} = k^{3 - 2\nu_{12}} \frac{1}{8 \pi^{3/2}} \frac{\Gamma ( \frac{3}{2} - \nu_1) \Gamma ( \frac{3}{2} - \nu_2) \Gamma ( \nu_{12} - \frac{3}{2} ) }{ \Gamma( \nu_1) \Gamma ( \nu_2) \Gamma ( 3 - \nu_1 - \nu_2) } \,  . 
\end{equation}
{Explicitly, we have:
\begin{align}
P_{13} ( k ) &= \sum_i B_{13,i} \, k^3 \sum_{m_1, m_2} c_{m_1} k^{-2 \nu_1} M_{13,i} ( \nu_1 ) \bar M_{13}(\nu_1) k^{-2 \nu_2} c_{m_2} \,  ,  \label{eq:BM1} \\
P_{22} ( k ) &= \sum_i B_{22,i} \, k^3 \sum_{m_1, m_2} c_{m_1} k^{-2 \nu_1} M_{22,i} ( \nu_1,\nu_2 ) \bar M_{22}(\nu_1, \nu_2)  k^{-2 \nu_2} c_{m_2} \,  , \label{eq:BM2}
\end{align}
where:
\begin{align}
\bar M_{13}(\nu) = & \frac{1}{14 \pi} \frac{\tan (\nu \pi) }{ (-3 + \nu)(-2 + \nu)(-1 + \nu) \nu} \, ,\\
\bar M_{22}(\nu_1, \nu_2) = & \frac{1}{8 \pi^{3/2}} \frac{\Gamma ( \frac{3}{2} - \nu_1) \Gamma ( \frac{3}{2} - \nu_2) \Gamma ( \nu_{12} - \frac{3}{2} ) }{ \Gamma( \nu_1) \Gamma ( \nu_2) \Gamma ( 3 - \nu_1 - \nu_2) } \, .
\end{align}
The various terms appearing in Eqs.~\eqref{eq:BM1}~and~\eqref{eq:BM2} are given by: 
{\tiny
\begin{align*}
B_{13,0} & = b_1^2 			& M_{13,0}(\nu) & = \frac{9}{8} \, , \\
B_{13,1} & = b_1 b_3 		& M_{13,1}(\nu) & = -\frac{1}{1+\nu} \, , \\
B_{13,2} & = b_1^2 f \mu^2 	& M_{13,2}(\nu) & = \frac{9}{4} \, , \\
B_{13,3} & = b_1 f \mu^2 		& M_{13,3}(\nu) & = \frac{3}{4} \frac{-1 + 3\nu}{1+\nu} \, , \\
B_{13,4} & = b_3 f \mu^2 		& M_{13,4}(\nu) & =  -\frac{1}{1+\nu} \, , \\
B_{13,5} & = b_1 f^2 \mu^2 	& M_{13,5}(\nu) & = - \frac{9}{4} \frac{1}{1+\nu} \, , \\
B_{13,6} & = b_1 f^2 \mu^4 	& M_{13,6}(\nu) & = \frac{9}{4} \frac{1 + 2\nu}{1 + \nu} \, , \\
B_{13,7} & = f^2 \mu^4 		& M_{13,7}(\nu) & = \frac{3}{8} \frac{-5 + 3\nu}{1+\nu} \, , \\
B_{13,8} & = f^3 \mu^4 		& M_{13,8}(\nu) & = -\frac{9}{4}\frac{1}{1+\nu}  \, , \\
B_{13,9} & = f^3 \mu^6 		& M_{13,9}(\nu) & = \frac{9}{4} \frac{\nu}{1+\nu} \, , \\
\end{align*}
}
and:
{\tiny
\begin{align*}
B_{22,0} & = b_1^2 			& M_{22,0}(\nu_1, \nu_2) & = \frac{\splitfrac{2 \left(\nu_1 \left(8 \nu_1^2-4 \nu_1-5\right)+2\right)
   \nu_2^3+\left(2 \nu_1 \left(-4 \nu_1^2+6 \nu_1+19\right)-13\right)
   \nu_2^2+\nu_1 (2 \nu_1+1) \left(2 \nu_1^2+\nu_1-7\right)+8
   \nu_1 \nu_2^5}{+4 (1-6 \nu_1) \nu_2^4+2 \nu_1 (\nu_1 (\nu_1
   (4 (\nu_1-3) \nu_1-5)+19)-3) \nu_2-7 \nu_2+6}}{4 \nu_1
   (\nu_1+1) (2 \nu_1-1) \nu_2 (\nu_2+1) (2 \nu_2-1)} \, , \\
B_{22,1} & = b_1 b_2 			& M_{22,1}(\nu_1, \nu_2) & = \frac{\left(2 \nu_1^3 (7 \nu_2+5)+\nu_1^2 (1-11 \nu_2)+\nu_1
   (\nu_2 (\nu_2 (14 \nu_2-11)-38)-12)+(2 \nu_2-3) (\nu_2 (5
   \nu_2+8)+6)\right) \Gamma (\nu_1) \Gamma (\nu_2)}{7 \Gamma (\nu_1+2)
   \Gamma (\nu_2+2)}  \, , \\
B_{22,2} & = b_1 b_4 			& M_{22,2}(\nu_1, \nu_2) & =  \frac{2 \nu_1-3}{\nu_2}+\frac{2 \nu_2-3}{\nu_1} \, , \\
B_{22,3} & = b_2^2 			& M_{22,3}(\nu_1, \nu_2) & = \frac{2 \left((7 \nu_1 (7 \nu_1+3)-20) \nu_2^2+3 \nu_1 (7 \nu_1+17) \nu_2-2 \nu_1 (10 \nu_1+1)-2
   \nu_2+48\right) \Gamma (\nu_2)}{49 \nu_1 (\nu_1+1) \Gamma (\nu_2+2)}  \, , \\
B_{22,4} & = b_2 b_4 			& M_{22,4}(\nu_1, \nu_2) & = \frac{4 (\nu_1 (7 \nu_2-2)-2 \nu_2+3)}{7 \nu_1 \nu_2} \, , \\
B_{22,5} & = b_4^2 			& M_{22,5}(\nu_1, \nu_2) & = 2 \, , \\
B_{22,6} & = b_1^2 f \mu^2			& M_{22,6}(\nu_1, \nu_2) & = (2 \nu_1+2 \nu_2-3) \frac{\splitfrac{4 \nu_1^4 \nu_2+2 \nu_1^3 (\nu_2-1) (2 \nu_2+1)+\nu_1^2 (2 \nu_2 (2
   (\nu_2-1) \nu_2-5)+3)+\nu_1 (2 \nu_2+1) (2 (\nu_2-2) \nu_2 (\nu_2+1)+3)}{+(3-2 \nu_2) \nu_2^2+3
   \nu_2-2}}{2 \nu_1 (\nu_1+1) (2 \nu_1-1) \nu_2 (\nu_2+1) (2 \nu_2-1)} \, , \\
B_{22,7} & = b_1 b_2 f \mu^2 			& M_{22,7}(\nu_1, \nu_2) & = \frac{(2 \nu_1+2 \nu_2-3) \left(\nu_1^2 (7 \nu_2+5)+\nu_1 (\nu_2 (7 \nu_2+10)+4)+\nu_2 (5
   \nu_2+4)+2\right)}{7 \nu_1 (\nu_1+1) \nu_2 (\nu_2+1)} \, , \\
B_{22,8} & = b_1 b_4 f \mu^2 			& M_{22,8}(\nu_1, \nu_2) & = \frac{(\nu_1+\nu_2) (2 \nu_1+2 \nu_2-3)}{\nu_1 \nu_2} \, , \\
B_{22,9} & = b_1 f \mu^2 			& M_{22,9}(\nu_1, \nu_2) & = (2 \nu_1+2 \nu_2-3) \frac{ \splitfrac{28 \nu_1^4 \nu_2+\nu_1^3 \left(28 \nu_2^2-46 \nu_2+2\right)+\nu_1^2 (2
   \nu_2 (14 (\nu_2-1) \nu_2-19)+5)+(\nu_2-2)(\nu_2+5) (2 \nu_2-1)}{+\nu_1 (2 \nu_2 (\nu_2 (\nu_2 (14 \nu_2-23)-19)+47)-23)}}{14 \nu_1 (\nu_1+1) (2 \nu_1-1) \nu_2 (\nu_2+1) (2 \nu_2-1)} \, , \\
B_{22,10} & = b_2 f \mu^2 			& M_{22,10}(\nu_1, \nu_2) & = \frac{(2 \nu_1+2 \nu_2-3) \left(7 \nu_1^2 (7 \nu_2+5)+\nu_1 (7 \nu_2 (7 \nu_2+2)+4)+\nu_2 (35
   \nu_2+4)-58\right)}{49 \nu_1 (\nu_1+1) \nu_2 (\nu_2+1)} \, , \\
B_{22,11} & = b_4 f \mu^2 			& M_{22,11}(\nu_1, \nu_2) & = \frac{(2 \nu_1+2 \nu_2-3) (7 \nu_1+7 \nu_2-8)}{7 \nu_1 \nu_2} \, , \\
B_{22,12} & = b_1^2 f^2 \mu^2 			& M_{22,12}(\nu_1, \nu_2) & = \frac{(2 \nu_1+2 \nu_2-3) (2 \nu_1+2 \nu_2-1) \left(-(2 \nu_1+1) \nu_2^2-2 \nu_1 (\nu_1+1)
   \nu_2+(\nu_1-1) \nu_1 (2 \nu_1+1)+2 \nu_2^3-\nu_2+2\right)}{8 \nu_1 (\nu_1+1) (2 \nu_1-1)
   \nu_2 (\nu_2+1) (2 \nu_2-1)} \, , \\
B_{22,13} & = b_1^2 f^2 \mu^4 			& M_{22,13}(\nu_1, \nu_2) & = \frac{(\nu_1+\nu_2+1) (\nu_1+\nu_2+2) \Gamma (\nu_1) \Gamma (\nu_2) \Gamma
   \left(\nu_1+\nu_2+\frac{1}{2}\right)}{2 \Gamma (\nu_1+2) \Gamma (\nu_2+2) \Gamma
   \left(\nu_1+\nu_2-\frac{3}{2}\right)} \, , \\
B_{22,14} & = b_1 f^2 \mu^2 			& M_{22,14}(\nu_1, \nu_2) & = \frac{(2 \nu_1+2 \nu_2-3) \left(-2 \nu_1^2+\nu_1-2 \nu_2^2+\nu_2+6\right)}{8 \nu_1 (\nu_1+1) \nu_2
   (\nu_2+1)} \, , \\
B_{22,15} & = b_1 f^2 \mu^4 			& M_{22,15}(\nu_1, \nu_2) & = (2 \nu_1+2 \nu_2-3) (2 \nu_1+2 \nu_2-1) \frac{112 \nu_1^3 \nu_2+2 \nu_1^2 (2 \nu_2 (28
   \nu_2-9)-33)+\nu_1 (4 \nu_2 (\nu_2 (28 \nu_2-9)-58)+41)+(41-66 \nu_2) \nu_2+38}{56 \nu_1
   (\nu_1+1) (2 \nu_1-1) \nu_2 (\nu_2+1) (2 \nu_2-1)} \, , \\
B_{22,16} & = b_2 f^2 \mu^2 			& M_{22,16}(\nu_1, \nu_2) & = -\frac{(2 \nu_1+2 \nu_2-3) (7 \nu_1 \nu_2+3 \nu_1+3 \nu_2+9) \Gamma (\nu_2)}{14 \nu_1 (\nu_1+1)
   \Gamma (\nu_2+2)} \, , \\
B_{22,17} & =  b_2 f^2 \mu^4 			& M_{22,17}(\nu_1, \nu_2) & = \frac{(2 \nu_1+2 \nu_2-3) (2 \nu_1+2 \nu_2-1) (7 \nu_1 \nu_2+5 \nu_1+5 \nu_2+5) \Gamma (\nu_2)}{14
   \nu_1 (\nu_1+1) \Gamma (\nu_2+2)} \, , \\
B_{22,18} & = b_4 f^2 \mu^2 			& M_{22,18}(\nu_1, \nu_2) & = \frac{-2 \nu_1-2 \nu_2+3}{2 \nu_1 \nu_2} \, , \\
B_{22,19} & = b_4 f^2 \mu^4 			& M_{22,19}(\nu_1, \nu_2) & = \frac{(2 \nu_1+2 \nu_2-3) (2 \nu_1+2 \nu_2-1)}{2 \nu_1 \nu_2} \, , \\
B_{22,20} & = f^2 \mu^4 			& M_{22,20}(\nu_1, \nu_2) & = \frac{(2 \nu_1+2 \nu_2-3) (2 \nu_1+2 \nu_2-1) \left(98 \nu_1^3 \nu_2+7 \nu_1^2 (2 \nu_2 (14
   \nu_2-9)-5)+\nu_1 (2 \nu_2 (7 \nu_2 (7 \nu_2-9)-33)-9)-\nu_2 (35 \nu_2+9)+50\right)}{196 \nu_1
   (\nu_1+1) (2 \nu_1-1) \nu_2 (\nu_2+1) (2 \nu_2-1)} \, , \\
B_{22,21} & = b_1 f^3 \mu^4 			& M_{22,21}(\nu_1, \nu_2) & = \frac{(2 \nu_1+2 \nu_2-3) (2 \nu_1+2 \nu_2-1) \left(4 \nu_1^3-8 \nu_1^2 \nu_2-8 \nu_1 \nu_2
   (\nu_2+1)+\nu_1+4 \nu_2^3+\nu_2+2\right)}{8 \nu_1 (\nu_1+1) (2 \nu_1-1) \nu_2 (\nu_2+1) (2
   \nu_2-1)} \, , \\
B_{22,22} & = b_1 f^3 \mu^6 			& M_{22,22}(\nu_1, \nu_2) & = \frac{(\nu_1+\nu_2+2) \Gamma (\nu_1) \Gamma (\nu_2) \Gamma \left(\nu_1+\nu_2+\frac{3}{2}\right)}{\Gamma
   (\nu_1+2) \Gamma (\nu_2+2) \Gamma \left(\nu_1+\nu_2-\frac{3}{2}\right)} \, , \\
B_{22,23} & = f^3 \mu^4 			& M_{22,23}(\nu_1, \nu_2) & = -\frac{(2 \nu_1+2 \nu_2-3) (2 \nu_1+2 \nu_2-1) (7 \nu_1+7 \nu_2-2)}{56 \nu_1 (\nu_1+1) \nu_2
   (\nu_2+1)} \, , \\
B_{22,24} & = f^3 \mu^6			& M_{22,24}(\nu_1, \nu_2) & = \frac{(2 \nu_1+2 \nu_2-3) (2 \nu_1+2 \nu_2-1) \left(56 \nu_1^3 \nu_2+2 \nu_1^2 (2 \nu_2 (28
   \nu_2-9)-19)+\nu_1 (4 \nu_2 (\nu_2 (14 \nu_2-9)-21)+9)+(9-38 \nu_2) \nu_2+26\right)}{56 \nu_1
   (\nu_1+1) (2 \nu_1-1) \nu_2 (\nu_2+1) (2 \nu_2-1)} \, , \\
B_{22,25} & = f^4 \mu^4 			& M_{22,25}(\nu_1, \nu_2) & = \frac{3 (2 \nu_1+2 \nu_2-3) (2 \nu_1+2 \nu_2-1)}{32 \nu_1 (\nu_1+1) \nu_2 (\nu_2+1)} \, , \\
B_{22,26} & = f^4 \mu^6 			& M_{22,26}(\nu_1, \nu_2) & = \frac{(2 \nu_1+2 \nu_2-3) (2 \nu_1+2 \nu_2-1) (2 \nu_1+2 \nu_2+1) \left(2 \left(\nu_1^2-4 \nu_1
   \nu_2+\nu_2^2\right)+1\right)}{16 \nu_1 (\nu_1+1) (2 \nu_1-1) \nu_2 (\nu_2+1) (2 \nu_2-1)} \, , \\
B_{22,27} & = f^4 \mu^8 			& M_{22,27}(\nu_1, \nu_2) & = \frac{(2 \nu_1+2 \nu_2-3) (2 \nu_1+2 \nu_2-1) (2 \nu_1+2 \nu_2+1) (2 \nu_1+2 \nu_2+3)}{32 \nu_1
   (\nu_1+1) \nu_2 (\nu_2+1)} \, . \\
\end{align*}
}

The one-loop power spectrum counterterms can be written as:
\begin{equation}
P_{\rm ct} ( k ) \propto \sum_{m} c_{m} k^{-2(\nu_m -1)} \ .  
\end{equation}
Noting that:
\begin{equation}
\frac{1}{2\pi^2} \int_0^\infty d x \, j_\ell ( x ) x^{2 - 2 \nu} = \frac{2^{-2\nu}}{\pi^{3/2}} \frac{\Gamma \left[ (3 + \ell -2 \nu) /2 ) \right]}{\Gamma \left[ ( \ell + 2 \nu ) /2 \right]} \equiv \tilde M^\ell(\nu) \ , 
\end{equation}   
and using Eq.~\eqref{eq:xi_to_ps_multipole}, the correlation function multipoles can easily be expressed as simple matrix multiplications as well (see e.g.~\cite{Lewandowski:2018ywf} for explicit expressions for the real-space one-loop matter correlation function).

Explicitly, the linear terms, counterterms, and one-loop terms $\sigma \in \{13,22\}$, of the galaxy correlation function multipoles, are given by, respectively:
\begin{align}
\xi^\ell_{11}(s) &= L^{\ell}_{11} \frac{i^{\ell}}{s^3} \sum_m c_m s^{2\nu_m} \tilde M^\ell(\nu_m)  \, , \\
\xi^\ell_{\rm ct}(s) &= L^{\ell}_{\rm ct} \frac{i^{\ell}}{s^5} \sum_m c_m  s^{2\nu_m} \tilde M^\ell(\nu_m-1)  \, , \\
\xi^\ell_{\sigma,i}(s) &= L^{\ell}_{\sigma,i} \frac{i^{\ell}}{s^6} \sum_{m_1, m_2} c_{m_1} c_{m_2} s^{2\nu_1+2\nu_2} \tilde M^\ell(\nu_1+\nu_2-3/2) M_{\sigma,i}(\nu_1, \nu_2) \bar M_{\sigma}(\nu_1, \nu_2) \, , 
\end{align}
where $L^{\ell}_{11/ {\rm ct}/\sigma,i}$ are functions of the EFT parameters and $f$ whose $\mu$-dependence is projected onto the multipole $\ell$. For the linear terms, we have: 
\begin{equation}\label{eq:L11}
L^{\ell}_{11} = \frac{2\ell+1}{2} \int_{-1}^{+1} d \mu \, \mathcal{L}_\ell(\mu) \, \left(b_1 + f\mu^2\right)^2. 
\end{equation}
 For the counterterms, we have: 
 \begin{equation}\label{eq:Lct}
 L^{\ell}_{\rm ct} =  \frac{2\ell+1}{2} \int_{-1}^{+1} d \mu \,  \mathcal{L}_\ell(\mu) \, 2 \left(b_1 + f\mu^2\right) \left( \frac{c_{\rm ct}}{k_{\rm M}^2} + \frac{c_{r,1}}{k_{\rm M}^2} \mu^2 +  \frac{c_{r,2}}{k_{\rm M}^2} \mu^4  \right) \, .
 \end{equation}
 Finally, for the loop terms {(that are not counterterms)}, we have: 
  \begin{align}
   L^{\ell}_{13,i} = \frac{2\ell+1}{2} \int_{-1}^{+1} d \mu \,  \mathcal{L}_\ell(\mu) \, B_{13,i} \, , \label{eq:L13}\\
 L^{\ell}_{22,i} = \frac{2\ell+1}{2} \int_{-1}^{+1} d \mu \,  \mathcal{L}_\ell(\mu) \,  B_{22,i} \, , \label{eq:L22}
 \end{align}
 where $B_{13,i}$ and $B_{22,i}$ are defined for and below Eqs.~\eqref{eq:BM1}~and~\eqref{eq:BM2}. 
 In practice, the integrals in $d\mu$ are performed analytically on the powers of $\mu^{2j}$, $j = 0, 1, 2, \dots$, and not on the whole integrands appearing in Eqs.~\eqref{eq:L11}-\eqref{eq:L22}. 
 In particular, the integrands in Eqs.~\eqref{eq:L11}~and~\eqref{eq:Lct} are expanded first. 
 We can then take the EFT parameters and the powers of $f$ out of the integrals of Eqs.~\eqref{eq:L11}-\eqref{eq:L22} before integrating the remaining parts in $\mu^{2j}$ analytically. 
 Such evaluation strategy allows us to marginalize analytically, at the level of the likelihood, over the EFT parameters appearing only linearly in the correlation function prediction, as described below in this section.  
 \\

Then, the IR-resummation is performed as follows.
In Fourier space, the IR-resummation in redshift space for galaxies up to the $N$-loop order reads~\cite{Senatore:2014vja, Lewandowski:2015ziq}:
\begin{align}
P^\ell(k)_{|N} & = \sum_{j=0}^N \sum_{\ell'}  4\pi (-i)^{\ell'} \int dq \, q^ 2 \, Q_{||N-j}^{\ell \ell'}(k,q) \, \xi^{\ell'}_j (q), \label{eq:resumConvol}\\
\xi^{\ell'}_j (q) & = i^{\ell'}  \int  \frac{dp\, p^2}{2\pi^2} P^{\ell'}_j(p)  \, j_{\ell'}(pq),
\end{align}
where $P^\ell(k)_{|N}$ denotes the resummed power spectrum, and $P^{\ell}_j(k)$ and $\xi^{\ell}_j (k)$ are the $j$-loop order pieces of the Eulerian ({\it i.e.} non-resummed) power spectrum and correlation function, respectively. The effects from the bulk displacements are encoded in $Q_{||N-j}^{\ell \ell'}(k,q)$, given by:
\begin{align}
Q_{||N-j}^{\ell \ell'}(k,q) & = \frac{2\ell+1}{2} \int_{-1}^{1}d\mu_k \,\frac{i^{\ell'}}{4 \pi} \int d^2 \hat{q} \, e^{-i\q \cdot \k} \, F_{||N-j}(\k,\q) \mathcal{L}_\ell(\mu_k) \mathcal{L}_{\ell'}(\mu_q) \,, \label{eq:resumQ}\\
F_{||N-j}(\k,\q) & = T_{0,r}(\k,\q)\times T_{0,r}^{-1}{}_{||N-j}(\k,\q) \, , \nonumber\\
T_{0,r}(\k,\q) & = \exp \left\lbrace -\frac{k^2}{2} \left[ \Xi_0(q) (1+2 f \mu_k^2 + f^2 \mu_k^2) + \Xi_2(q) \left( (\hat k \cdot \hat q)^2 + 2 f \mu_k \mu_q (\hat k \cdot \hat q) + f^2 \mu_k^2 \mu_q^2 \right) \right] \right\rbrace  , \nonumber
\end{align}
where $\Xi_0(q)$ and $\Xi_2(q)$ are given by: 
\begin{align}
\begin{aligned}\label{eq:Xi_02}
\Xi_0(q) & = \frac{2}{3} \int \frac{dp}{2\pi^2} \, \exp \left(-\frac{p^2}{\Lambda_{\rm IR}^2} \right)  P_{11}(p) \, \left[1 - j_0(pq) - j_2(pq) \right] \, , \\
\Xi_2(q) & = 2 \int \frac{dp}{2\pi^2}\, \exp \left( -\frac{p^2}{\Lambda_{\rm IR}^2} \right)  P_{11}(p) \, j_2(pq) .
\end{aligned}
\end{align}

Ref.~\cite{DAmico:2020kxu} showed that by Taylor expanding the effects from the bulk displacements ({\it i.e.} the exponential in $Q_{||N-j}^{\ell \ell'}(k,q)$), the IR-resummed power spectrum can be written as a sum of the non-resummed one plus `IR-corrections'. 
Those IR-corrections can be gathered and easily inverse-Fourier transformed to configuration space. 
Thus, the IR-resummed correlation function $\xi^\ell(s)_{|N}$ is simply the sum of the non-resummed one $\xi^\ell(s)$ plus the configuration-space IR-corrections:
\begin{align}\label{eq:masterIRformula}
&\xi^\ell(s)_{|N} = \xi^\ell(s)  \\
&+\int \, \frac{dk}{2\pi^2} k^2 j_\ell(ks) \sum_{j=0}^N \sum_{\ell'} \sum_{n=1} \sum_{\alpha} 4\pi (-i)^{\ell'}   k^{2n}  \,\mathcal{Q}_{||N-j}^{\ell \ell'}(n, \alpha) \, \int dq \, q^ 2 \,  \left[ \Xi_i(q) \right]^n   \xi_j^{\ell'}(q) \, j_{\alpha}(kq) \, , \nonumber
\end{align}
where $j=0,1{,\ldots,N}$ is the loop order, $n$ is the integer controlling the expansion in powers of $k^2$ of the exponential of the bulk displacements, $\left[ \Xi_i(q) \right]^n$ denotes a product of the form $\Xi_0(q) \times ...  \times \Xi_0(q) \times \Xi_2(q) \times ... \times \Xi_2(q)$ such that the total number of terms in the product is $n$, and $\mathcal{Q}_{||N-j}^{\ell \ell'}(n, \alpha)$ is a number that depends on $N-j$, $\ell$, $\ell'$, $n$, $\alpha$ (and $f$). $\alpha$ represents the orders of the spherical Bessel functions generated in the Taylor expansion and runs over $\lbrace 0, 2, 4, ... \rbrace$~\footnote{All spherical Bessel functions of odd order can be expressed as functions of spherical Bessel functions of even order.}.
Notice that, in Eq.~\eqref{eq:masterIRformula}, both the $k$ and $q$ integrals are performed numerically through the FFTLog. One could be tempted to perform the $q$ integrals analytically since $\xi_j(q) \sim q^{\nu_1} \dots q^{\nu_j}$; however, the factor $\left[ \Xi_i(q) \right]^n$ adds $n$ more sums, $\sim q^{m_1} q^{m_2} \dots q^{m_n}$, from the FFTLog's of $\Xi_{0/2}(q)$ in Eq.~\eqref{eq:Xi_02}, resulting in $n+j$ matrix multiplications, which make the integration slow. We leave the study of further optimizations to future work.

With the exception of one additional (inverse-)Fourier transform (per EFT-parameter-independent term), such evaluation represents almost no extra operation with respect to the one of the power spectrum, that was shown to be extremely fast within \code{PyBird}~\cite{DAmico:2020kxu}.
The evaluation time of the correlation function is somewhat increased with respect to the power spectrum, because of the fact that, generally, the IR-resummation needs to be Taylor expanded to higher order to grasp, in Fourier space, all the BAO wiggles up to $k\sim 0.6\hinvMpc$, in order to reconstruct at best the BAO peak in configuration space. 
Within \code{PyBird}, if one power spectrum evaluation takes about $\lesssim 0.3$ second on a laptop (for $\ell = 0, 2$ and up to $\kmax \sim 0.25 \hinvMpc$), the correlation function is evaluated in $\lesssim 1$ second (for $\ell=0,2,4$~\footnote{We here quote the time for $\ell=0,2,4$ as even when analyzing only the two first multipoles, the hexadecapole contributes significantly {to} the resummation of the BAO peak in configuration space (correspondingly in Fourier space, for the BAO wiggles above $k \gtrsim 0.25 \hinvMpc$), and thus needs to be computed anyway.} and for all~$s$).

Finally, we apply the Alcock-Paczynski effect to correct for the choice of the fiducial cosmology ($\Omega_m = 0.310$) used to transform the galaxy coordinates into distances \cite{Alcock:1979mp} and bin the theory model in $s$ as we bin the data. Explicitly, we introduce the distortion parameters
\begin{equation}
  q_{\perp} = \frac{D_A(z) H_0}{D_A^{\rm ref}(z) H_0^{\rm ref}} \, , \qquad
  q_{\parallel} = \frac{H^{\rm ref}(z) / H_0^{\rm ref}}{H(z) / H_0} \, ,
\end{equation}
where $D_A(z)$, $H(z)$, $H_0$ are, respectively, the true angular diameter distance at redshift $z$, the true Hubble function at redshift $z$, and the true Hubble parameter at present time.
The quantities with superscript `ref' are the same quantities for the reference cosmology.
In terms of these, the relations between the components of the separation $\vs$ in the true cosmology and $\vs^{\rm ref}$ in the reference cosmology is:
\begin{equation}
  \vs_{\perp} = \vs_{\perp}^{\rm ref} q_{\perp} \, , \qquad
  s_{\parallel} = s_{\parallel}^{\rm ref} q_{\parallel} \, .
\end{equation}
Finally, the multipoles of the correlation function in the reference cosmology are computed as:
\begin{equation}
  \xi_{\ell}(s^{\rm ref}) = \frac{2 \ell + 1}{2} \int_{-1}^{1} d \mu^{\rm ref} \, \xi\left(s(s^{\rm ref}, \mu^{\rm ref}), \mu(\mu^{\rm ref})\right) \mathcal{L}_{\ell}(\mu^{\rm ref}) \, ,
\end{equation}
where $\mathcal{L}_{\ell}$ are the Legendre polynomials.
The true $s, \mu$ are related to the reference $s^{\rm ref}, \mu^{\rm ref}$ by
\begin{equation}
  s = s^{\rm ref} G \, , \qquad
  \mu = \mu^{\rm ref} q_{\parallel} / G \, , \qquad
  G = \sqrt{(\mu^{\rm ref})^2 q_{\parallel}^2 + (1 -(\mu^{\rm ref})^2) q_{\perp}^2} \, .
\end{equation}

\paragraph{Priors and Likelihood:} For all runs presented here we run with the same physical priors on the EFT parameters chosen in \cite{DAmico:2019fhj} (but with no stochastic term): flat prior $[0, 4]$ on $b_1$ and $[-4, 4]$ on $c_2 = (b_2+b_4)/\sqrt{2}$, setting $b_2-b_4 = 0$, and Gaussian prior centered on $0$ of width $2$ ($8$) on $b_3$, $c_{\rm ct}$ ($c_{r,1}$). 
As we fit only the monopole and the quadrupole, we set $c_{r,2}=0$ as it is degenerate with $c_{r,1}$~(\footnote{We have checked that the hexadecapole does not bring significant extra information for the BOSS survey.}).
We choose $k_{\rm M}=0.7 \hinvMpc$.
We {analytically} marginalize over the EFT parameters $b_3$, $c_{\rm ct}$, $c_{r,1}$, that appear only linearly in the power spectrum (so at most quadratically in the likelihood) at the level of the likelihood by performing the Gaussian integrals as in~\cite{DAmico:2019fhj}. 
We use one set of EFT parameters $\lbrace b_1, c_2, b_3, c_{\rm ct}, c_{r,1} \rbrace$ per skycut when fitting the BOSS DR12 correlation function.
In addition to the non-marginalized EFT parameters $b_1, c_2$, we sample over the cosmological parameters $\omega_b$, $\omega_{cdm}$, $h$, $\ln (10^{10}A_s)$, $n_s$ and~$\sum m_\nu$. For the neutrinos we take the normal hierarchy.

\section{PS scale cut with NNLO}\label{app:nnlo_ps}

\begin{figure}[h]
\centering
\includegraphics[width=.9\textwidth]{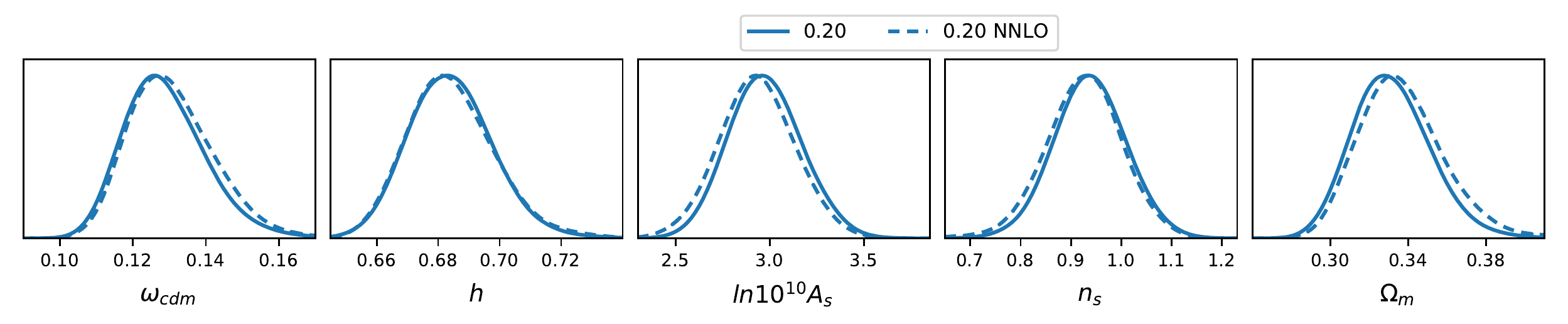}
\includegraphics[width=.9\textwidth]{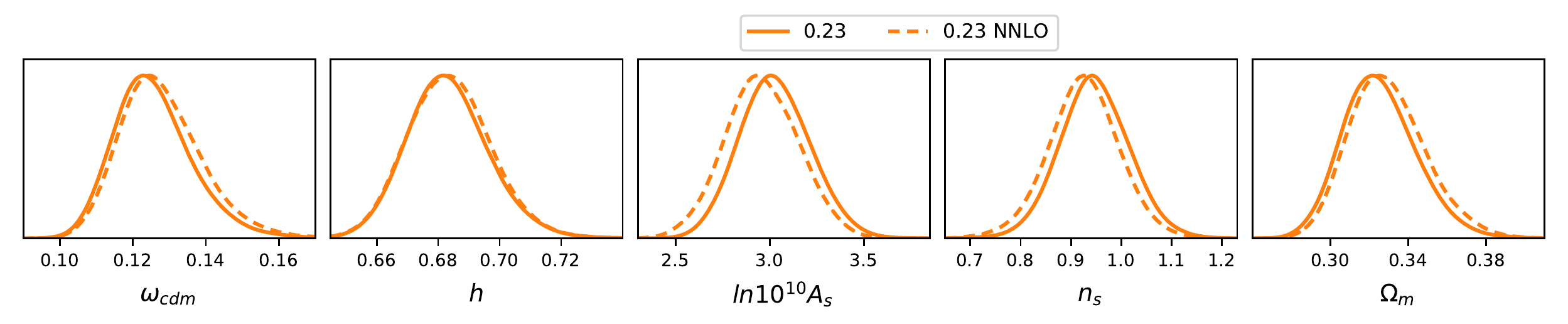}
\includegraphics[width=.9\textwidth]{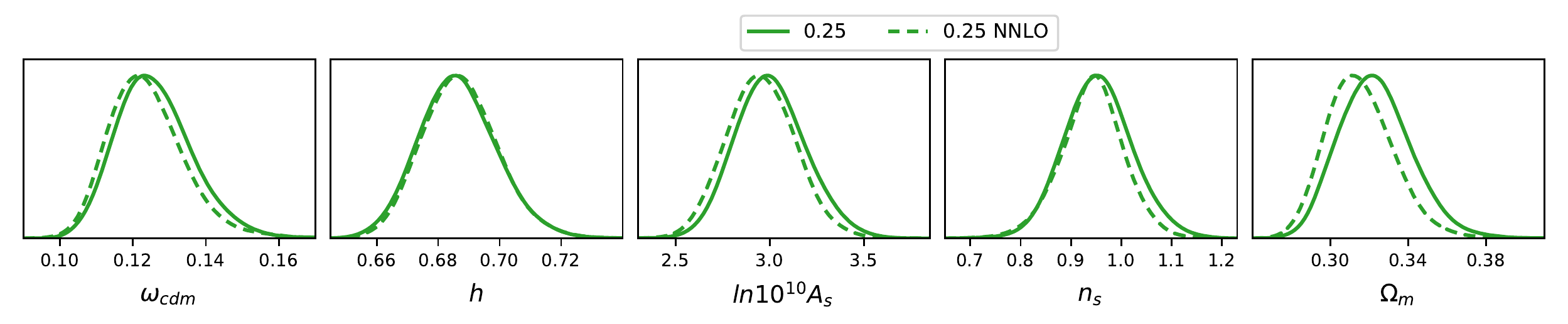}
\caption{\small Posteriors obtained fitting BOSS PS on $\nu\Lambda$CDM with a BBN prior, with or without NNLO term, at $\kmax = 0.20, 0.23, 0.25 \hinvMpc$. }
\label{fig:nnlo_ps}
\end{figure}

We present in Fig.~\ref{fig:nnlo_ps} the analogous scale cut study of Sec.~\ref{sec:nnlo} by varying the NNLO term, Eq.~\eqref{eq:nnlo}, for BOSS PS FS on $\nu\Lambda$CDM with a BBN prior, at $\kmax = 0.20, 0.23, 0.25 \hinvMpc$. 
We find that the shifts in the cosmological parameters are negligible at $\kmax = 0.20 \hinvMpc$ and $0.23 \hinvMpc$, $\lesssim \sigma/3$, but start to become significant at $\kmax = 0.25 \hinvMpc$: we find about $0.5\sigma$ on $\Omega_m$. 
As this result is further corroborated by the answer we get from tests on simulations~\cite{Colas:2019ret,DAmico:2020kxu}, this tells that we should stop fitting at $\kmax = 0.23 \hinvMpc$ in order to get cosmological constraints free from uncontrolled theory-systematic errors.

\section{Best fits and systematics}
\label{app:bestfit}

In Fig.~\ref{fig:challenge_bestfit} and Fig.~\ref{fig:boss_bestfit}, we show the best fits on simulations and on the BOSS data, respectively. 

\begin{figure}
\centering
\includegraphics[width=0.49\textwidth]{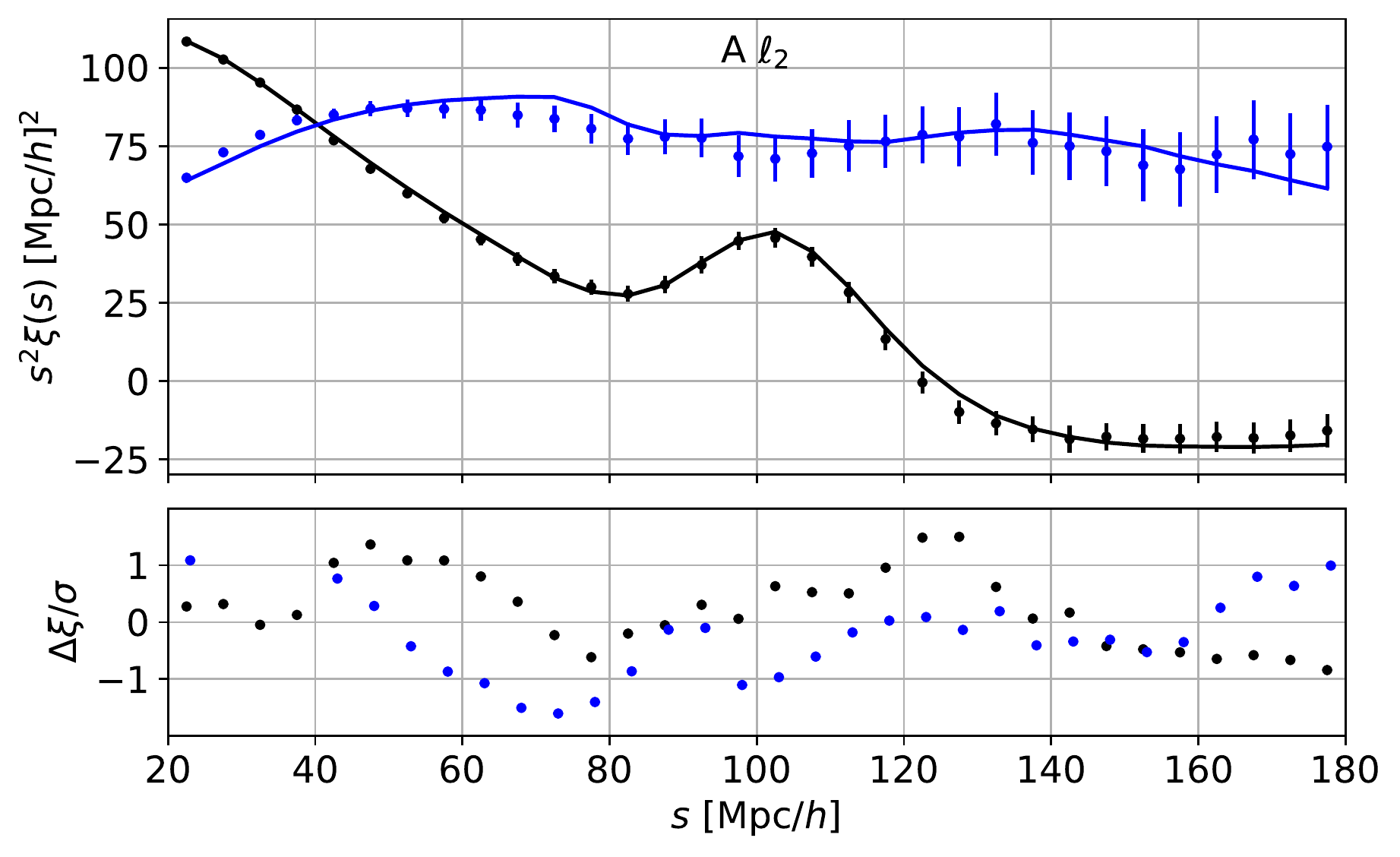}
\includegraphics[width=0.49\textwidth]{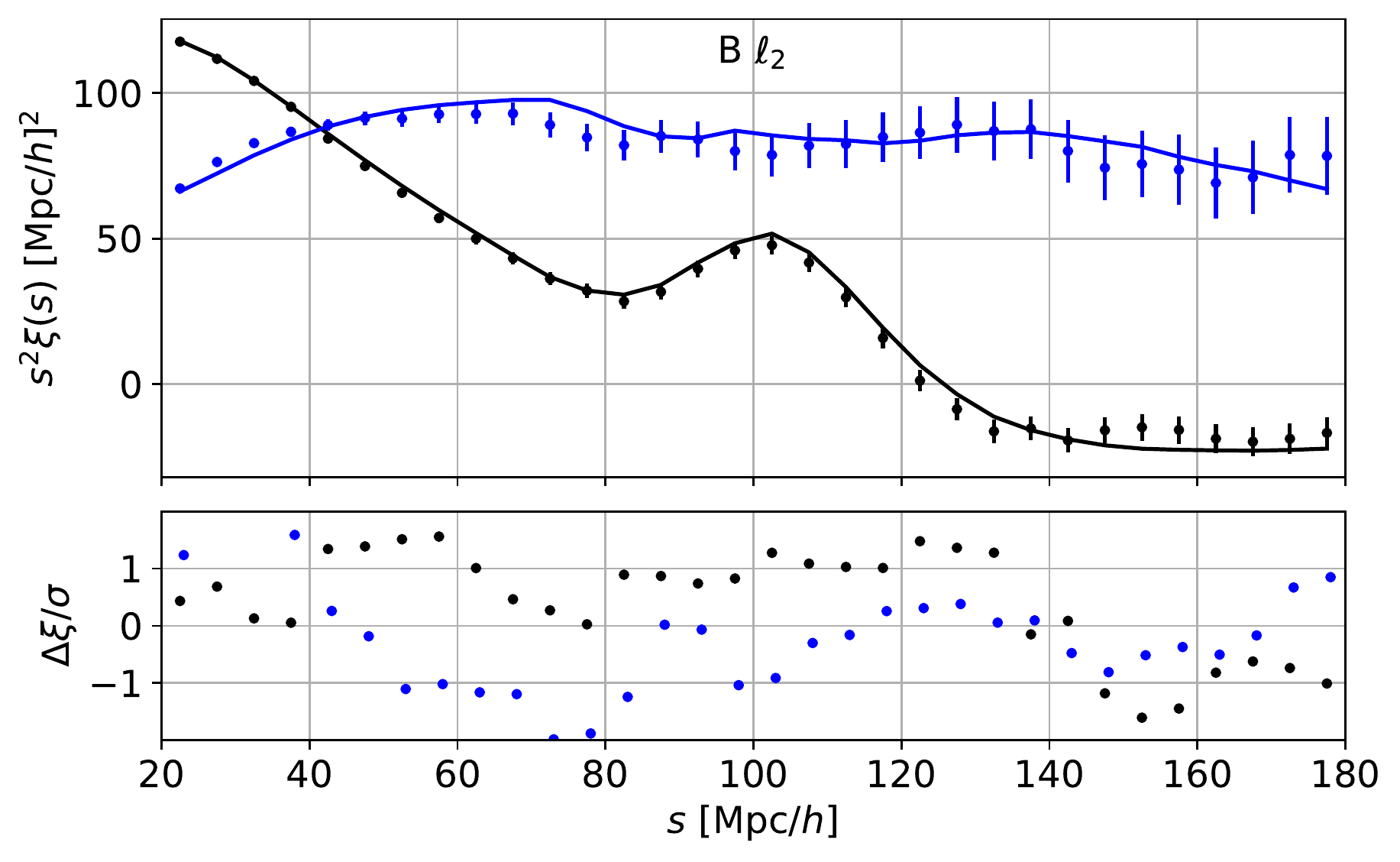}
\includegraphics[width=0.49\textwidth]{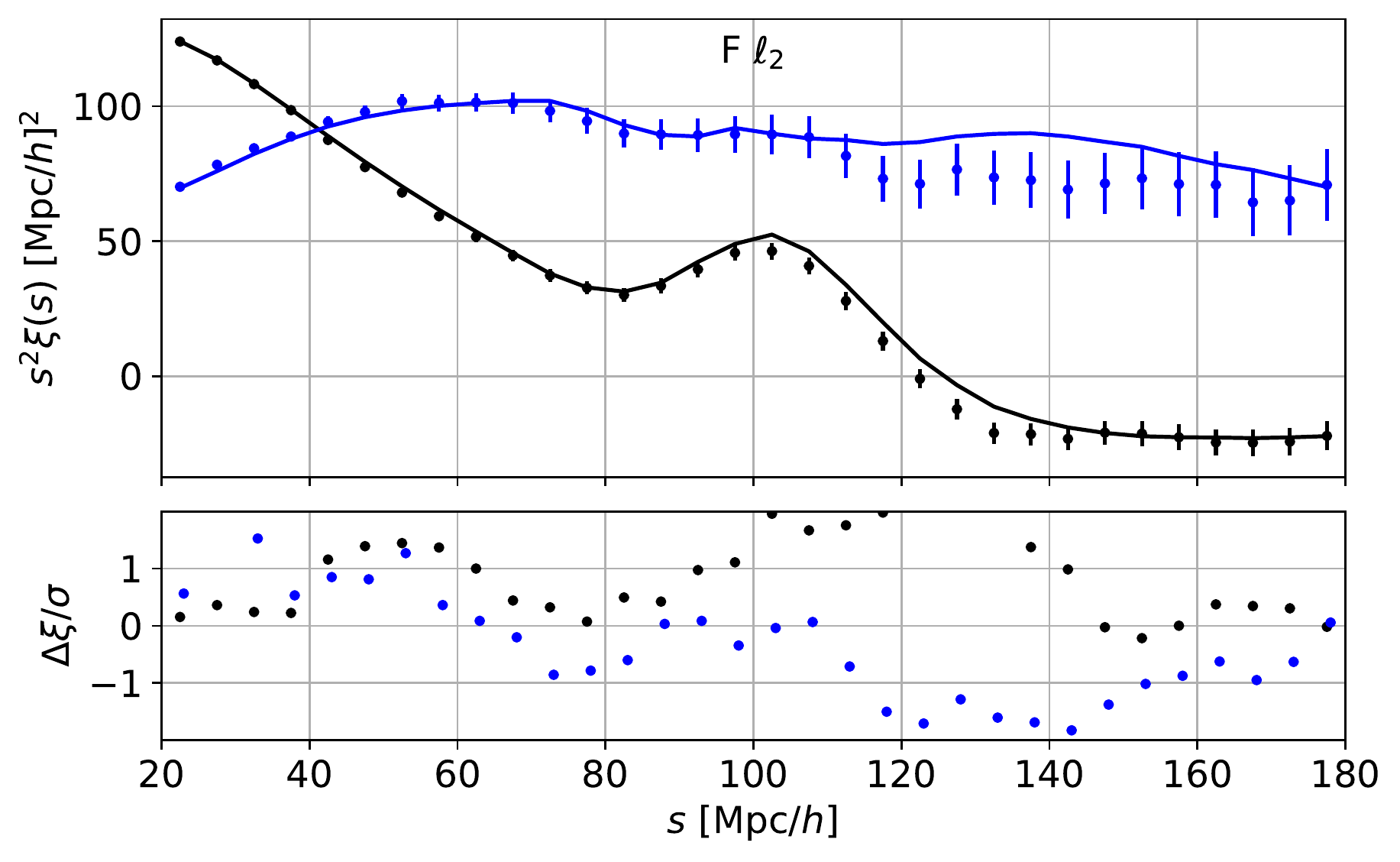}
\includegraphics[width=0.49\textwidth]{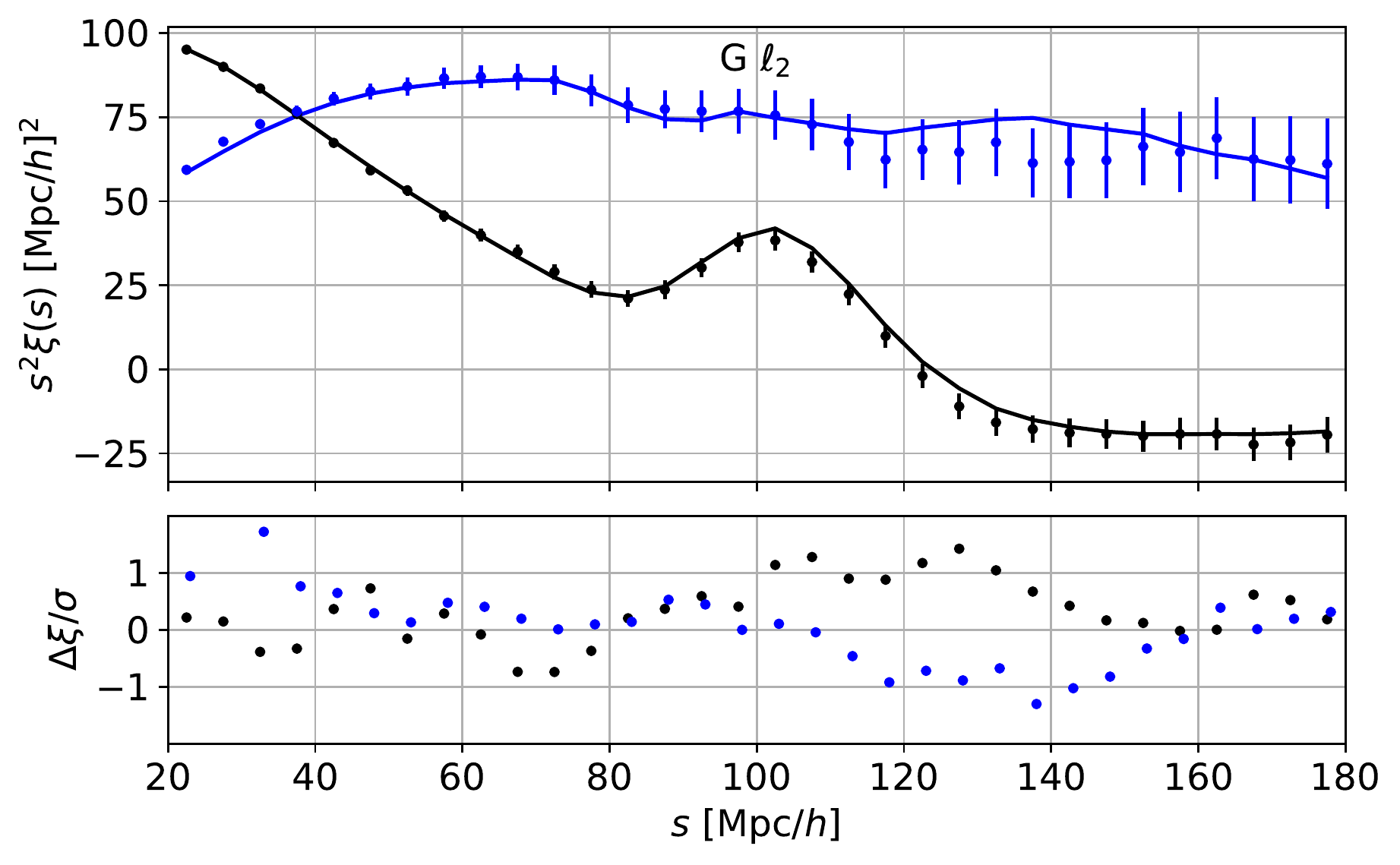}
\includegraphics[width=0.49\textwidth]{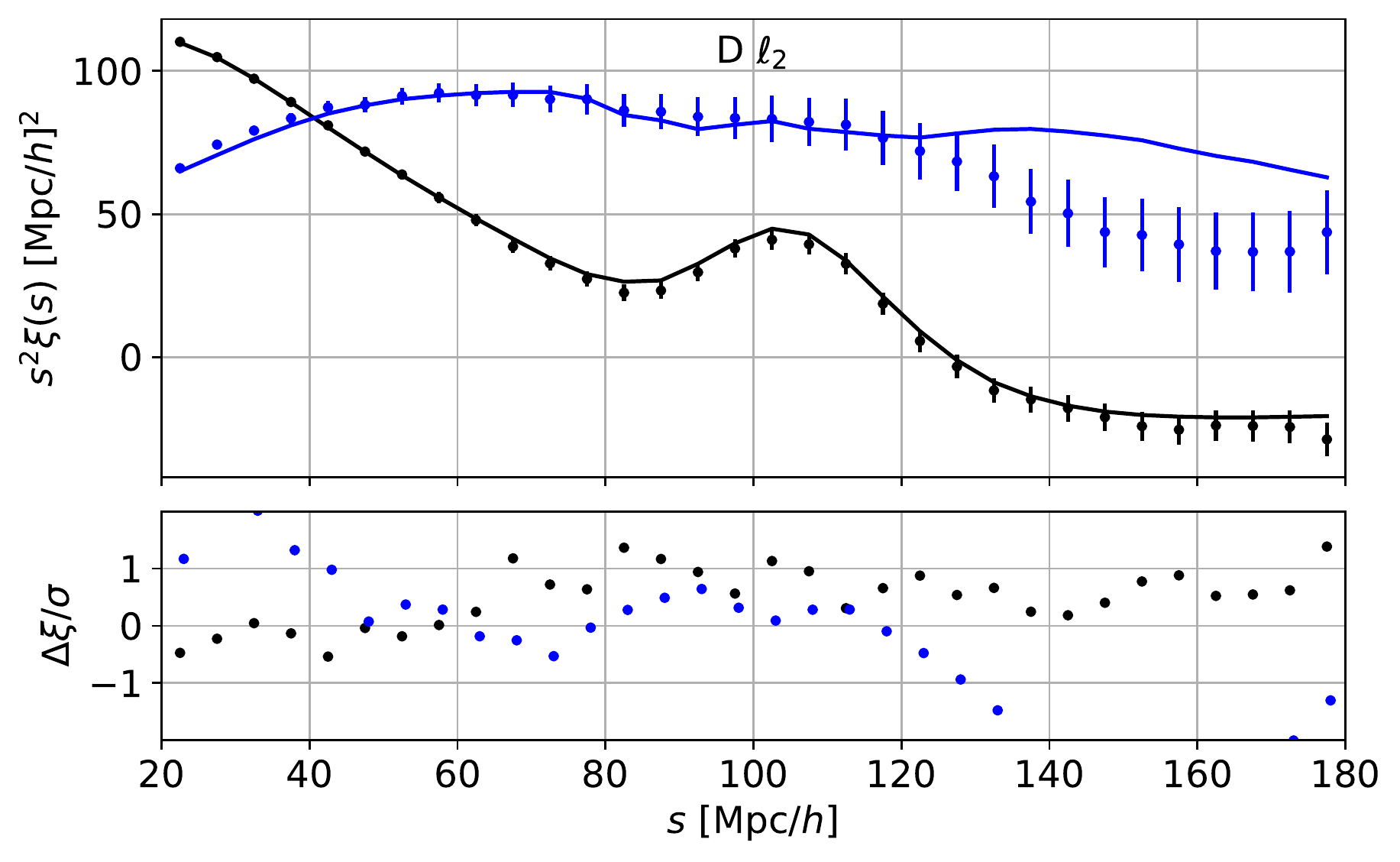}
\includegraphics[width=0.49\textwidth]{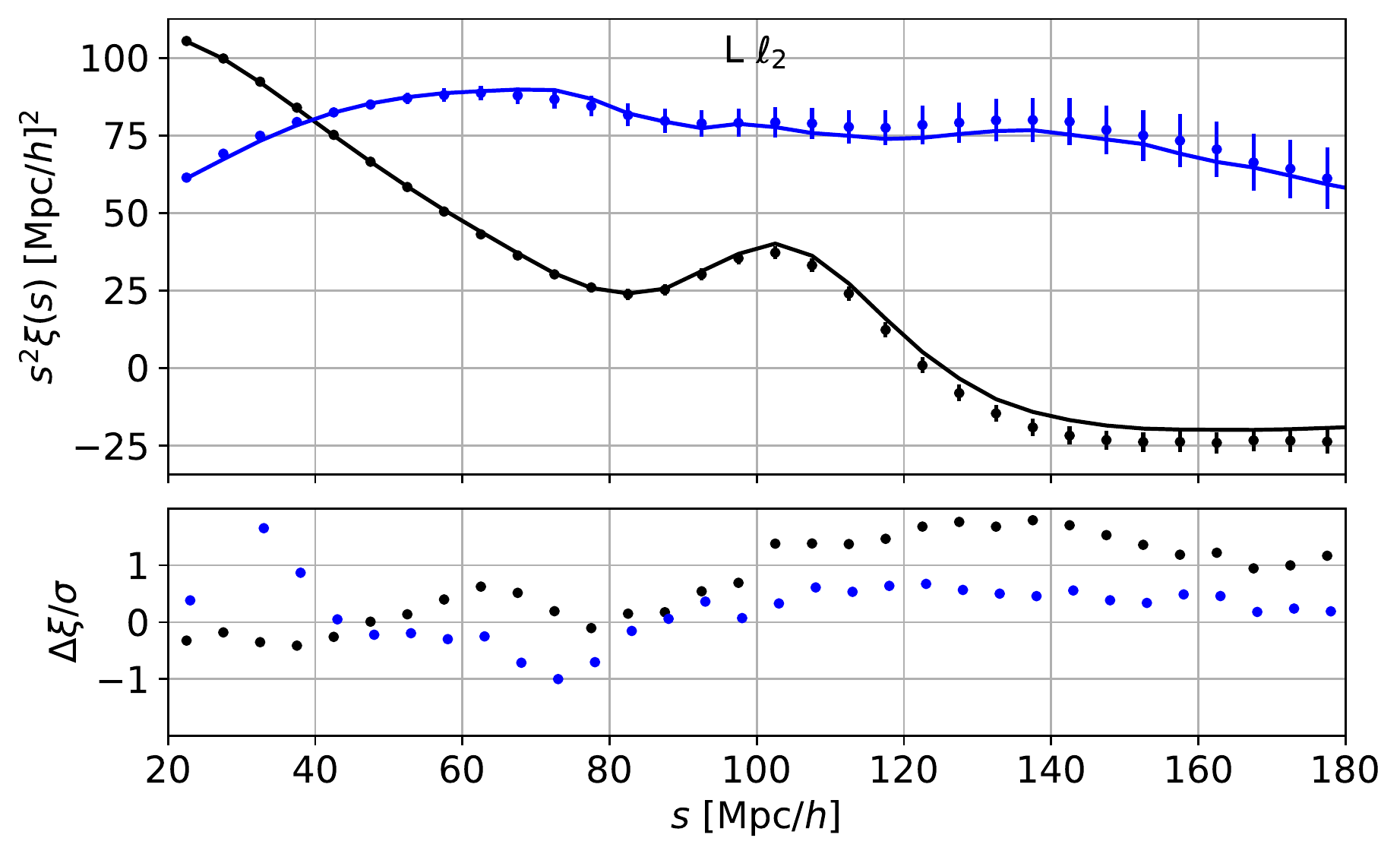}
\caption{\small Best fits and residuals of simulation multipoles: monopole (black) and quadrupole (blue). The quadrupole is shown with a minus sign. The error bars are the square root of the covariance diagonal elements. From top to bottom: lettered challenge box A, B, F, G, D, and mean of 1000 patchy mock Lightcones. The residuals are shown with a slight shift in $s$ for clarity.
} 
\label{fig:challenge_bestfit}
\end{figure}

\begin{figure}[h]
\centering
\includegraphics[width=0.49\textwidth]{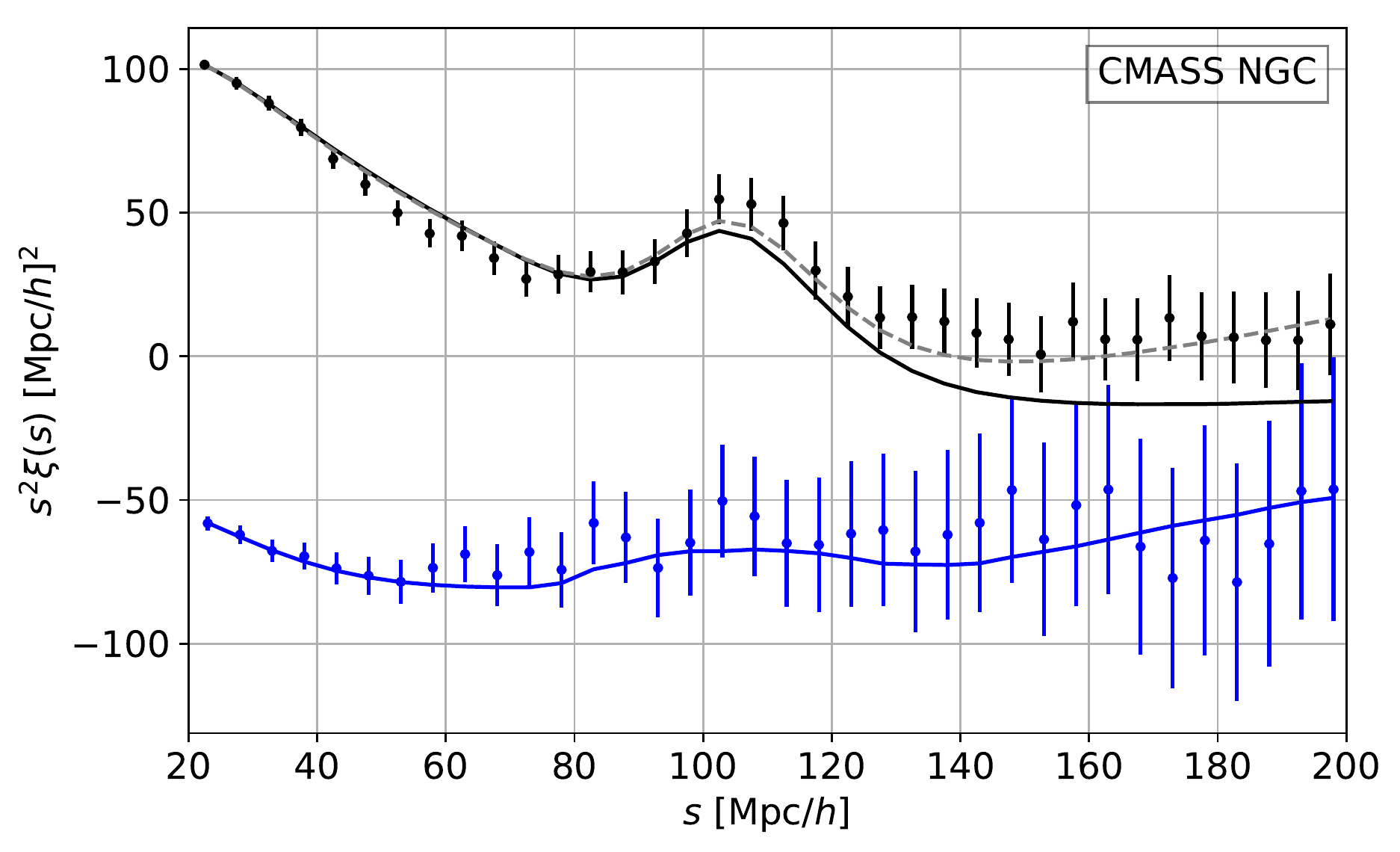}
\includegraphics[width=0.49\textwidth]{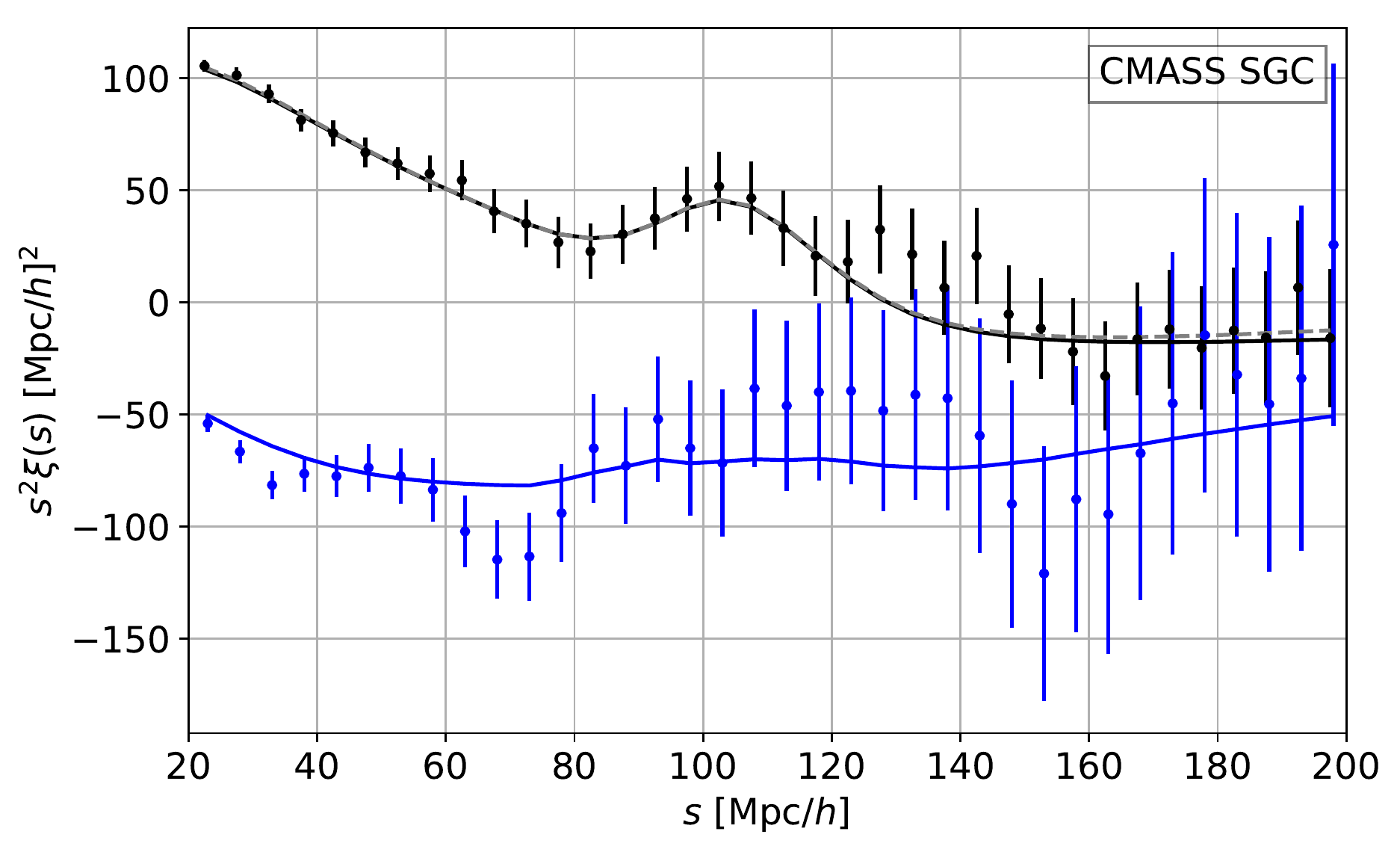}
\includegraphics[width=0.49\textwidth]{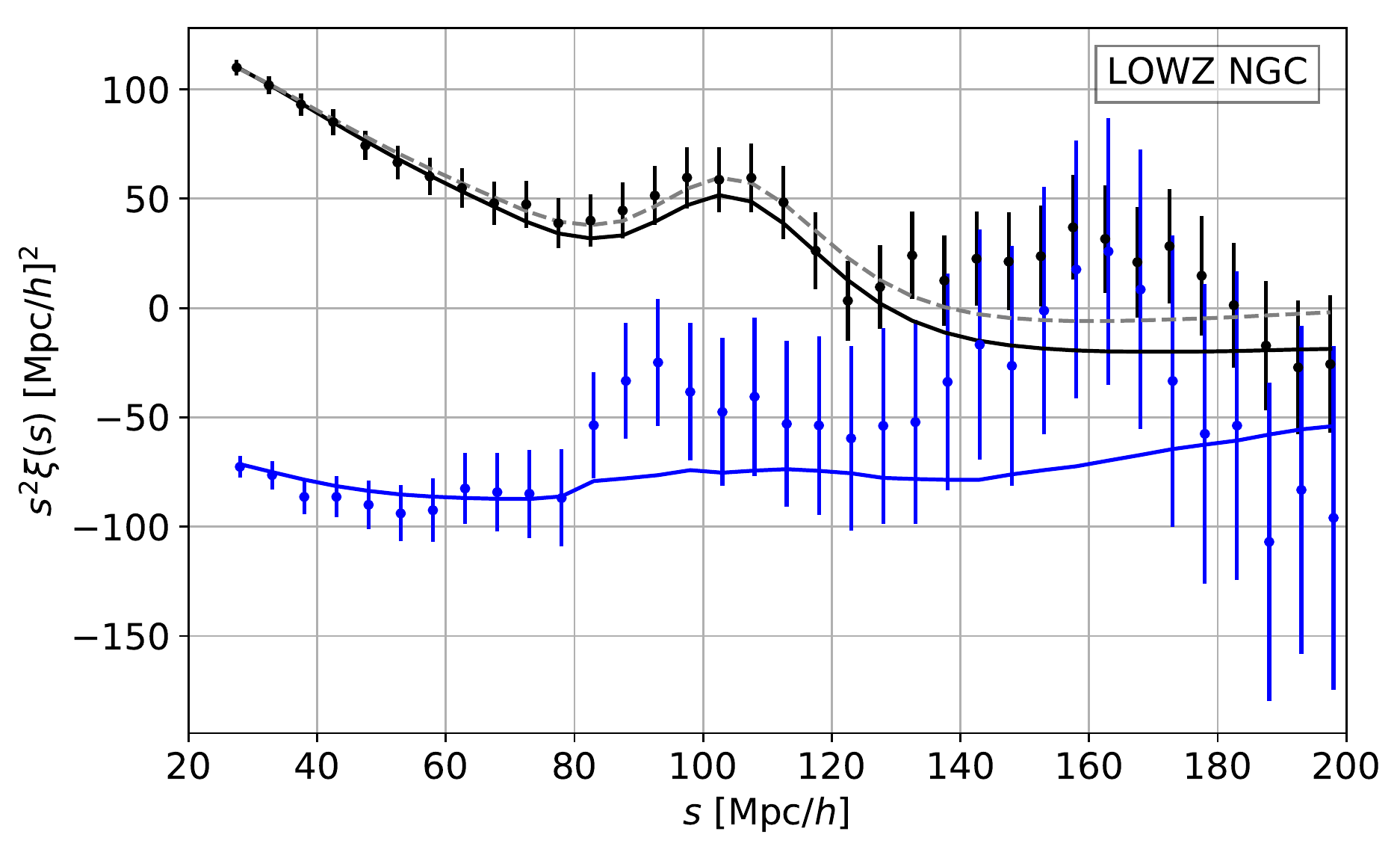}
\includegraphics[width=0.49\textwidth]{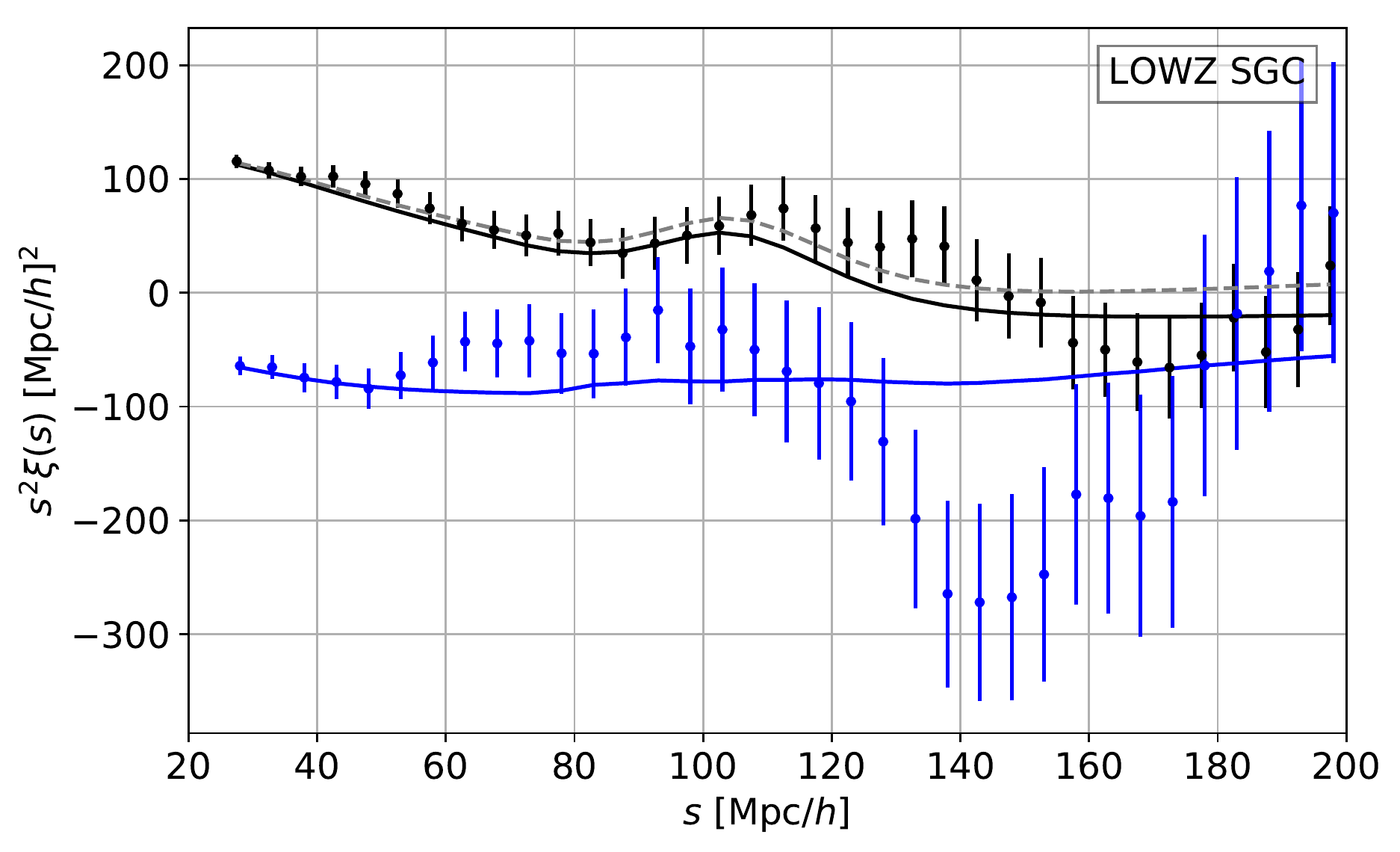}
\caption{\small Best fits of BOSS multipoles. The individual min $\chi^2$ are respectively, for CMASS NGC, CMASS SGC, LOWZ NGC and LOWZ SGC: $64$, $88$, $61$ and $68$. The total min $\chi^2 / {\rm d.o.f.}$ is $280/(2 \cdot 72 + 2 \cdot 70 - 5 - 4 \cdot 6)=1.10$, for a corresponding $p$-value of $0.14$. The multipoles are shown with a slight shift in $s$ for clarity. 
In dashed grey line is the fit adding the smooth broadband systematic corrections to the monopole, that does not affect the cosmological constraints. 
} 
\label{fig:boss_bestfit}
\end{figure}

Overall, the fits on simulations look acceptable.
The monopole displays small residuals. 
Around the BAO peak, we notice a similar trend, although not so significant, in the residuals of boxes A, B, F, G.
First, we remind that those boxes are not independent but they actually originate from the same realization, while indeed the shape of the residuals is different in box D, which originates from a different realization.
Second, we observe that the correlations are important, therefore one should interpret the residuals with care.
There are strong positive correlations among the $s$ bins. On average, in the monopole, we find about $90\%$, $80\%$ and $70\%$ for the first, second, and third diagonals above the main, respectively.
As such, the best fit stays consistently on one side of the data for consecutive bins. We checked that using only the diagonal of the covariance instead produces a closer fit around the BAO peak. This suggests that there are no significant theory systematics.
Turning to the quadrupole, we notice large deviations on large scales, especially for box D.
To dissipate doubts about the potential role of systematics coming from this regime (EFTofLSS systematics are unexpected on theoretical grounds in this regime), we have fitted this box with a large-scale cut -- $s_{\rm max} = 150 \, \Mpcinvh$ -- finding no appreciable shifts on the cosmological parameters. Last, we present the best fit on patchy lightcone mocks, which appears to show no significant residuals. 

On the data, the first aspect to notice is that the monopole data seems to be systematically high with respect to the best fit model on large scales.
For instance, for CMASS NGC, the monopole is always positive and the correlation function cannot satisfy the integral constraint.
This is a known broadband observational systematic. As suggested in~\cite{Chuang:2013wga}, it can be corrected by adding to the correlation function monopole the function $A(s) = a_0 + a_1 / s + a_2 / s^2$.
We have checked that, by adding $A(s)$, the $\chi^2$ of the fit is marginally improved by about $6.2$, while the  posteriors on the cosmological parameters are not affected.
Actually, the improvement on $\chi^2$ comes largely from the CMASS NGC skycut, with a $\Delta \chi^2 = 3.9$, while the other skycuts give $\Delta \chi^2 \simeq 0.7$ each.
In fact, the monopole data of the other skycuts seem to display oscillating features on large scales, rather than a smooth broadband effect.

More worrisome residuals come from features in the quadrupole, in particular on CMASS SGC, whose best fit has a high $\chi^2 / d.o.f. = 87.5 / (72-5-6)=1.43$, corresponding to a rather low $p$-value of $\sim 0.015$~(\footnote{We count the degrees of freedom as the number of data points minus 5 cosmological parameters (as fixing $\omega_b$ instead of letting it vary inside the BBN prior does not change the $\chi^2$) minus 6 EFT parameters. Note that we calculate the $p$-value taking the data points as independent, which is an approximation.}).
To check that this feature (which could represent a systematic error) does not affect our determination of the cosmological parameters, we combined at the level of the pair counts CMASS NGC and CMASS SGC, which reduces the oscillating residuals, as done in~\cite{Ross:2016gvb}. 
Running MCMC chains on the single CMASS sample plus the LOWZ NGC and LOWZ SGC skycuts, we find the same posteriors on the cosmological parameters, with only negligible shifts: the largest one is on $\omega_{cdm}$, of $0.2 \sigma$. 
For the multipoles, the $\chi^2 / d.o.f. $ on the CMASS sample is now $67.5/(72-5-6)=1.10$, with corresponding $p$-value of $0.26$, while on the whole BOSS data set, CMASS plus LOWZ NGC and LOWZ SGC, we get $\chi^2 / d.o.f. = 1.04$, with corresponding $p$-value of $0.35$.
We therefore conclude that these potential undetected systematics on the data do not affect the determination of the cosmological parameters, and the best fit provides a good description of the data.

\section{Line-of-sight selection effects of galaxies}\label{app:selection}

As first recognized in~\cite{Hirata:2009qz}, there is a potential systematic effect when measuring the galaxy correlations in redshift space.
This is due to the alignment of galaxies with large-scale tidal fields, and a selection effect which prefers galaxies viewed down the long axis: as a result, Fourier modes along the line of sight are suppressed.
This is a small effect, which is expected to contaminate the dependence on $f$ at linear order at the $4-8\%$ level or less~\cite{Hirata:2009qz}. Since it mimics the angular dependence of linear term in $f\mu^2$, it is difficult to remove at the level of the dataset.

This, and similar effects that take the name of line-of-sight selection effects, can be understood by realizing that the true observable is not just the density in redshift space, but a combination of the density in redshift space and other quantities that carry vector indices, such as the ellipticity, with the indices contracted with the line of sight unit vector. Of course, how large {they} are and how much weight is in these additional quantities strongly depends on the experimental setting. However, these effects can be described extending the bias expansion by allowing for all the ordinary EFTofLSS-operators carrying a vector index to have that vector index contracted with the line of sight unit vector. At linear level, one has:
\begin{equation}
  \delta_{g, r, {\rm obs}}(\vk) = (b_1 + f \mu^2) \delta_m(\vk) +  A s_{ij}(\vk) \hat{z}^i \hat{z}^j = (b_1 + f \mu^2) \delta_m(\vk) +A \((\hat{k} \cdot \hat{z})^2 - \frac{1}{3} \) \delta_m(\vk) \, ,
\end{equation}
where $\hat{z}$ is unit vector in the direction of the line of sight and $s_{ij}=\left(\partial_i\partial_j-\frac{1}{3}\delta_{ij}\partial^2\right)\phi$, with $\phi$ being the Newtonian gravitational potential.
Here, $\delta_{g,r,{\rm obs}}$ is the redshift-space {\it observed} galaxy overdensity, which is, in principle, different than the  redshift-space galaxy overdensity, $\delta_{g,r}$,  and $\delta_m$  is the matter overdensity.
The term in $A$ represents the contribution at linear order of line-of-sight selection effects. For extensions at non-linear order, see for example~\cite{Desjacques:2018pfv}.
Accounting for the term in $A$, the observed galaxy power spectrum in redshift space is therefore, at linear level,
\begin{equation}
  P_{g, r, {\rm obs}}(k) = \left[ b_1 - \frac{A}{3} + (f + A) \mu^2 \right]^2 P_m(k) \, .
\end{equation}
It follows that there is a potential systematic effect of relative size $A/f$.

One can estimate the size of $A$ following~\cite{Hirata:2009qz}.
The coefficient $A$ can be written as $A = 2 B \eta \chi$, where $B$ is a coefficient that measures the intrinsic alignment of galaxies (due to a large-scale tidal field), and $\eta \chi$ is a selection-dependent coefficient depending on the galaxy orientation.
A recent measurement on BOSS data performed in~\cite{Martens:2018uqj} finds that $B \simeq - 0.03$, in agreement with the estimate of~\cite{Hirata:2009qz}, which also estimates that $\eta \chi \sim 0.2$.
Given that the growth rate at the BOSS redshifts is $f(0.3) \simeq 0.68$ and $f(0.5) \simeq 0.75$, the relative systematic error due to these selection effects can be of order $A /f \sim 2\%$.

It is clear that this potential systematics needs to be included in our data analysis, since we have seen that it is very difficult to subtract from the data and therefore it is not included in the mock catalogs. 
We extend the model to non-linear scales by adding the line-of-sight selection terms order by order in perturbation theory:
\begin{equation}\label{eq:non-linear-selection}
    \delta_{g,r,{\rm obs}}^{(n)}(\vk) =  \delta_{g,r}^{(n)}(\vk) + A \( (\hat{k} \cdot \hat{z})^2 - \frac{1}{3} \) \delta_m^{(n)}(\vk)  \, ,
\end{equation}
where $n$ denotes the $n$-th order in perturbation theory.
We do the data analysis adding $A$ as an additional parameter~\footnote{More exactly, we add one additional parameter per skycut. In the following, when quoting results on $A$, we refer to the one of CMASS NGC.}. Considering the size of errors in~\cite{Martens:2018uqj} and the estimates of~\cite{Hirata:2009qz}, we impose a Gaussian prior on $A$ with mean $-0.05$ and standard deviation $0.05$, allowing for a ``pessimistic'' case in which the systematics amounts to a $\sim 7\%$ correction of the dependence on $f$.
In principle, we should add to Eq.~\eqref{eq:non-linear-selection} the contributions from the non-linear line-of-sight selection biases~\cite{Desjacques:2018pfv}.
However, assuming, as expected, that the physical prior on these additional parameters is comparable to the one on $A$, the effect of the non-linear corrections is very small, and we neglect it.

\begin{figure}[h!]
    \centering
    \includegraphics[width=0.9\textwidth]{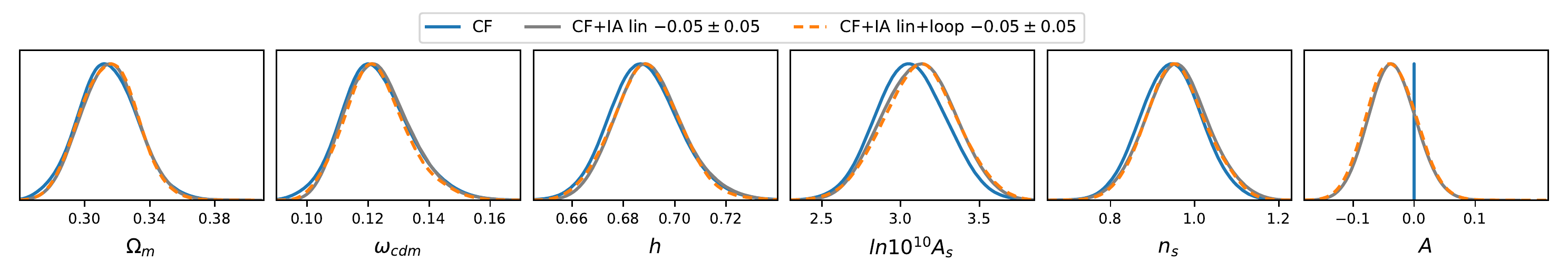}
    \includegraphics[width=0.9\textwidth]{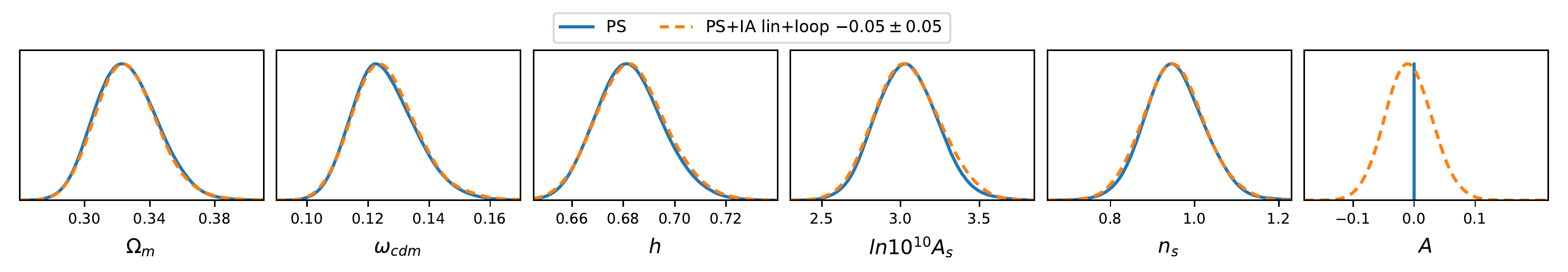}\\

    \caption{\small Results of the fit to the BOSS CF and PS including line-of-sight selection effects. In the CF, we also show the posteriors obtained when including the line-of-sight selection effects only at tree-level for reference, which is the dominant one. }
    \label{fig:IA}
\end{figure}

We show our results in Fig.~\ref{fig:IA}, for both the power spectrum and correlation function fitting all BOSS skycuts.
We show the model without any line-of-sight selection effect compared to the models in which the selection effect, measured by the $A$ parameter, is added only at linear or at linear plus loop level.
First, we notice that the contribution of the term in $A$ at loop level is negligible, justifying us to neglect the inclusion of the full set of non-linear line-of-sight selection effects.
Second, it is apparent that in both analyses with selection effects, we measure the $A$ parameter without being completely prior-dominated, which shows that the linear-level degeneracy is mildly broken: we have $A = -0.037 \pm 0.040$ for the correlation function and $A = -0.009 \pm 0.039$ for the power spectrum. 
This can be understood as follow. 
Without selection effects, the growth rate $f$ is determined mainly by $\Omega_m$, that is well measured thanks to the shape and geometrical information (see e.g.~\cite{DAmico:2019fhj}). 
This leaves us with two `broadband' parameters, the amplitude $A_s$ and the galaxy bias $b_1$, that are measured by fitting both the monopole and the quadrupole. 
Including selection effects, a degeneracy is introduced by the new parameter $A$ at linear level, that is mildly broken by the loop correction that depends on $A_s$, $b_1$ and $A$, with a different functional form.

More importantly, the shift in cosmological parameters is negligible.
In particular, the change in the posteriors for $h,\; n_s$ and $\omega_{\rm cdm}$ are barely visible.
For $\ln (10^{10} A_s)$, the peak is shifted by $0.24\sigma$ for the CF while is barely shifted for the PS, and the error bar is increased by about $5\%$ and $10\%$, respectively.
We can estimate the shift in $A_s$ in the following way.
The presence of $A$ allows $A_s$ to shift by about $2 f A$. 
Given the measured value of $A$, the relative shift on $A_s$ for the CF (for the PS) is then around $5\%$ ($1\%$), which is in agreement with the observed shift in $\ln (10^{10} A_s)$ of about $0.05$ ($0.01$). 
We conclude that the FS analysis of BOSS data is robust to line-of-sight selection effects, assuming, as we expect, that the additional nonlinear line-of-sight selection biases have similar physical priors.

\bibliographystyle{JHEP}
\bibliography{references}

\end{document}